\documentclass[11pt,a4paper]{article}
\pdfoutput=1
\usepackage{jheppub}
\usepackage{braket}
\usepackage[english]{babel}
\usepackage[utf8]{inputenc}
\usepackage{amsmath}
\usepackage{array, url}
\usepackage{fancyhdr}
\usepackage{amsfonts}
\usepackage{amsthm}
\usepackage{mathtools}
\usepackage{bm}
\usepackage{titlesec}
\usepackage{graphicx}
\usepackage{color}
\usepackage{tensor}
\usepackage{mathrsfs}
\usepackage{etoolbox}
\usepackage{tikz}
\usetikzlibrary{positioning}
\usepackage{comment}
\usepackage{tensor}
\usepackage[title]{appendix}
\usepackage{enumerate}
\usepackage{dsfont}
\usepackage{import}
\usepackage{svg}
\usepackage{makecell}
\usepackage{float}
\usepackage{fancybox}

\numberwithin{equation}{section}

\usepackage{amsmath}

\DeclareMathOperator\arcosh{arcosh}
\DeclareMathOperator\tr{Tr}

\newcommand{\1}{\mathds{1}}

\usepackage{xcolor}

\renewcommand*\thesection{\arabic{section}}
\definecolor{mathematica1}{rgb}{0.368417, 0.506779, 0.709798}
\definecolor{mathematica2}{rgb}{0.880722, 0.611041, 0.142051}
\definecolor{mathematica3}{rgb}{0.560181, 0.691569, 0.194885}
\definecolor{mathematica4}{rgb}{0.922526, 0.385626, 0.209179}
\definecolor{mathematica6}{rgb}{0.772079, 0.431554, 0.102387}
\definecolor{pink}{rgb}{1, 0.5, 0.5}

\newcommand{\be}{\begin{equation}}
\newcommand{\ee}{\end{equation}}

\renewcommand{\d}{\mathrm{d}}

\newcommand{\Secref}[1]{Section \ref{#1}}
\newcommand{\Appref}[1]{Appendix \ref{#1}}
\newcommand{\Figref}[1]{Figure \ref{#1}}

\usepackage{tabu}

\newcommand{\overbar}[1]{\mkern 1.5mu\overline{\mkern-1.5mu#1\mkern-1.5mu}\mkern 1.5mu}

\title{Topological Einstein gravity as Kodaira--Spencer gravity}

\author[a]{Johanna Erdmenger,}
\author[a,b]{Jonathan Karl,}
\author[a]{Jani Kastikainen,}
\author[a]{Ren\'e Meyer}
\author[a]{and \newline Henri Scheppach}

\affiliation[a]{Institute for Theoretical Physics and Astrophysics and Würzburg-Dresden Cluster of Excellence
ct.qmat, Julius-Maximilians-Universit\"at W\"urzburg, Am Hubland, 97074 Würzburg, Germany}

\affiliation[b]{Department of Physics, Princeton University, Princeton, NJ 08544}

\emailAdd{jonathan.karl@uni-wuerzburg.de}
\emailAdd{jani.kastikainen@uni-wuerzburg.de}
\emailAdd{henri.scheppach@uni-wuerzburg.de}

\abstract{As a contribution towards quantizing three-dimensional gravity, we show at the classical level that Euclidean three-dimensional Einstein gravity with a negative cosmological constant is uplifted to the $SU(2)$-invariant sector of Kodaira--Spencer gravity on a Calabi--Yau three-fold. Kodaira--Spencer gravity appears in the target space description of the B-model topological string theory and describes deformations of a complex structure. We prove that given a reference solution of Einstein gravity in the first-order formulation, a second off-shell configuration uplifts to a unique complex structure deformation in six dimensions. If the configuration satisfies Einstein's equations, the complex structure deformation is integrable, i.e.~a solution of Kodaira--Spencer gravity. We demonstrate the uplift explicitly for Ba\~nados solutions. Our construction embeds three-dimensional gravity into topological string theory and AdS$_3$/CFT$_2$ duality into twisted holography.}

\begin{document}
\maketitle

\section{Introduction}

A central goal in theoretical physics is the formulation of a consistent theory of quantum gravity. In recent years, significant progress in this regard was achieved in lower-dimensional models such as Jackiw--Teitelboim (JT) gravity \cite{TEITELBOIM198341,JACKIW1985343} and three-dimensional (3D) Einstein gravity, both of which are free of dynamical gravitons, making them more amenable to quantization than their higher-dimensional counterparts. While the path integral of JT gravity was exactly solved and is holographically related to a matrix model \cite{Saad:2019lba,Stanford:2019vob}, an analogous treatment of 3D Einstein gravity remains elusive so far. Progress in this direction was reported in \cite{Eberhardt:2019ywk} in a string theory approach to AdS$_3$/CFT$_2$. Recently, a quantization of 3D gravity with negative cosmological constant was performed, leading to Virasoro TQFT \cite{Collier:2023fwi,Collier:2024mgv}, which is also related to Turaev--Viro CFT \cite{Hartman:2025cyj,Hartman:2025ula}.

In the presence of a negative cosmological constant, 3D Einstein gravity can be reformulated as a Chern--Simons (CS) theory at the classical level \cite{Achucarro:1986uwr,Witten:1988hc}. This provides an useful avenue to the quantization of gravity since the quantization of CS theory is better understood than the general case \cite{Witten:1988hf,Elitzur:1989nr, Axelrod:1989xt}, given that it is a standard gauge theory, as well as topological. The gauge theory formulation has proven to be power\-ful for decoding universal properties of the AdS$_3$\slash CFT$_2$ duality \cite{Cotler:2018zff} and of quantum gravity in general \cite{Carlip:2004ba}. However, it is important to keep in mind that the relation between gravity and gauge theory at the quantum level comes with well known caveats concerning the invertibility of the dreibein (non-degeneracy of the metric) and the sum over topologies \cite{Witten:1988hc,Witten:1989ip}.

A promising avenue is to embed 3D Einstein gravity into a gravitational setup with well-understood quantization. In this work, we follow this avenue and reformulate three-dimensional Einstein gravity with a ne\-ga\-tive cosmological constant as Kodaira--Spencer (KS) gravity on a six-dimensional principal bundle. KS gravity appears as the target space theory of the B-model topological string theory, where its loop expansion captures the genus expansion of the topological string on a Calabi--Yau (CY) three-fold \cite{Bershadsky:1993ta,Bershadsky:1993cx,Costello:2012cy}. In addition, KS gravity on the principal bundle has a precise holographic dual description as the large-$N$ limit of a gauged $U(N)$ beta-gamma system \cite{Beem:2013sza}, a chiral Euclidean conformal field theory (CFT) in two dimensions, via twisted holography \cite{costelloGaiotto2018twisted}. Twisted holography is a conjectured duality between the topological B-model on the deformed conifold $\text{AdS}_3\times S^3$ and the beta-gamma system.

Our analysis provides a map between both the off-shell and on-shell degrees of freedom of 3D Einstein gravity on the base $B_3$ and KS gravity on the principal bundle $B_3\times S^3$ in Euclidean signature. This mapping is a first necessary step toward relating the two theories at the quantum level and toward embedding 3D quantum gravity into the framework of the topological string and twisted holography. In this way, a precise holographic dual of pure 3D gravity, realized as a possible subsector of the beta-gamma system via twisted holography \cite{costelloGaiotto2018twisted}, may potentially be used to address foundational questions about AdS$_3$/CFT$_2$ duality, such as the need for a sum over topologies and ensemble averaging \cite{Cotler:2020ugk,DiUbaldo:2023qli,Belin:2023efa,deBoer:2023vsm,Jafferis:2024jkb,Boruch:2025ilr}.

Relations between 3D Einstein gravity in its CS formulation and KS gravity were established in the past. For example, in \cite{Dijkgraaf:2007sx} two-dimensional KS gravity on the Riemann surface $\Sigma_2$ was related to 3D Einstein gravity on $\Sigma_2\times \mathbb{R}$. Similarly, a web of closely connected results appear in the context of the unification of different topological gravities into the topological M-theory in seven dimensions \cite{Dijkgraaf:2004te} (see \cite{Nekrasov:2005bb} for a review). Topological M-theory was argued to be described by seven-dimensional Hitchin gravity \cite{Hitchin:2000jd}, which in turn is dimensionally reduced to the B-model topological string theory in six dimensions and to Einstein gravity in three dimensions \cite{Dijkgraaf:2004te} (see also \cite{Chagoya:2020uog} for a discussion of the latter). In a similar vein, 3D Einstein gravity coupled to a two-form was obtained as the dimensional reduction of 6D Hitchin gravity \cite{Hitchin:2000jd} in \cite{Herfray:2016std}. These results indicate a direct, but as yet unformulated, relationship between six-dimensional KS gravity and 3D Einstein gravity, which we make precise in this work.

KS gravity is a diffeomorphism invariant theory that describes deformations of a complex structure on a differentiable manifold. An almost complex structure $J$ is a real automorphism of the tangent bundle, which squares to minus the identity. A complex structure requires that $J$ satisfies an additional integrability condition, the vanishing of the Nijenhuis tensor. The equation of motion of KS gravity simply states that the deformation $\alpha$ is integrable or equivalently that the resulting deformed complex structure $J + 4\,\text{Im}\,\alpha$ is integrable. The equation of motion is a first-order differential equation for $\alpha$ \cite{newlander1957complex}, the Kodaira--Spencer (KS) equation. Because the complex structure is independent of a Riemannian structure, KS gravity does not involve the metric tensor. However, similarly to Einstein gravity in three dimensions, KS gravity is also a topological theory, i.e.~all of its on-shell degrees of freedom are pure diffeomorphisms. Hence there are no propagating degrees of freedom and all solutions are locally diffeomorphic to each other.

The starting point of our reformulation of 3D Einstein gravity as KS gravity in Euclidean signature is its CS formulation using an $\mathfrak{sl}(2,\mathbb{C})$-valued 1-form gauge field $A$ \cite{Achucarro:1986uwr,Witten:1988hc,Witten:1989ip}. This gauge field encodes all degrees of freedom of the three-dimensional metric and of an independent, a priori torsionful, connection. The flatness of $A$, i.e.~the equation of motion of CS theory, is equivalent to three-dimensional Einstein's equations together with a torsion-free constraint on the connection. The local $SO(3)$ rotation symmetry of Einstein gravity is realized as an $SU(2)$ gauge symmetry acting on $A$.

We relate the fundamental fields of 3D Einstein gravity and KS gravity by uplifting the CS gauge field on the three-dimensional manifold $B_3$ to a six-dimensional 1-form $\mathcal{A}$ on $B_3\times S^3$ whose real part is a $SU(2)$ principal bundle connection. In this bundle description of gauge theory, the symmetries of CS theory, three-dimensional diffeomorphisms and gauge transformations of $A$, are combined into a subgroup of six-dimensional diffeomorphisms preserving $B_3\times S^3$ (see \cite{Nakahara:2003nw} for a review). The uplift $\mathcal{A}$ of the gauge field $A$ can be used to define a canonical almost complex structure on the principal bundle which is integrable exactly when $A$ is flat, i.e.~when the metric derived from $A$ satisfies Einstein's equations in three dimensions \cite{zentner2013integrable,Herfray:2016std}.\footnote{There are many constructions of almost complex structures on fiber bundles in the literature. For a base with metric $g$,
they were
introduced in the seminal work \cite{Dombrowski1962} and later extended in \cite{aguilar1996isotropic}. The integrability conditions for these structures were shown to 
imply constant curvature for the metric $g$, and vice-versa.} We identify deformations $\alpha$ of this complex structure as the field of KS gravity.

Our main result is an explicit off-shell uplift of the pair of CS gauge fields $(A,A')$ on $B_3$ to a complex structure deformation $\alpha$ on $B_3\times S^3$. The resulting $\alpha$ is invariant under six-dimensional diffeomorphisms, which correspond to the right-action of $SU(2)$ on the bundle. The first gauge field $A$ is required to be on-shell in order to define a background complex structure on $B_3\times S^3$, while $A'$ and its uplift $\alpha$ remain off-shell. We show that when $A'$ additionally satisfies the flatness conditions of 3D CS theory, $\alpha$ solves the Kodaira--Spencer equation, thereby matching the solution space of Euclidean 3D Einstein gravity with a negative cosmological constant with an $SU(2)$-invariant subsector of the solution space of KS gravity. Moreover, we show how the symmetries of 3D gravity are unified into the six-dimensional diffeomorphism symmetry of the Kodaira--Spencer equation.

Together with this uplift, we show that all $SU(2)$-invariant complex structure deformations on $B_3\times S^3$ can be consistently dimensionally reduced to a single $\mathfrak{sl}(2,\mathbb{C})$-valued gauge field $A$ on the base.\footnote{On top of $SU(2)$-invariance, there is an additional condition stating that the deformed almost complex structure must define a valid notion of horizontality for vector fields on the bundle.} In this sense, 3D Einstein gravity is the $SU(2)$-invariant zero-mode sector of KS gravity on $B_3\times S^3$. Following \cite{Zeng:2023qqp}, more general $SU(2)$-non-invariant deformations can be expanded in modes of the $S^3$ and lead to higher Kaluza--Klein (KK) modes on $B_3$. These KK modes are required for the UV completion of pure 3D Einstein gravity in topological string theory, and for the existence of a holographic dual of 3D Einstein gravity via twisted holography.\footnote{Anomaly cancellation of quantum KS gravity requires coupling to six-dimensional holomorphic Chern--Simons theory \cite{Costello:2019jsy} which produces further three-dimensional fields upon dimensional reduction. In this work, we focus solely on the classical relationship between pure KS and 3D Einstein gravities. The study of the coupling to holomorphic CS theory is left for future work.}

Our classical uplift provides necessary ingredients towards an extension of the relation between the two theories to the quantum level. It is well known that quantum KS gravity not only requires a background complex structure for its definition, as it does at the classical level, but a full background Calabi--Yau structure: a holomorphic volume form is needed to write down an action while a K\"ahler metric is used for gauge fixing in perturbation theory \cite{costelloGaiotto2018twisted}. We show that a Calabi--Yau structure on $B_3\times S^3$ emerges automatically from the reference solution $A$ of Einstein's equations on the base $B_3$. In particular, $A$ can be used to define a Ricci flat K\"ahler metric on $B_3\times S^3$ which coincides with the deformed conifold \cite{Candelas:1989js}. This sets up a route for a full path integral over the off-shell fields $A'$ and $\alpha$, as well as relating it to the genus expansion of the topological string.

\subsection{Summary of results}

Building on the results of \cite{Herfray:2016std,zentner2013integrable}, we prove a classical relation between Euclidean three-dimensional Einstein gravity with a negative cosmological constant in its first-order formulation and the $SU(2)$-invariant sector of KS gravity on a six-dimensional $SU(2)$ principal bundle.

More specifically, given a reference solution of three-dimensional gravity $(E^i,\omega^{ij})$, where $E^i$ denotes the dreibein and $\omega^{ij}$ the spin connection, we show there is an in\-ver\-ti\-ble map between off-shell degrees of freedom $(E'^i,\omega'^{ij})$ and $SU(2)$-invariant complex structure deformations $\alpha_{i}^{\;\;\,j}$ in six dimensions,
\begin{figure}[H]
\noindent
\makebox[\linewidth]{
  \hspace*{0.1125\linewidth}

  \fbox{
    \begin{minipage}{\dimexpr0.296\linewidth-2\fboxsep-2\fboxrule\relax}
\centering
off-shell dreibein $E'^i$\\
and spin connection $\omega'^{ij}$\\ in three dimensions
    \end{minipage}
  }\hfill
    \begin{minipage}{\dimexpr0.183\linewidth-2\fboxsep-2\fboxrule\relax}
\centering
$\longleftrightarrow$
    \end{minipage}
  \hfill
  \fbox{
    \begin{minipage}{\dimexpr0.296\linewidth-2\fboxsep-2\fboxrule\relax} 
\centering
$SU(2)$-invariant complex structure deformation $\alpha_{i}^{\;\;j}$ in six dimensions.
    \end{minipage}
  }

  \hspace*{0.1125\linewidth}
}
\end{figure}
Furthermore, when $(E'^i,\omega'^{ij})$ are on-shell and solve three-dimensional Einstein's equations with a negative cosmological constant, then $\alpha_{i}^{\;\;\,j}$ is a solution of Kodaira--Spencer gravity,
\begin{figure}[H]
\noindent
\makebox[\linewidth]{
  \hspace*{0.1125\linewidth}

  \fbox{
    \begin{minipage}{\dimexpr0.296\linewidth-2\fboxsep-2\fboxrule\relax}
\centering
Einstein's equations with a negative cosmological constant
    \end{minipage}
  }\hfill
    \begin{minipage}{\dimexpr0.183\linewidth-2\fboxsep-2\fboxrule\relax}
\centering
$\longleftrightarrow$
    \end{minipage}
  \hfill
  \fbox{
    \begin{minipage}{\dimexpr0.296\linewidth-2\fboxsep-2\fboxrule\relax}
\centering
    Kodaira--Spencer equation in six dimensions.
    \end{minipage}
  }

  \hspace*{0.1125\linewidth}
}
\end{figure}
\noindent Our approach is based on the CS formulation where the fields $(E^i,\omega^{ij})$ of 3D Einstein gravity on a three-manifold $B_3$ are combined into a single $\mathfrak{sl}(2,\mathbb{C})$-valued Chern--Simons gauge field $A$ such that Einstein's equations are equivalent to the flatness of $A$. We prove that given a single reference solution $A$ of Einstein's equations on the three-manifold $B_3$, a second off-shell gauge field $A'\neq A$ determines a unique complex structure deformation $\alpha$ on the $SU(2)$ principal bundle $B_3 \times S^3$.

\paragraph{Complex structures on principal bundles.} Our results are based on the definition of an integrable almost complex structure on any $SU(2)$ principal bundle from a flat $\mathfrak{sl}(2,\mathbb{C})$-valued connection \cite{zentner2013integrable,Herfray:2016std}. Any such connection $\mathcal{A}$ determines an almost complex structure $J$ on the bundle $B_3\times S^3$ such that $J(\mathcal{A}) = i\,\mathcal{A}$, in other words, $\mathcal{A}$ is a $(1,0)$-form with respect to $J$. When the $(1,0)$-form is flat $\d \mathcal{A} + \mathcal{A}\wedge \mathcal{A} = 0$, $J$ is integrable, which allows to cover the bundle with complex coordinates $w^i$. The implication also holds in the opposite direction when $J$ is invariant under the right-action of $SU(2)$ on the bundle: any $SU(2)$-invariant $J$ implies that the corresponding $(1,0)$-form $\mathcal{A}$ is flat.\footnote{There is a further technical condition on $J$ that requires it to have no invariant vertical subspace.} This equivalence between flatness and integrability has been proven in \cite{zentner2013integrable}, but we give a new streamlined proof of the fact.

Three-dimensional Einstein gravity on the base automatically determines an almost complex structure on the bundle. This is because a gauge field $A $ on $B_3$ uplifts to a unique connection $\mathcal{A}$ on $B_3\times S^3$ via the formula
\begin{equation}
    \mathcal{A} = G^{-1}AG + G^{-1}\d G\,,
    \label{eq:curly_A_summary}
\end{equation}
where $G \in SU(2)$. The flatness of $A$ uplifts to the flatness of the connection $\mathcal{A}$. In addition, because $\mathcal{A}$ arises from an uplift, it is $SU(2)$-equivariant, meaning it transforms homogeneously under the right-action diffeomorphism of $SU(2)$ (see \eqref{eq:equivariance_of_A} for the precise definition). This equivariance implies that the corresponding almost complex structure is $SU(2)$-invariant so that Einstein's equations on the base are equivalent to its integrability.

We show that a solution of 3D gravity on the base not only determines a complex structure on the bundle, but also a full Calabi--Yau structure consisting of a compatible holomorphic volume form $\Omega$ and a Ricci flat Kähler metric $h$ given by
\begin{equation}
\begin{gathered}
    \Omega = -4 \tr{(\mathcal{A}\wedge\mathcal{A}\wedge\mathcal{A})}\,,\\ h_{ij} = K''(\kappa)\tr{(\partial_i \mathcal{M} \,\mathcal{M}^\dagger)} \tr{(\mathcal{M}\,\overbar{\partial}_j\mathcal{M}^\dagger )}+K'(\kappa) \tr{(\partial_i \mathcal{M}\,\overbar{\partial}_j \mathcal{M}^\dagger )}\,,
\end{gathered}
\end{equation}
where $\mathcal{M}$ is defined by $\mathcal{A}=\mathcal{M}^{-1} \d \mathcal{M}$, $K$ is a function of $\kappa= \tr{(\mathcal{M} \mathcal{M}^\dagger)}$ and the trace is evaluated in the fundamental representation. The function $K$ can be chosen such that $h$ is Ricci flat (see \eqref{eq:Ricci_flat_K}) in which case the principal bundle $B_3\times S^3$ equipped with $(J,\Omega,h)$ is a Calabi--Yau manifold that coincides with the deformed conifold.

\paragraph{Relation between 3D Einstein and KS gravities.} Given a reference solution $A$ of Einstein's equations that determines a background complex structure on the bundle, a second off-shell gauge field $A'$ uplifts to a unique deformation $\alpha$ of the complex structure. Let $\mathcal{A}' = \mathcal{A}'^i\,\tau_i$ be the uplift of the gauge field $A'$ via the general formula \eqref{eq:curly_A_summary} where $\tau_i$ are generators of $\mathfrak{su}(2)$. Then the components $\alpha_{i}^{\;\,\, j}$ of the deformation in complex coordinates $(w^i,\overbar{w}^i)$ of the background are given by
\begin{equation}
    \mathcal{A}'^i= \Sigma'^i_j\,(\d w^j + \alpha_{k}^{\;\,\, j}\d\overbar{w}^{k})\,.
\end{equation}
Because $\mathcal{A}'$ arising from an uplift is $SU(2)$-equivariant, it follows that $\alpha$ is $SU(2)$-invariant (see \eqref{eq:alpha_SU2_invariance} for the precise definition). This provides an explicit off-shell mapping between 3D Einstein gravity and the $SU(2)$-invariant sector of 6D KS gravity on $B_3\times S^3$. It follows that the KS equation for $\alpha$ is equivalent to $\mathcal{A}'$ and $A'$ being flat
\begin{equation*}
    \partial_{[\mu}A'^i_{\nu]}+\frac{1}{2}\,\varepsilon^{i}_{\;\,jk}\,A'^j_{[\mu} \,A'^k_{\nu]}= 0\,\;\; \leftrightarrow \;\; \partial_{[a}\mathcal{A}'^i_{b]}+\frac{1}{2}\,\varepsilon^{i}_{\;\,jk}\,\mathcal{A}'^j_{[a} \,\mathcal{A}'^k_{b]} = 0\;\; \leftrightarrow \;\; \overbar{\partial}_{[j}\, \alpha_{k]}^{\;\;\,i}-\alpha_{[j}^{\;\;\,l}\partial_{\vert l \vert}\,\alpha_{k]}^{\;\;\,i}=0\,.
\end{equation*}
where $\mu = 1,2,3$ label coordinates on the base $B_3$, $a=1,\ldots, 6$ label real coordinates on the bundle $B_3\times S^3$ and $i=1,2,3$ label either $\mathfrak{su}(2)$ indices (for $A'$ and $\mathcal{A'}$) or complex coordinates $(w^i,\overbar{w}^i)$ on $B_3\times S^3$ (for $\alpha$). This establishes the relation between Einstein's equations on $B_3$ and KS equations on $B_3\times S^3$.

\paragraph{Matching of symmetries.} 3D Einstein gravity in the first-order formulation is invariant under three-dimensional diffeomorphism and local $SO(3)$ rotations. Under these symmetries, the $\mathfrak{sl}(2,\mathbb{C})$-valued gauge field $A$ transforms as a 1-form and as an $SU(2)$ gauge field respectively. At the level of the Einstein's equations $\d A + A\wedge A = 0$, these symmetries are realized as three-dimensional diffeomorphism covariance\footnote{By covariance we mean that if $A$ is a solution, then so is the pull-back of $A$ with a diffeomorphism.} and gauge invariance. The symmetry transformations on the base $B_3$ are unified into six-dimensional diffeomorphisms that preserve the principal bundle $B_3\times S^3$ and under which the KS equation is covariant \cite{Bandelloni:1998vp}. In this way, the symmetries of Einstein's equations in three dimensions are realized by the KS equation in six dimensions.

\paragraph{Uplift of Ba\~nados solutions.} We demonstrate our construction explicitly by uplifting a family of solutions of Einstein's equations with a negative cosmological constant known as Ba\~nados geometries \cite{Banados:1998gg}. These solutions are holographically dual to excited states of the dual CFT in which the expectation values of the holomorphic and antiholomorphic stress tensor operators are Schwarzian derivatives of the functions $f(z)$ and $\overbar{f}(\overbar{z})$ respectively.

We take our reference solution $A$ to be the AdS$_3$ solution $f(z) = z$ and $\overbar{f}(\overbar{z}) = \overbar{z}$ in Poincar\'e coordinates $(r,z,\overbar{z})$, corresponding to a CFT state with vanishing stress tensor expectation value. The uplift of this solution produces a complex structure with complex coordinates $(w_1,w_2,w_3)$ and $(\overbar{w}_1,\overbar{w}_2,\overbar{w}_3)$ which are functions of the Poincar\'e coordinates on $B_3$ and Euler angles on $S^3$. We take the second solution $A'$ to be a general Ba\~nados geometry with $f(z) \neq z$ and $\overbar{f}(\overbar{z}) \neq \overbar{z}$ and show that it uplifts to the complex structure deformation
\begin{align}
        \alpha_i^{\;\,\, j}(w,\overbar{w})\,\d \overbar{w}^i\otimes \partial_j &= \frac{T(z)}{||w||^{6}}\,\overbar{w}_1\overbar{w}_2\,\biggl(\d \overbar{w}_2 \otimes \partial_1+ \d\overbar{w}_1  \otimes \partial_2-\frac{\overbar{w}_2}{\overbar{w}_1}\,\d \overbar{w}_1  \otimes\partial_1-\frac{\overbar{w}_1}{\overbar{w}_2}\, \d \overbar{w}_2  \otimes\partial_2\biggr)\nonumber\\
        &+\frac{T(z)}{||w||^{8}}\frac{\overbar{w}_2^2}{w_1^2}\left(\overbar{w}_1\, \d \overbar{w}_2\otimes \partial_3-\overbar{w}_2 \,\d \overbar{w}_1\otimes\partial_3\right)\,,\label{eq:alpha_explicit_summary}
\end{align}
where $T(z) \equiv \frac{1}{2}\{f(z),z\}$ is the Schwarzian derivative and $||w||\equiv  \sqrt{|w_1|^2+|w_2|^2} $. We check explicitly that this deformation solves the KS equation, that it is $SU(2)$-invariant under the right-action of $SU(2)$ and that it has other properties required by KS gravity.

The paper is structured as follows. In \Secref{sec:review_3d_grav}, we review the relevant aspects of three-dimensional gravity, focusing on its formulation as a Chern--Simons theory. In \Secref{sec:KS_review}, we provide a self-consistent review of complex structure deformations and of Kodaira--Spencer gravity. In \Secref{sec:review_of_complex_structures_on_principal_bundles}, after introducing the basics of $SU(2)$ principal bundles, we show how they can be equipped with a natural almost complex structure following \cite{zentner2013integrable,Herfray:2016std}. Here, we provide a new streamlined proof of the equivalence between flatness of $\mathcal{A}$ and integrability. In \Secref{sec:uplift}, we present our main result: the off-shell uplift of 3D Einstein gravity to KS gravity. In this section, we also prove the equivalence of the equations of motion and derive necessary conditions for a consistent dimensional reduction. Next, in \Secref{sec:banados_uplift}, we demonstrate our results explicitly by uplifting Ba\~nados solutions to KS gravity. Lastly, in \Secref{sec:outlook}, we discuss our results and possible future directions. Moreover, various technical details are relegated to the appendices.

\section{Review of three-dimensional Einstein gravity}\label{sec:review_3d_grav}

We begin this paper by reviewing the relevant aspects of three-dimensional Einstein gravity and its relation to Chern--Simons theory in its first-order formulation. Then we present general solutions of the theory in both formulations when the cosmological constant is negative.

\subsection{Metric and Chern--Simons formulations}\label{subsec:3D_gravity_review}

Consider a three-dimensional manifold $B_3$ equipped with a Euclidean metric $g$ and coordinates $y^\mu$. In the metric formulation, the Euclidean action of Einstein(--Hilbert) gravity is given by
\begin{equation}
    I_{\text{EH}}[g] = -\frac{1}{2\kappa}\int_{B_3}\d^{3}y\sqrt{g}\,\biggl(R+\frac{2}{\ell^2}\biggr) \,,
\label{eq:3D_Einstein_hilbert}
\end{equation}
where $R$ is the Ricci scalar of the Euclidean metric $g_{\mu\nu}$ and $-1\slash \ell^2$ with $\ell>0$ is the negative cosmological constant. The equations of motion for the metric are given by $R_{\mu\nu} - \frac{1}{2}\,R\,g_{\mu\nu} -\frac{1}{\ell^2} g_{\mu\nu} = 0$ which can be compactly written as
\begin{equation}
    \delta^{\mu}_{[\nu}\delta^{\mu_1}_{\rho_1}\delta^{\nu_1}_{\sigma_1]}\,\biggl[R^{\rho_1\sigma_1}_{\mu_1\nu_1}+\frac{1}{\ell^2}\,(\delta_{\mu_1}^{\rho_1}\delta_{\nu_1}^{\sigma_1} - \delta_{\mu_1}^{\sigma_1}\delta_{\nu_1}^{\rho_1})\biggr] = 0\,.
\label{eq:3D_Einstein_equation}
\end{equation}
By using the identity $3!\,\delta^{\mu}_{[\nu}\delta^{\mu_1}_{\rho_1}\delta^{\nu_1}_{\sigma_1]} = \varepsilon^{\mu\mu_1\nu_1}\,\varepsilon_{\nu\rho_1\sigma_1}$ in three dimensions,\footnote{In our conventions, anti-symmetrization of $n$ indices contains a factor of $1\slash n!$ .} \eqref{eq:3D_Einstein_equation} implies
\begin{equation}
    R_{\rho\sigma}^{\mu\nu} = -\frac{1}{\ell^2}\,(\delta^{\mu}_\rho\delta^{\nu}_\sigma - \delta^{\mu}_\sigma\delta^{\nu}_\rho)\,.
    \label{eq:locally_AdS}
\end{equation}
Therefore all solutions to Einstein's equations with a negative cosmological constant in three dimensions are locally AdS$_3$ in an open neighborhood around any point on $B_3$. Einstein gravity in three dimensions is thus topological in this sense.

\paragraph{First-order formulation.} In the metric formulation, the connection $\Gamma^{\rho}_{\mu\nu}$ is not independent from the metric, but fixed to be the unique torsion-free Levi-Civita connection. In order to relate three dimensional gravity to Chern--Simons theory, we will treat $\Gamma^{\rho}_{\mu\nu}$ as an independent field and write the Euclidean metric as
\begin{equation}
	g_{\mu\nu} = \delta_{ij}\,E^{i}_\mu E^{j}_\nu\,,
    \label{eq:metric_vielbein}
\end{equation}
where $E^i_\mu$ is the dreibein. The Einstein's equation \eqref{eq:locally_AdS} for the metric \eqref{eq:metric_vielbein} is then equivalent to
\begin{equation}
    \mathcal{R}_{\rho\sigma}^{\mu\nu} +\frac{1}{\ell^2}\,(\delta^{\mu}_\rho\delta^{\nu}_\sigma - \delta^{\mu}_\sigma\delta^{\nu}_\rho) = 0\,,\quad \Gamma_{[\mu\nu]}^\rho = 0\,,
    \label{eq:Einstein_Palatini_equations}
\end{equation}
where $\mathcal{R}^{\rho}_{\;\;\sigma\mu\nu}=\partial_\mu\Gamma^\rho_{\nu\sigma}-\partial_\nu\Gamma^\rho_{\mu\sigma}+\Gamma^\rho_{\mu \lambda}\Gamma^\lambda_{\nu\sigma}-\Gamma^\rho_{\nu \lambda}\Gamma^\lambda_{\mu\sigma}$ is the Riemann tensor of the non-symmetric connection. The second equation is the torsion-free constraint whose solution is the Levi--Civita connection which when substituted to the first equation gives \eqref{eq:locally_AdS}.

We can trade the Christoffel symbols $\Gamma^{\rho}_{\mu\nu}$ for a spin connection which is defined as
\begin{equation}
    (\omega^i_{\;\;\,j})_\mu = E^i_\rho\, E_j^\nu\,\Gamma^{\rho}_{\mu\nu} - E^\nu_j\,\partial_\mu E^i_\nu\,,
    \label{eq:omega_gamma_formula}
\end{equation}
where we have defined the inverse dreibein via
\begin{equation}
    E_j^\mu\,E^i_\mu = \delta^i_j\,.\label{eq:inverse_vielbein}
\end{equation}
By writing $(\omega^i_{\;\;\,j})_\mu =\delta_{jk}\,\omega^{ik}_\mu$, the Riemann tensor $\mathcal{R}^{\rho}_{\;\;\sigma\mu\nu}$ of $\Gamma^{\rho}_{\mu\nu}$ can be written as
\begin{equation}
    (f^{ij})_{\mu\nu} = \partial_{[\mu}\omega^{ij}_{\nu]}+\delta_{kl}\,\omega^{ik}_{[\mu}\, \omega^{lj}_{\nu]}\,,\quad (f^{ij})_{\mu\nu} = \frac{1}{2}\,\mathcal{R}^{\rho\sigma}_{\mu\nu}\,E^i_\rho E^j_\sigma\,,
    \label{eq:curvature_2_form}
\end{equation}
where $(f^{ij})_{\mu\nu}$ is called the curvature of the spin connection. In terms of $(E,\omega)$, the equations \eqref{eq:Einstein_Palatini_equations} then take the form
\begin{equation}
    (f^{ij})_{\mu\nu}+ \frac{1}{\ell^2}\,E_{[\mu}^iE^j_{\nu]} = 0\,,\quad \partial_{[\mu}E^{i}_{\nu]}+\delta_{jk}\,\omega^{ij}_{[\mu}\,E^{k}_{\nu]} = 0\,,
    \label{eq:first_order_equations}
\end{equation}
which are the equations of motion of Einstein gravity in its first-order formulation.

\paragraph{Chern--Simons formulation.} The fields $(E,\omega)$ can be packaged into a single Chern--Simons gauge field $A$ as follows \cite{Achucarro:1986uwr,Witten:1988hc,Witten:1989ip}. We define a set of three 1-forms $W^i$ via\footnote{We set $\varepsilon_{123} = 1$.}
\begin{equation}
	W^{i} \equiv -\frac{1}{2}\,\varepsilon_{kl}^{\;\;\;\,i}\,\omega^{kl}\,,\quad \omega^{ij} = -\varepsilon^{ij}_{\;\;\;\,k}\,W^{k}\,.
	\label{eq:omega_w_defs}
\end{equation}
Similarly from the curvature we define
\begin{equation}
	f^{i} \equiv -\frac{1}{2}\,\varepsilon_{kl}^{\;\;\;\,i}\,f^{kl}\,,\quad f^{ij} = -\varepsilon^{ij}_{\;\;\;\,k}\,f^{k}\,.
    \label{eq:f_defs}
\end{equation}
The equations \eqref{eq:first_order_equations} thus take the form
\begin{equation}
    \partial_{[\mu}W^{i}_{\nu]} +\frac{1}{2}\,\varepsilon_{jk}^{\;\;\;\,i}\,W^j_{[\mu}\,W^k_{\nu]} -\frac{1}{\ell^2}\frac{1}{2}\,\varepsilon_{jk}^{\;\;\;\,i}\, E_{[\mu}^jE^k_{\nu]} = 0 \,,\quad \partial_{[\mu}E^{i}_{\nu]}+\varepsilon_{jk}^{\;\;\;\;i}\,W^{j}_{[\mu}\,E^{k}_{\nu]}=0\,.
    \label{eq:palatini_EOM_version_2}
\end{equation}
We then define the Lie algebra $\mathfrak{s}\mathfrak{u}(2)$-valued 1-forms
\begin{equation}
    W_\mu \equiv W^i_\mu\,\tau_i\,,\quad E_\mu \equiv E^i_\mu\,\tau_i\,,
    \label{eq:W_E_defs}
\end{equation}
where $\tau_i = -\tau_i^\dagger$ are the three generators of $\mathfrak{su}(2)$ with commutation and normalization relations
\begin{equation}
    [\tau_i,\tau_j] = \varepsilon_{ij}^{\;\;\;k}\,\tau_k\,,\quad \tr{(\tau_i\,\tau_j)} = -\frac{1}{2}\,\delta_{ij}\,,
    \label{eq:su2_lie_algebra}
\end{equation}
and the trace is taken over the fundamental representation of $\mathfrak{su}(2)$.\footnote{While on the level of the Lie algebra $\mathfrak{su}(2)$ is isomorphic to $\mathfrak{so}(3)$ via $\tau_i = -\frac{i}{2}\,\sigma_i$, where $\sigma_i$ are Pauli matrices, the Lie group $SU(2)$ is the double cover of $SO(3)$} Therefore, \eqref{eq:palatini_EOM_version_2} can be compactly written as
\begin{equation}
    \d W+W\wedge W -\frac{1}{\ell^2}\,E\wedge E = 0\,,\quad \d E + W\wedge E + E\wedge W = 0\,,
    \label{eq:Palatini_EOM_forms}
\end{equation}
where we define the exterior derivative and the wedge product of $\mathfrak{su}(2)$-valued 1-forms as
\begin{equation}
    (\d C)_{\mu\nu} \equiv \partial_{[\mu}C_{\nu]}^i\,\tau_i \,,\quad (C\wedge D)_{\mu\nu} \equiv (C^i\wedge D^j)_{\mu\nu}\,\tau_i\tau_j = \frac{1}{2}\,C_{[\mu}^iD^j_{\nu]}\,[\tau_i,\tau_j]\,.
\end{equation}
We define the complex Lie algebra valued 1-forms
\begin{equation}
	A \equiv W + \frac{i}{\ell}\, E\,,\quad \overbar{A} \equiv  W - \frac{i}{\ell}\,E\,,
    \label{eq:A_Abar_definitions}
\end{equation}
which have the expansions
\begin{equation}
    A_\mu = A^i_\mu\,\tau_i\,,\quad \overbar{A}_\mu = \overbar{A}^i_\mu\,\tau_i\,,
\end{equation}
where $\overbar{A}^i_\mu = (A^i_\mu)^*$ is the complex conjugate. 
It follows that the two equations \eqref{eq:Palatini_EOM_forms} are equivalent to
\begin{equation}
    \d A + A\wedge A = 0\,,\quad \d\overbar{A} + \overbar{A}\wedge \overbar{A} = 0\,.
    \label{eq:CS_equation_motion}
\end{equation}
We note that $A$ and $\overbar{A}$ are $\mathfrak{sl}(2,\mathbb{C})$-valued 1-forms, because $\mathfrak{sl}(2,\mathbb{C})$ is the complexification of $\mathfrak{su}(2)$ i.e.~elements of $\mathfrak{su}(2)$ and $\mathfrak{sl}(2,\mathbb{C})$ are linear combinations of the generators $\tau_i$ with real and complex valued coefficients respectively. The number of real degrees of freedom in $A$ is six and $\overbar{A}$ is not an independent field. By making $(E^i,W^i)$ complex valued one may double the number of degrees of freedom to 12 making $(A^i,\overbar{A}^i)$ independent fields, but we will not do this in the following.

\paragraph{Local symmetries.} Einstein gravity is invariant under the local rotation group $ \Lambda(x) \in SO(3)$ that acts on the dreibein as
\begin{equation}
    \widehat{E}^i =  \Lambda_{\;\;j}^i\,E^j\,,\quad \Lambda_{\;\;i}^k\,\Lambda_{\;\;j}^l\,\delta_{kl} = \delta_{ij}\,.
    \label{eq:e_transformation}
\end{equation}
The action \eqref{eq:3D_Einstein_hilbert} is trivially invariant, because the metric \eqref{eq:metric_vielbein} is invariant. Einstein's equations \eqref{eq:first_order_equations} in the first-order formulation are also invariant if the spin connection transforms as
\begin{equation}
    (\widehat{\omega}^i_{\;\;\,j})_\mu = \Lambda^i_{\;\;k}\,(\Lambda^{-1})^l_{\;\;j}\,(\omega^k_{\;\;\,l})_\mu -(\Lambda^{-1})^k_{\;\;j}\,\partial_\mu\Lambda^i_{\;\;k}\,.
    \label{eq:omega_transformation}
\end{equation}
This descends from the invariance of the Einstein's equation \eqref{eq:locally_AdS} in the metric formulation upon using the formula \eqref{eq:omega_gamma_formula} and the fact that the Levi-Civita connection is trivially invariant since the metric is invariant.

The transformations \eqref{eq:e_transformation} and \eqref{eq:omega_transformation} can be implemented by an adjoint action of $SU(2)$ on the $\mathfrak{su}(2)$ Lie algebra introduced above. For local $U(x)\in SU(2)$ the adjoint action is given by
\begin{equation}
    U^{-1}\,\tau_i\,U = \Lambda^j_{\;\;i}\,\tau_j\,,
    \label{eq:adjoint_action}
\end{equation}
where $\Lambda^j_{\;\;i}(x) \in SO(3)$. The reason why the double cover $SU(2)$ of $SO(3)$ appears is because the $\pm 1$ factor cancels in the adjoint action. Using the identities\footnote{The first identity follows by taking a derivative of \eqref{eq:adjoint_action} and using \eqref{eq:su2_lie_algebra}.}
\begin{equation}
    (\Lambda^{-1})_{\;\;i}^k\,\d\Lambda^j_{\;\;k} = \varepsilon_{ik}^{\;\;\;\,j}\,(U^{-1}\d U)^k\,,\quad \Lambda^l_{\;\;i}\,\Lambda^m_{\;\;j}\,\Lambda^n_{\;\;k}\,\varepsilon_{lmn} = \varepsilon_{ijk}\,,
    \label{eq:maurer_cartan_identity}
\end{equation}
we can see that \eqref{eq:e_transformation} and \eqref{eq:omega_transformation} are respectively equivalent to
\begin{equation}
    \widehat{W} =  U^{-1}\,W\,U+ U^{-1}\d U\,,\quad \widehat{E} =  U^{-1}\,E\,U\,,
    \label{eq:W_hat_E_hat}
\end{equation}
which can be combined into
\begin{equation}
    \widehat{A} =  U^{-1}\,A\,U+ U^{-1}\d U\,.
    \label{eq:gauge_transformation_A}
\end{equation}
Therefore the $\mathfrak{sl}(2,\mathbb{C})$-valued 1-form $A$ transforms as an $SU(2)$ gauge field and the invariance of Einstein's equations under local rotations in the first-order formulation amounts to the gauge invariance of the Chern--Simons equations of motion. We note that we obtain the gauge group $SU(2)$, because we work in Euclidean signature where $SU(2)$ is the double cover of the symmetry group $SO(3)$ of the flat Euclidean metric $\delta_{ij}$.

\subsection{Solutions of three-dimensional Einstein gravity} \label{sec:solutions_in_FG_coordinates}

Let us now classify solutions of three-dimensional Einstein--Hilbert gravity both in its metric and Chern--Simons formulations.

\paragraph{Metric formulation.} In the metric formulation, all solutions are locally AdS$_3$ around any point. In Fefferman--Graham coordinates $y^\mu = (r,\hat{y}^\alpha)$ with $r >0$, the most general solution of \eqref{eq:locally_AdS} is given by
\begin{equation}
    g_{\mu\nu}\,\d y^\mu \d y^\nu = \ell^2\d r^2  +  \ell^2e^{2r}\,\bigl(\hat{\gamma}_{\alpha\beta} + \gamma_{(2)\alpha\beta}\,e^{-2r} + \gamma_{(4)\alpha\beta}\,e^{-4r}\bigr)\,\d \hat{y}^\alpha \d \hat{y}^\beta\,,
    \label{eq:FG_metric}
\end{equation}
where the subleading coefficients satisfy \cite{Skenderis:1999nb}
\begin{equation}
    \hat{\gamma}^{\alpha\beta}\,\gamma_{(2)\alpha\beta} = -\frac{1}{2}\,\hat{R}\,,\quad \hat{\nabla}^\beta\gamma_{(2)\alpha\beta} =-\frac{1}{2}\,\partial_\alpha\hat{R} \,,\quad \gamma_{(4)\alpha\beta} = \frac{1}{4}\,\hat{\gamma}^{\delta\gamma}\,\gamma_{(2)\alpha\delta}\,\gamma_{(2)\beta\gamma}\,.
    \label{eq:subleading_coefficients}
\end{equation}
Here $\hat{\gamma}_{\alpha\beta}(\hat{y})$ is the metric on the conformal boundary located at $r\rightarrow \infty$ and $\hat{R}$ is its Ricci scalar.

Assume now that the boundary metric is flat, $\hat{R} = 0$, and takes the form
\begin{equation}
    \hat{\gamma}_{\alpha\beta}\,\d y^\alpha \d y^\beta = \d z \d \overbar{z}
    \label{eq:flat_metric}
\end{equation}
in boundary complex coordinates $z = z(\hat{y})$, $\overbar{z} =\overbar{z}(\hat{y}) = z(\hat{y})^*$. Then the solutions are parametrized by two functions $(T,\overbar{T})$ as
\begin{equation}
    \gamma_{(2)zz}(z,\overbar{z}) = T(z)\,,\quad \gamma_{(2)\overbar{z}\overbar{z}}(z,\overbar{z}) = \overbar{T}(\overbar{z})\,,\quad \gamma_{(2)z\overbar{z}}(z,\overbar{z}) = 0\,,
\end{equation}
and take the form
\begin{equation}
    g_{\mu\nu}\,\d y^\mu \d y^\nu = \ell^2\d r^2  +  \ell^2\,(\d z - e^{-2r}\,\overbar{T}(\overbar{z})\,\d\overbar{z})(\d\overbar{z} - e^{-2r}\,T(z)\,\d z)\,.
    \label{eq:Banados_metric}
\end{equation}
These solutions are known as Ba\~nados geometries \cite{Banados:1998gg} and the functions $(T,\overbar{T})$ can be identified with the expectation values of the stress tensor operator of the dual CFT. The Poincar\'e AdS geometry corresponds to $T = \overbar{T} = 0$.

\begin{comment}
By using the last equation of \eqref{eq:subleading_coefficients}, the metric can be written as
\begin{equation}
    g_{\mu\nu}\,dx^\mu dx^\nu =  \ell^2dr^2  +  \ell^2e^{2r}\,\hat{\gamma}_{\alpha_1\beta_1}\biggl( \delta^{\alpha_1}_{\alpha} + \frac{1}{2}\,\gamma_{(2)\alpha}^{\alpha_1}\,e^{-2r} \biggr)\biggl( \delta^{\beta_1}_{\beta} + \frac{1}{2}\,\gamma_{(2)\beta}^{\beta_1}\,e^{-2r} \biggr)\,dy^\alpha dy^\beta\,.
\end{equation}
Defining the boundary zweibein $\hat{\gamma}_{\alpha\beta} = \delta_{AB}\,\hat{e}_\alpha^A \hat{e}_{\beta}^B$, the bulk dreibein becomes
\begin{equation}
	e^{r}_r = \ell\,,\quad \frac{1}{\ell}\,e^{A}_{\alpha} =  \hat{e}^a_{\alpha}\,e^r + \frac{1}{2}\,\gamma_{(2)\alpha}^{\beta}\,\hat{e}^a_{\beta}\,e^{-r} \,.
\end{equation}
The Levi--Civita spin connection on the conformal boundary is given by X. In two dimensions, the Ricci scalar is a total derivative $\hat{R} = -2\,\varepsilon^{ij}\,\partial_i \hat{\omega}_j$.
\end{comment}

\paragraph{Chern--Simons formulation.} We will now consider solutions of the equations of motion \eqref{eq:CS_equation_motion} in the CS formulation. In an open neighborhood around any point, the equations of motion \eqref{eq:CS_equation_motion} are solved by an $\mathfrak{sl}(2,\mathbb{C})$ gauge field of the form\footnote{To define global solutions, the holonomy needs to be specified, which we neglect.}
\begin{equation}
    A = M^{-1}\d M\,,\quad \overbar{A} = \overbar{M}^{-1}\d\overbar{M}\,,\label{eq:3d_flatness_solution}
\end{equation}
where $M(y),\overbar{M}(y) \in SL(2,\mathbb{C})$. In terms of the gauge fields, the metric \eqref{eq:metric_vielbein} is given by
\begin{equation}
    g_{\mu\nu} = \frac{\ell^2}{2}\tr{[(A_\mu-\overbar{A}_\mu) (A_\nu-\overbar{A}_\nu)]}\,.
    \label{eq:metric_gauge_fields}
\end{equation}
From the six $\mathfrak{sl}(2,\mathbb{C})$ generators $(\tau_{1,2,3},i\tau_{1,2,3})$ it is useful to define the generators
\begin{equation}
	J_{\pm} = i\,(\tau_1 \pm i \tau_2)\,,\quad J_3 = i\tau_3\,,\label{eq:j_generators}
\end{equation}
which by \eqref{eq:su2_lie_algebra} satisfy the $\mathfrak{sl}(2,\mathbb{R})$ algebra
\begin{gather}
	[J_3,J_{\pm}] = \pm J_{\pm}\,,\quad [J_+,J_-] = 2J_3\,,\quad \tr{(J_3J_3)} = \frac{1}{2}\,,\quad \tr{(J_+J_-)} = 1\,.
\end{gather}
We will focus only on gauge fields for which the metric \eqref{eq:metric_gauge_fields} takes the Fefferman--Graham form \eqref{eq:FG_metric}. This requires that the matrices $M$ and $\overbar{M}$ factorize as
\begin{equation}
     M(r,\hat{y}) = h(\hat{y})\,e^{r J_3}\,,\quad \overbar{M}(r,\hat{y}) = \overbar{h}(\hat{y})\,e^{-r J_3}\,,\label{eq:M_base}
\end{equation}
where $h,\overbar{h}\in SL(2,\mathbb{C})$ depend only on the boundary coordinates $\hat{y}$. The solution \eqref{eq:3d_flatness_solution} then takes the form
\begin{equation}
    A = e^{-r J_3}\,B\,e^{r J_3} + \d r\,J_3\,,\quad \overbar{A} = e^{r J_3} \,\overbar{B}\,e^{-r J_3}  -\d r\,J_3\,,
    \label{eq:3d_flatness_solution_2}
\end{equation}
where we have defined $\mathfrak{sl}(2,\mathbb{C})$-valued 1-forms $B \equiv h^{-1}\d h$ and $\overbar{B} \equiv \overbar{h}^{-1}\d \overbar{h}$ which have components only in the boundary directions. In terms of the $\mathfrak{sl}(2,\mathbb{R})$ generators \eqref{eq:j_generators}, we can parametrize the $SL(2,\mathbb{C})$ elements as \cite{Banados:2002ey}
\begin{equation}
    h(y) = e^{F\,J_-}e^{\varphi\,J_3}e^{\sigma\,J_+}\,,\quad \overbar{h}(y) = e^{-\overbar{F}\,J_+}e^{-\overbar{\varphi}\,J_3}e^{-\overbar{\sigma}\,J_-}\,,
    \label{eq:h_hbar}
\end{equation}
where $F,\varphi,\sigma$ are arbitrary complex valued functions of the boundary complex coordinates $(z,\overbar{z})$ and $\overbar{F},\overbar{\varphi},\overbar{\sigma}$ are their complex conjugates which ensures that the expansions $B^i_\alpha, \overbar{B}^i_\alpha$ in $\mathfrak{su}(2)$ generators $\tau_i$ are complex conjugates of each other. This is required also for $A^i_\mu,\overbar{A}^i_\mu$ in \eqref{eq:3d_flatness_solution_2} to be complex conjugates.

The metric \eqref{eq:metric_gauge_fields} corresponding to the parametrization \eqref{eq:h_hbar} does not take the standard Fefferman--Graham form \eqref{eq:FG_metric}, but contains cross terms $\text{Im}\,B^3_\alpha\,\d r \d \hat{y}^\alpha$ \cite{Banados:2002ey}. Therefore it can be put into Fefferman--Graham form by imposing the condition $\text{Im}\,B^3_\alpha = 0$ which fixes $\sigma,\overbar{\sigma}$ in terms of $F,\overbar{F}$ and $\varphi,\overbar{\varphi}$. The metric on the conformal boundary takes the form
\begin{equation}
    \hat{\gamma}_{\alpha\beta}\,\d y^\alpha \d y^\beta = e^{2\omega}\,(\d z+\mu\,\d\overbar{z}) (\d \overbar{z}+\overbar{\mu}\,\d z)\,,
\end{equation}
where we have defined
\begin{equation}
    e^{2\omega}= e^{\varphi + \overbar{\varphi}}\,\partial_z F\,\partial_{\overbar{z}} \overbar{F}\,,\quad \mu = \frac{\partial_{\overbar{z}} F}{\partial_z F}\,,\quad \overbar{\mu} = \frac{\partial_{z} \overbar{F}}{\partial_{\overbar{z}} \overbar{F}}\,. 
\end{equation}
The flat metric \eqref{eq:flat_metric} on the conformal boundary is obtained by setting $\omega = \mu = \overbar{\mu} = 0$ which sets
\begin{equation}
    F(z,\overbar{z}) = f(z)\,,\quad \overbar{F}(z,\overbar{z}) = \overbar{f}(\overbar{z})\,,\quad \varphi(z,\overbar{z}) = -\log{f'(z)}\,,\quad \overbar{\varphi}(z,\overbar{z}) = -\log{\overbar{f}'(\overbar{z})}\,.\label{eq:banados_sol1}
\end{equation}
Imposing Fefferman--Graham gauge $\text{Im}\,B^3_\alpha = 0$ gives
\begin{equation}
    \sigma(z,\overbar{z}) = -\frac{1}{2}\frac{f''(z)}{f'(z)}\,,\quad \overbar{\sigma}(z,\overbar{z}) = -\frac{1}{2}\frac{\overbar{f}''(\overbar{z})}{\overbar{f}'(\overbar{z})}\,. \label{eq:banados_sol2}
\end{equation}
With these functions in \eqref{eq:h_hbar}, the gauge field takes the form
\begin{equation}
    A = J_3\,\d r - e^{-r}\,T(z)\,J_+\,\d z + e^{r}\,J_-\,\d z\,,\quad \overbar{A} = -J_3\,\d r- e^{r}\,J_+\,\d \overbar{z} + e^{-r}\,\overbar{T}(\overbar{z})\,J_-\,\d \overbar{z} \label{eq:banados_sol_gauge_field}
\end{equation}
with functions
\begin{equation}
    T(z) = \frac{1}{2}\,\{f(z),z\}\,,\quad \overbar{T}(z) = \frac{1}{2}\,\{\overbar{f}(\overbar{z}),\overbar{z}\}\,,
    \label{eq:Schwarzian_T_Tbar}
\end{equation}
where $\{f(z),z\}$ is the Schwarzian derivative. The metric \eqref{eq:metric_gauge_fields} takes the Ba\~nados form \eqref{eq:Banados_metric} with the same functions \eqref{eq:Schwarzian_T_Tbar}. It follows that the Poincar\'e AdS geometry is obtained by setting $f(z) = z$ and $\overbar{f}(\overbar{z}) = \overbar{z}$ together with $\varphi = \overbar{\varphi} = \sigma = \overbar{\sigma} = 0$.

\section{Review of Kodaira--Spencer gravity}\label{sec:KS_review}

In this Section, we introduce the Kodaira--Spencer theory of gravity whose fundamental field is a deformation of an underlying background almost complex structure of a given differentiable manifold. In \Secref{sec:uplift}, we will relate this theory to Einstein gravity in its Chern--Simons formulation which we reviewed in the previous section.

To setup the notation and present facts relevant for later sections, we will first in \Secref{subsec:complex_structures_intro} review the basics of almost complex structures. We present the definition of integrability, that determines when an almost complex structure becomes a complex structure, and diffeomorphism transformation properties that determine when two structures are equivalent. In the second part in \Secref{subsec:deformations_review}, we review deformations of almost complex structures and derive the Kodaira--Spencer (KS) equation as equation of motion of KS gravity. It determines how the deformation gives rise to an integrable complex structure. 

\subsection{Introduction to almost complex structures}\label{subsec:complex_structures_intro}

As a review geared towards a physics audience, we  begin by introducing the basics of almost complex structures. To this end, let $n =1,2,\ldots$ be a positive integer and consider a $2n$-dimensional differentiable manifold $P$ with coordinates $x^a$ where the $a = 1,\ldots,2n$. An \textit{almost complex structure} on $P$ is defined to be a real vector valued 1-form $J^a_b(x)$ that satisfies\footnote{More precisely, $J$ is a real automorphism $J\colon TP\rightarrow TP$ of the tangent bundle of $P$. We use the notation $T_pP$ for the tangent space at point $p$ and its elements are tangent vectors. The tangent bundle $TP$ is disjoint union of tangent spaces of all points $p\in P$ and its elements are vector fields.}
\begin{equation}
    J^a_c\,J^c_b = -\delta^a_b\,.\label{eq:almost_complex_structure_def}
\end{equation} 
As we review in \Appref{app:J_DOFs}, $J$ has conjugate pairs of eigenvectors with eigenvalues $\pm i$. Thus, the existence of an almost complex structure provides a decomposition of the complexification $T_p^{\hspace{1pt}\mathbb{C}}P$ of the tangent space $T_p P$ at every point $p\in P$ as $T_p^{\hspace{1pt}\mathbb{C}}P= T^{(1,0)}_pP\oplus T^{(0,1)}_pP $. Here $T^{(1,0)}_pP$ and $T^{(0,1)}_pP$ are called holomorphic and anti-holomorphic tangent spaces of $P$ that are spanned by the eigenvectors with eigenvalues $\pm i$ respectively. The decomposition is achieved by the projectors
\begin{equation}
    \Pi^a_b = \frac{1}{2}\,(\delta^a_b - i\,J^a_b)\,,\quad \overbar{\Pi}^a_b = \frac{1}{2}\,(\delta^a_b + i\,J^a_b)
    \label{eq:projectors}
\end{equation}
that project vectors to $T^{(1,0)}_pP$ and $T^{(0,1)}_pP$ respectively. The same applies to the cotangent space which at point $p$ decomposes analogously $\Omega_p^{\hspace{1pt}\mathbb{C}} = \Omega^{(1,0)}_p\oplus \Omega^{(0,1)}_p $. This can be generalized to the space of $(q,r)$-forms, which is defined by taking $q$ wedge products of $ \Omega^{(1,0)}_p$ and $r$ wedge products of $\Omega^{(0,1)}_p$, \begin{equation}
    \Omega_p^{(q,r)}= \Omega_p^{(1,0)}\wedge\dots \wedge \Omega_p^{(1,0)} \wedge \Omega_p^{(0,1)}\wedge\dots\wedge\Omega_p^{(0,1)}\,.
\end{equation}
Consider a set of $n$ linearly independent non-vanishing $(1,0)$-forms $\mathcal{A}^i\in \Omega^{(1,0)}$ and $n$ independent $(0,1)$-forms $\overbar{\mathcal{A}}^i\in \Omega^{(0,1)}$ which always exist in open neighborhoods around any point. By definition, they satisfy
\begin{equation}
    \overbar{\Pi}^b_a\,\mathcal{A}^i_b =0\,,\quad \Pi^b_a\,\overbar{\mathcal{A}}^i_b =0
    \label{eq:10_forms}
\end{equation}
and produce bases for the cotangent spaces. Let $\mathcal{V}^{a}_i$ and $\overbar{\mathcal{V}}^{a}_i$ be the corresponding dual vector fields such that
\begin{equation}
    \mathcal{A}_a^i\,\mathcal{V}^{a}_j =\overbar{\mathcal{A}}_a^i\,\overbar{\mathcal{V}}^{a}_j = \delta^{i}_j\,,\quad \mathcal{A}_a^i\,\overbar{\mathcal{V}}^{a}_j = \overbar{\mathcal{A}}_a^i\,\mathcal{V}^{a}_j=0\,.
    \label{eq:orthonormality}
\end{equation}
Then the almost complex structure can be written as
\begin{equation}
    J^b_a = i\,(\mathcal{A}^{i}_a\,\mathcal{V}^{b}_i-\overbar{\mathcal{A}}^{i}_a\,\overbar{\mathcal{V}}^{b}_i)\,.
    \label{eq:complex_structure_A_V}
\end{equation}

\paragraph{Integrability.} An almost complex structure $J$ can be used to define complex coordinates which are linearly independent scalar fields $w^i(x)$ and $\overbar{w}^i(x)$, where $i = 1,2\ldots, n$, that solve the equations
\begin{equation}
    \overbar{\pounds}_a w^i(x)\equiv (\partial_a + i\,J^b_a(x)\,\partial_b)\,w^i(x) = 0\,,\quad \pounds_a \overbar{w}^i(x) \equiv(\partial_a- i\,J^b_a(x)\,\partial_b)\,\overbar{w}^i(x) = 0\,.
    \label{eq:definition_holomorphic_coordinates}
\end{equation}
If $w^i(x)$ solves the first equation, then $\overbar{w}^i(x) = (w^i(x))^* $ solves the second equation. However, the existence of solutions to these equations in an open neighborhood is non-trivial. If solutions do exist, then the almost complex structure $J$ is called \textit{integrable}. Clearly a necessary condition for a solution to exist is\footnote{Notice that the differential operators $\pounds = \Pi \circ \d$ and $\overbar{\pounds} = \overbar{\Pi}\circ \d$, where $\d$ is the exterior derivative, are the Dolbeault operators. Here we understand $\Pi,\overbar{\Pi}$ to act on arbitrary differential forms by projecting all form indices. The integrability condition is then equivalent to $\pounds^2 = \overbar{\pounds}^2 = 0$, or, to $\d = \pounds+\overbar{\pounds}$.}
\begin{equation}
    \overbar{\pounds}_{[a}\overbar{\pounds}_{b]}\,w^i(x) =0=\pounds_{[a}\pounds_{b]}\,\overbar{w}^i(x)\,.
\end{equation}
In \Appref{app:integrability_conditions}, we show that these two equations are equivalent to the vanishing of the Nijenhuis tensor
\begin{equation}
    N_{ab}^c \equiv 2\,\bigl(J^c_d\,\partial_{[a}J_{b]}^d-J_{[a}^d\,\partial_{\vert d\vert} J_{b]}^c\bigr) = 0\,.
    \label{eq:vanishing_Nijenhuis}
\end{equation}
The fact that this is also a sufficient condition for solutions $(w^i,\overbar{w}^i)$ to exist is the statement of the Newlander--Nirenberg theorem \cite{newlander1957complex}: an almost complex structure is integrable if and only if $N_{ab}^c = 0$. If the almost complex structure is integrable at a point $p$ then complex coordinates exist in an open neighborhood around that point. The complex coordinates can then be defined globally by matching them on patch overlaps using transition functions of the underlying differentiable structure. An integrable almost complex structure $J$ is called a \textit{complex structure} and a manifold $P$ which has a complex structure is a complex manifold denoted by $(P,J)$.

The integrability \eqref{eq:vanishing_Nijenhuis} of $J$ has implications for the associated $(1,0)$-forms and $(0,1)$-forms. Notice that the equations \eqref{eq:definition_holomorphic_coordinates} can be written as
\begin{equation}
    \overbar{\Pi}^b_a\,\partial_bw^i = 0 = \Pi^b_a\,\partial_b\overbar{w}^i\,.
    \label{eq:complex_coordinates_Pi}
\end{equation}
Since $(w^i,\overbar{w}^i)$ exist by integrability, this implies that $\d w^i \in \Omega^{(1,0)}$ and $\d\overbar{w}^i\in \Omega^{(0,1)}$. Because $\partial_aw^i$, $\partial_a\overbar{w}^i$ are linearly independent they form a basis and we can expand the 1-forms satisfying \eqref{eq:10_forms} as
\begin{equation}
    \mathcal{A}^i_a = \Sigma_j^i\,\partial_aw^j\,,\quad \overbar{\mathcal{A}}^i_a = \overbar{\Sigma}_j^i\,\partial_a\overbar{w}^j
\end{equation}
with some local $GL(3,\mathbb{C})$ matrices $\Sigma^i_j(w,\overbar{w})$ and $\overbar{\Sigma}^i_j(w,\overbar{w})$. Here, invertibility of $\Sigma^i_j ,\overbar{\Sigma}^i_j$ ensures linear independence of the form fields $\mathcal{A}^i$. Moreover, note that $\det{\Sigma} \det{\overbar{\Sigma}} = \vert\det{\Sigma}\, \vert ^2> 0$. Thus, as long as we impose that $\overbar{\mathcal{A}}^i_a$ is the complex conjugate of $\mathcal{A}^i_a$, the orientation of the basis spanned by $\d w^i$ is preserved.

Defining $\sigma^i_j $ as the inverse $\sigma_k^i \,\Sigma^k_j = \delta^i_j$, we can thus write
\begin{equation}
     \mathcal{A}^i = \Sigma_j^i\,\d w^j\,,\quad  \mathcal{V}_i = \sigma^j_i\,\frac{\partial}{\partial w^j}\label{eq:connection_sigma_def}
\end{equation}
and similarly for the barred fields. Using \eqref{eq:complex_structure_A_V}, the integrable almost complex structure in the complex coordinates takes the simple form
\begin{equation}
    J_b^a(x)\,\d x^b\otimes \partial_a = i\,(\d w^i\otimes \partial_{i} - \d\overbar{w}^{i}\otimes \overbar{\partial}_{i})\,,
    \label{eq:J_simple_form}
\end{equation}
where $\Sigma,\overbar{\Sigma}$ and $\sigma,\overbar{\sigma}$ being inverses of one another have canceled each other. We have also introduced the short-hand
\begin{equation}
    \partial_i \equiv \frac{\partial}{\partial w^i}\,,\quad \overbar{\partial}_i \equiv \frac{\partial}{\partial \overbar{w}^i}\,.
    \label{eq:Dolbeault_index_notation}
\end{equation}
In the $(w^i, \overbar{w}^i)$ coordinates, the differential operators $\pounds_a,\overbar{\pounds}_a$ defined in \eqref{eq:definition_holomorphic_coordinates} are simply
\begin{equation}
    \pounds_i = \partial_i \,,\quad \overbar{\pounds}_i = \overbar{\partial}_i\,.\label{eq:del_delbar_definition}
\end{equation}
Equation \eqref{eq:J_simple_form} implies that in an open neighborhood around any point $p \in P$, $J$ is a diagonal matrix with entries $\pm i$ which is the standard complex structure of $\mathbb{C}^n$. Therefore all integrable almost complex structures of the same complex dimension $n$ are locally diffeomorphic to this standard structure and to each other. This fact is in contrast to Riemannian structures on a manifold: not all Riemannian metrics are diffeomorphic to each other in an open neighborhood around any point even though they are diffeomorphic to first order in an expansion in geodesic distance (Riemann normal coordinates). The obstruction is the existence of the Riemann tensor which transforms homogeneously under diffeomorphisms: two metrics are locally diffeomorphic at point $p$ if and only if the Riemann tensor of the two metrics coincide at $p$. For an almost complex structure, the only obstruction is the Nijenhuis tensor, but it is equal (to zero) for all integrable structures.

\paragraph{Transformation under diffeomorphisms.} The notion of an almost complex structure and its integrability are coordinate invariant statements. Let $\widetilde{J}$ be defined by the action of an active diffeomorphism $\mathcal{D}\colon P\rightarrow P$ on $J$ as
\begin{equation}
    \widetilde{J}^a_b(x) = \frac{\partial (\mathcal{D}^{-1})^a(\mathcal{D}(x))}{\partial x^c}\frac{\partial \mathcal{D}^d(x)}{\partial x^b}\,J^c_d(\mathcal{D}(x))\,,
    \label{eq:J_tilde}
\end{equation}
where $\mathcal{D}^{-1}$ denotes the inverse diffeomorphism that satisfies $\mathcal{D}^{-1}(\mathcal{D}(x)) = x$. Since $\widetilde{J}^a_c\,\widetilde{J}^c_b = -\delta^a_b$, it follows that if $J$ is an almost complex structure then so is $\widetilde{J}$. In addition, integrability \eqref{eq:vanishing_Nijenhuis} is a diffeomorphism invariant statement, because the Nijenhuis tensor $N^c_{ab}$ transforms homogeneously as a vector valued 2-form under diffeomorphisms: if $J$ is integrable then so is $\widetilde{J}$. The complex coordinates associated to $\widetilde{J}$ are given by
\begin{equation}
    \widetilde{w}^i(x) = w^i(\mathcal{D}(x))\,,\quad \widetilde{\overbar{w}}^i(x) = \overbar{w}^i(\mathcal{D}(x))\,.
    \label{eq:complex_coordinates_are_scalars}
\end{equation}
In other words, complex coordinates transform as scalar fields under diffeomorphisms.\footnote{This is also emphasized in \cite{Skliros:2019bqr}.}

As shown above, any two integrable almost complex structures are locally diffeomorphic. To make this more explicit, denote the two structures by $J,J'$ and let $(w^i,\overbar{w}^i), (\hat{w}^i,\hat{\overbar{w}}^i)$ be their respective complex coordinates that both exist in an open neighborhood $L$ around a point $p$. We can always find a diffeomorphism $\mathcal{F}\colon L\rightarrow L$ such that $\hat{w}^i(x) = w^i(\mathcal{F}(x)) $ and $\hat{\overbar{w}}^i(x) = \overbar{w}^i(\mathcal{F}(x)) $ in $L$, because $\mathcal{F}^a(x)$ are six real functions and so are $(\hat{w}^i,\hat{\overbar{w}}^i)$. It follows that $J' = \widetilde{J}$ where $\widetilde{J}$ is obtained by the action of the diffeomorphism $\mathcal{D} = \mathcal{F}$ on $J$ via the formula \eqref{eq:J_tilde}.

\paragraph{Equivalence of complex structures.} Notice that given a single solution $(w^i,\overbar{w}^i)$ of \eqref{eq:definition_holomorphic_coordinates} for $J$, then $(f^i(w),\overbar{f}^i(\overbar{w}))$ is also a solution for the same $J$ as long as the maps $f^i,\overbar{f}^i\colon \mathbb{C}^3\rightarrow \mathbb{C}^3$ have invertible Jacobians $\partial_if^j $ and $\overbar{\partial}_i\overbar{f}^j $. As a result, for a given complex structure $J$ there is a family $(f^i(w),\overbar{f}^i(\overbar{w}))$ of complex coordinates parametrized by invertible functions. In other words, there is a one-to-one correspondence between complex structures $J$ and equivalence classes of complex coordinates under left-action with $(f^i,\overbar{f}^i)$. Therefore two complex structures $J$ and $\widetilde{J}$ with complex coordinates $(w^i,\overbar{w}^i)$ and $(\widetilde{w}^i,\widetilde{\overbar{w}}^i)$ respectively are the same $\widetilde{J} = J$ if their complex coordinates lie within the same equivalence class
\begin{equation}
    \widetilde{w}^i(x) = f^i(w(x))\,,\quad  \widetilde{\overbar{w}}^i(x) = \overbar{f}^i(\overbar{w}(x))\,.
\end{equation}
It follows that some diffeomorphisms do not actually produce new complex structures via \eqref{eq:J_tilde}. These are diffeomorphisms $\mathcal{C}_J\colon P\rightarrow P$ called holomorphic-antiholomorphic diffeomorphisms with respect to the complex structure $J$ that satisfy
\begin{equation}
    w^i(\mathcal{C}_J(x)) = f^i(w(x))\,,\quad \overbar{w}^i(\mathcal{C}_J(x)) = \overbar{f}^i(\overbar{w}(x))\,.\label{eq:hol_antihol_def}
\end{equation}
Under the action of $\mathcal{C}_J$ on $J$ via the formula \eqref{eq:J_tilde}, $J$ is invariant which is also evident from its form \eqref{eq:J_simple_form} in complex coordinates.

\subsection{Complex structure deformations and Kodaira--Spencer gravity}\label{subsec:deformations_review}

Above, we introduced the concept of an almost complex structure on a manifold and how its integrability defines a complex structure. It is also useful to consider their deformations and have conditions when such a deformation produces another integrable structure. This is described by the theory of deformations of almost complex structures and their integrability is controlled by the Kodaira--Spencer equation which is the equation of motion of Kodaira--Spencer gravity. We will now give an introduction to these concepts.

\paragraph{Deformations of almost complex structures.} Let us consider two almost complex structures $J$ and $J'$. We will assume that $J$ is integrable implying the existence of complex coordinates $(w^i,\overbar{w}^i)$ in which it takes the form \eqref{eq:J_simple_form}. We assume the second complex structure $J'$ is not a priori integrable and has the general form \eqref{eq:complex_structure_A_V} in terms of a basis of 1-forms $\mathcal{A}'^i$ and dual vector fields $\mathcal{V}_{i}'$. In the complex coordinates of $J$, we can always write without loss of generality (see also \cite{Bandelloni:1998vp})
\begin{equation}
    \mathcal{A}'^i = \Sigma'^i_j\,(\d w^j + \alpha_{k}^{\;\,\, j}\d\overbar{w}^{k})\,,\quad \overbar{\mathcal{A}}'^i = \overbar{\Sigma}'^i_j\,(\d\overbar{w}^j + \overbar{\alpha}_{k}^{\;\,\, j}\d w^{k})\,,
    \label{eq:Aprime_definition}
\end{equation}
where $\Sigma'^i_j,\overbar{\Sigma}'^i_j$ are local $GL(3,\mathbb{C})$ matrices with positive determinant, the field $\alpha_{j}^{\;\,\,i}$ is called the deformation of the almost complex structure $J$ and $\overbar{\alpha}_{j}^{\;\,\, i} = (\alpha_{j}^{\;\,\,i})^*$ is its complex conjugate.\footnote{In two dimensions, $\alpha,\overbar{\alpha}$ are called Beltrami differentials.} Similarly, the dual vector fields can be written as
\begin{equation}
 \mathcal{V}_{i}' = \sigma'^j_i\,(\partial_{j} - \overbar{\alpha}_{j}^{\;\;\, k}\,\overbar{\partial}_k)\,,\quad \overbar{\mathcal{V}}_{i}' = \overbar{\sigma}'^j_i\,(\overbar{\partial}_{j}-\alpha_{j}^{\;\;\, k}\,\partial_k)\,,
 \label{eq:dualvectors_primed}
\end{equation}
where the derivatives with respect to the complex coordinates are defined in \eqref{eq:Dolbeault_index_notation}. The orthonormality relations \eqref{eq:orthonormality} require
\begin{equation}
    \Sigma'^i_k\,\sigma'^l_j\,(\delta^k_l-\overbar{\alpha}_{l}^{\;\,\, n}\,\alpha_{n}^{\;\,\, k})=\delta^i_j\,,\quad \overbar{\Sigma}'^i_k\,\overbar{\sigma}'^l_j\,(\delta^k_l-\alpha_{l}^{\;\,\, n}\,\overbar{\alpha}_{n}^{\;\,\, k})=\delta^i_j\,,
    \label{eq:orthonormality_deformed}
\end{equation}
which are equivalent to $J'$ squaring to minus the identity. Moreover, we impose that the vector fields \eqref{eq:dualvectors_primed} are linearly independent and preserve the orientation of the basis spanned by $\partial_i$ and $\overbar{\partial}_i$. This is equivalent to
\begin{equation}
    \det{\Sigma'} \det{\overbar{\Sigma}'}\det{(\delta^k_l-\alpha_{l}^{\;\,\, n}\,\overbar{\alpha}_{n}^{\;\,\, k})} > 0\,.
\end{equation}
By construction, $\Sigma'$ and $\overbar{\Sigma}'$ satisfy $\det\Sigma' \det\overbar{\Sigma}' = \vert \det\Sigma'\,\vert^2 > 0$, such that the above condition becomes
\begin{equation}
    \det (\delta^k_l-\alpha_{l}^{\;\,\, n}\,\overbar{\alpha}_{n}^{\;\,\, k}) > 0\,.\label{eq:determinant_condition}
\end{equation}
This condition ensures that we can express the vector fields $(\partial_i,\overbar{\partial}_i)$ in terms of $(\mathcal{V}_i,\overbar{\mathcal{V}}_i )$.

\paragraph{Diffeomorphisms.} Let us consider a diffeomorphism $\mathcal{D}^a = \mathcal{D}^a(x)$ which in complex coordinates of $J$ has components $ \mathcal{D}=(D^i(w,\overbar{w}),\overbar{D}^i(w,\overbar{w}))$. As shown in \Appref{app:diffeos}, the deformation $\alpha$ transforms under a diffeomorphism $\mathcal{D}$ as (see also \cite{Bandelloni:1998vp})
\begin{equation}
    \widetilde{\alpha}_i^{\;\,\, j}(w,\overbar{w}) = (N^{-1})^{j}_{k}(w,\overbar{w})\,\overbar{N}^{k}_{i}(w,\overbar{w})\,,
    \label{eq:alpha_transformation_D}
\end{equation}
where $(N^{-1})^i_k\,N^k_j = \delta^i_j$ and we have defined the matrices
\begin{equation}
	N_{j}^{i} \equiv \partial_{j}D^{i}+(\alpha_{k}^{\;\;\,i}\circ \mathcal{D})\,\partial_{j}\overbar{D}^{k}\,,\quad \overbar{N}^{i}_{j} \equiv \overbar{\partial}_{j}D^{i}+(\alpha_{k}^{\;\;\,i}\circ \mathcal{D})\,\overbar{\partial}_{j}\overbar{D}^{k}\,,
\end{equation}
and $\circ$ denotes composition of functions $(\alpha_{i}^{\;\;\,j}\circ \mathcal{D})(w,\overbar{w}) = \alpha_{i}^{\;\;\,j}(D(w,\overbar{w}),\overbar{D}(w,\overbar{w}))$. This follows from the transformation of $\mathcal{A}'^i$ as a 1-form, or equivalently, from the transformation of $\mathcal{V}_i'$ as a vector field.

For a holomorphic-antiholomorphic diffeomorphism $\mathcal{C}_J$ of $J$, which in complex coordinates takes the form $C^i_J(w,\overbar{w}) = f^i(w) $ and $\overbar{C}^i_J(w,\overbar{w}) = \overbar{f}^i(\overbar{w}) $, the formula \eqref{eq:alpha_transformation_D} reduces to
\begin{equation}
    \widetilde{\alpha}_i^{\;\,\, j}(w,\overbar{w}) = \overbar{\partial}_i\overbar{f}^k(w)\,(\partial_l f^{-1})^j(f(w))\,\alpha_k^{\;\,\, l}(f(w),\overbar{f}(\overbar{w}))\,.
    \label{eq:alpha_transformation_C}
\end{equation}
Since two almost complex structures $J'$ and $\widetilde{J}'$ related by $\mathcal{C}_J$ are equal $J' = \widetilde{J}'$, we see that two deformations $\alpha$ and \eqref{eq:alpha_transformation_C} of $J$ produce the same structure. In this sense, $\widetilde{\alpha}$ defined in \eqref{eq:alpha_transformation_C} is equivalent to $\alpha$ as a deformation. Two deformations related by a general diffeomorphism $\mathcal{D}$ via \eqref{eq:alpha_transformation_D} produce different structures in general.

Similarly, the transformation of $\Sigma'^i_j$ is given by
\begin{equation}
    \widetilde{\Sigma}'^i_j(w,\overbar{w}) = N^{k}_{j}(w,\overbar{w})\,\Sigma'^i_k(D(w,\overbar{w}),\overbar{D}(w,\overbar{w}))\,,
\end{equation}
and in particular, under holomorphic-antiholomorphic diffeomorphisms
\begin{equation}
    \widetilde{\Sigma}'^i_j(w,\overbar{w}) = \partial_j f^k(w)\,\Sigma'^i_k(f(w),\overbar{f}(\overbar{w}))\,.
    \label{eq:Sigma_transformation_C}
\end{equation}

\paragraph{Integrability of the deformation.} The complex structure deformation $\alpha$ is defined to be integrable if and only if $J'$ is integrable. In this case, the complex coordinates $(\hat{w}^i,\hat{\overbar{w}}^i)$ of $J'$ satisfy equations \eqref{eq:definition_holomorphic_coordinates} with $J'$. By contracting the equations with the dual vector fields \eqref{eq:dualvectors_primed} and using $\mathcal{V}'^{b}_iJ'^a_b = i\,\mathcal{V}'^{a}_i$, $\overbar{\mathcal{V}}'^{ib}J'^a_b = -i\,\overbar{\mathcal{V}}'^{ia}$, we find $(\hat{w}^i,\hat{\overbar{w}}^i)$ are solutions of the equations \cite{newlander1957complex}
\begin{equation}
     \overbar{\mathcal{V}}_{j}'^a\partial_a\hat{w}^i=\overbar{\sigma}'^k_j\,(\overbar{\partial}_{k}-\alpha_{k}^{\;\;\, l}\,\partial_l)\,\hat{w}^i = 0\,,\quad \mathcal{V}_{j}'^a\partial_a\hat{\overbar{w}}^l=\sigma'^j_i\,(\partial_{j} - \overbar{\alpha}_{j}^{\;\;\, k}\,\overbar{\partial}_k)\,\hat{\overbar{w}}^l = 0\,,
    \label{eq:definition_holomorphic_coordinates_2}
\end{equation}
where $(\hat{w}^i,\hat{\overbar{w}}^i)$ are understood to be functions of $(w^i,\overbar{w}^i)$.
Clearly a necessary condition for the existence of a solution to these equations is therefore
\begin{equation}\label{eq:necessary_integrability_commutator}
    \Pi'^a_b\,[\overbar{\mathcal{V}}_{i}',\overbar{\mathcal{V}}_{j}']^b=\overbar{\Pi}'^a_b\,[\mathcal{V}_{i}',\mathcal{V}_{j}']^b = 0\,,
\end{equation}
where we have defined the Lie bracket of two vector fields $[X,Y]^a = X^b\,\partial_b Y^a - Y^b\,\partial_b X^a$ and the projectors appear since $\partial_a\hat{w}^i$ and $\partial_a \hat{\overbar{w}}^i$ appearing in \eqref{eq:definition_holomorphic_coordinates_2} span the holomorphic and antiholomorphic cotangent spaces of $J'$ respectively. Equation \eqref{eq:necessary_integrability_commutator} indicates that the complex structure is integrable when the algebra of $(1,0)$- and $(0,1)$-vector fields closes (see \Appref{app:integrability_conditions} for a proof that this is equivalent to the vanishing of the Nijenhuis tensor).

By computing the Lie brackets \eqref{eq:necessary_integrability_commutator}, we find that the complex structure $J'$ associated with the complex structure deformation $\alpha$ is integrable if it satisfies the equation \cite{newlander1957complex} (see \Appref{app:integrability_conditions})
\begin{equation}
		E_{jk}^{i} \equiv \overbar{\partial}_{[j}\, \alpha_{k]}^{\;\;\,i}-\alpha_{[j}^{\;\;\,l}\partial_{\vert l \vert}\,\alpha_{k]}^{\;\;\,i}=0\,,\quad \overbar{E}_{jk}^{i} \equiv \partial_{[j}\, \overbar{\alpha}_{k]}^{\;\;\,i}-\overbar{\alpha}_{[j}^{\;\;\,l}\overbar{\partial}_{\vert l \vert}\,\overbar{\alpha}_{k]}^{\;\;\,i}=0\,,
        \label{eq:KS_equation_definition}
\end{equation}
which is known as the Kodaira--Spencer equation. 
Naturally, the Kodaira--Spencer equation is equivalent to the vanishing of the Nijenhuis tensor of $J'$ as proven in \Appref{app:integrability_conditions}.

Since integrability of $J'$ is a diffeomorphism invariant statement so should be the integrability of $\alpha$. Due to the complicated transformation law \eqref{eq:alpha_transformation_D}, it is not immediately obvious whether integrability of $\alpha$ is preserved under diffeomorphisms, however, in \Appref{app:diffeos} (see also \cite{Bandelloni:1998vp}), we show explicitly that this is the case: if $\alpha$ is a solution of the KS equation then so is $\widetilde{\alpha}$ given by \eqref{eq:alpha_transformation_D}.

In the previous section, we showed that any integrable almost complex structure $J'$ can be obtained from another one $J$ by the action of a certain diffeomorphism $\mathcal{F}$ in an open neighborhood. This implies that the associated $(1,0)$-forms are also related by the same diffeomorphism as
\begin{equation}
    \mathcal{A}'^i_a(x) = \frac{\partial \mathcal{F}^b(x)}{\partial x^a}\,\mathcal{A}^i_b(\mathcal{F}(x))\,.
\end{equation}
Writing the diffeomorphism as $\mathcal{F} = (F^i,\overbar{F}^i)$ in the complex coordinates $(w^i,\overbar{w}^i)$ of $J$, the complex structure deformation takes the form
\begin{equation}
    \alpha_i^{\;\,\, j}(w,\overbar{w}) = \overbar{\partial}_i F^k(w,\overbar{w})\,(\partial_k F^{-1})^j(F(w,\overbar{w}),\overbar{F}(w,\overbar{w}))\,.
    \label{eq:alpha_diffeo}
\end{equation}
where the inverse $(F^{-1})^i$ is defined as $(F^{-1})^i(F(w,\overbar{w}),\overbar{F}(w,\overbar{w})) = w^i$. One can explicitly check that \eqref{eq:alpha_diffeo} indeed solves the KS equation \eqref{eq:KS_equation_definition}.

Lastly, we note that integrability of $J'$ implies additional relations on the coefficients of $\mathcal{A}'$. We obtain that
\begin{align}
    \Sigma'^i_j(w,\overbar{w}) &= \partial_j F^k(w,\overbar{w})\,\Sigma^i_k(F(w,\overbar{w}),\overbar{F}(w,\overbar{w}))\nonumber\\
    \sigma'^i_j(w,\overbar{w}) &= \partial_j F^k(w,\overbar{w})\,\sigma^i_k(F(w,\overbar{w}),\overbar{F}(w,\overbar{w}))\,.
    \label{eq:Sigma_diffeo_F}
\end{align}
Since $\partial_j F^k$ is invertible and $\sigma = \Sigma^{-1}$, we obtain that $\Sigma' = (\sigma')^{-1} $. Substituting to \eqref{eq:orthonormality_deformed}, we obtain that an integrable deformation satisfies
\begin{equation}
    \alpha_{k}^{\;\,\, i}\,\overbar{\alpha}_{j}^{\;\,\, k} = 0\,,\quad \alpha_{j}^{\;\,\, k}\,\overbar{\alpha}_{k}^{\;\,\, i} = 0\,.\label{eq:alpha_squared}
\end{equation}
If this condition holds then \eqref{eq:determinant_condition} is automatically satisfied.
By using \eqref{eq:complex_structure_A_V}, the complex structure $J'$ and these relations, we obtain\footnote{Note that $J'$ takes the simple form \eqref{eq:Jprime_explicitly} only when it is integrable.}
\begin{equation}
    J' = J+2i\,(\alpha_{j}^{\;\;\,i}\,d\overbar{w}^{j}\otimes \partial_i-\overbar{\alpha}_{j}^{\;\;\,i}\,dw^{j}\otimes \overbar{\partial}_i)
    \label{eq:Jprime_explicitly}
\end{equation}
and $\overbar{\alpha}_{j}^{\;\,\, i} = (\alpha_{j}^{\;\,\,i})^*$ ensures that $J'$ is real as required.

\paragraph{Kodaira--Spencer gravity.} The Kodaira--Spencer theory of gravity concerns integrable almost complex structure deformations $\alpha$ on a differentiable complex manifold. The deformation $\alpha$ is the fundamental field of the theory and its equation of motion is the Kodaira--Spencer equation
\begin{equation}
    \overbar{\partial}_{[j}\, \alpha_{k]}^{\;\;\,i}-\alpha_{[j}^{\;\;\,l}\partial_{\vert l \vert}\,\alpha_{k]}^{\;\;\,i}=0\,.
    \label{eq:KS_EOM}
\end{equation}
It is a topological theory in the sense that all solutions are diffeomorphic to each other in open neighborhoods around any point. To see this, note that all complex structure deformations are diffeomorphic to the trivial solution $\alpha=0$. This can be checked by writing $\alpha$ in the form \eqref{eq:alpha_diffeo} using the diffeomorphism $\mathcal{F}$ and subsequently transforming it under the inverse diffeomorphism $\mathcal{F}^{-1}$ defined by $\mathcal{F}\circ \mathcal{F}^{-1} = \mathrm{id}$. This is analogous to how all solutions of three-dimensional Einstein gravity are diffeomorphic to a constant negative-curvature metric and to each other. Thus both theories are topological in the same sense.

In order to write an action for Kodaira--Spencer gravity whose variational principle yields the Kodaira--Spencer equation \eqref{eq:KS_EOM} additional structure is needed. Part of this additional structure is the holomorphic volume form with respect to a background complex structure $J$, which we denote by $\Omega = \d w^1 \wedge\dots\wedge \d w^n$. The action has a non-local kinetic term in $\alpha$ involving the inverse of the Dolbeault derivative $\partial_i$ of the background \cite{Bershadsky:1993cx}. In order for the inverse to be well defined, $\alpha$ has to satisfy an additional cohomology constraint \cite{costelloGaiotto2018twisted}
\begin{equation}
    \partial_i \alpha_j^{\;\;\, i} = 0\,.\label{eq:cohomology_condition}
\end{equation}
This is an extra condition which does not arise from a variational principle and must be imposed by hand. 

KS gravity on a three-dimensional complex manifold (six-dimensional real manifold) is a part of the target space effective theory of the topological B-model string theory \cite{Bershadsky:1993cx}. For first-order bosonic strings, which can be seen as the bosonic part of the B-model worldsheet theory, the derivation of the KS equation from the vanishing of the beta functions is shown in perturbation theory in \cite{Gamayun:2009hy,Gamayun:2009sp} where also the cohomology constraint \eqref{eq:cohomology_condition} is needed to ensure that the worldsheet stress tensor is a quasi-primary operator of weight two.

As discussed above and shown in \cite{Bandelloni:1998vp} (see also Appendix \ref{app:diffeos}), the KS equation is covariant under diffeomorphisms. As in ordinary Einstein gravity, this symmetry becomes a gauge symmetry at the linearized level in both the diffeomorphism and the deformation $\alpha$. From \eqref{eq:alpha_transformation_D} it follows that an infinitesimal diffeomorphism $\mathcal{D}^a(x) = x^a + \xi^a(x)$ acts as
\begin{equation}
    \delta\alpha_{j}^{\;\;\,i} =\xi^{k}\,\partial_k\,\alpha_{j}^{\;\;\,i}+\overbar{\xi}^{k}\,\overbar{\partial}_k\,\alpha_{j}^{\;\;\,i} -\alpha_{j}^{\;\;\,k}\,(\partial_k\xi^{i} +\alpha_{l}^{\;\;\,i}\,\partial_k \overbar{\xi}^{l})+\overbar{\partial}_j\xi^{i}+\alpha_{k}^{\;\;\,i}\,\overbar{\partial}_{j}\overbar{\xi}^{k}\,,
\end{equation}
where $\xi^i$ and $\overbar{\xi}^i$ are the holomorphic and antiholomorphic components of the vector field $\xi$ respectively. In particular, at leading order in $\alpha\rightarrow 0$, it reduces to
\begin{equation}
    \delta\alpha_{j}^{\;\;\,i} = \overbar{\partial}_j\xi^{i}\,.
\end{equation}
This is a gauge symmetry of the linearized KS equation $\overbar{\partial}_{[i}\alpha_{j]}^{\;\;\, k} = 0$ in which the quadratic interaction is neglected. The gauge can be fixed by an additional condition outlined in \cite{Bershadsky:1993cx} which is analogous to the gauge fixing conditions (such as the harmonic gauge) in linearized Einstein gravity. Note that this gauge fixing condition uses a K\"ahler metric \cite{costelloGaiotto2018twisted}.

\section{Complex structures on \texorpdfstring{$SU(2)$}{SU(2)} principal bundles}\label{sec:review_of_complex_structures_on_principal_bundles}

Before presenting the relation between Einstein and Kodaira--Spencer gravities, we will first review the construction of almost complex structures on six-dimensional $SU(2) =S^3$ principal fiber bundles over a three-dimensional base $B_3$ first presented in \cite{zentner2013integrable} and later discussed in \cite{Herfray:2016std}. Integrability of this almost complex structure (vanishing of the Nijenhuis tensor) is equivalent to equations on the base \cite{zentner2013integrable} which we now recognize as Einstein's equations in the first-order formulation. We will provide a new clear proof of this statement in the notation of \cite{Herfray:2016std} which has so far been missing from the literature.

To setup the notation, we begin in \Secref{subsec:principal_bundles} by reviewing $SU(2)$ principal bundles and their associated structures such as the notion of verticality, the definition of a connection and horizontal 1-forms. Then we use this data to construct the almost complex structure of \cite{zentner2013integrable} following \cite{Herfray:2016std}. As an example, we discuss the group manifold of the Lie group $SL(2,\mathbb{C})$ which is an $SU(2)$ principal bundle whose base is a hyperbolic space $\mathbb{H}_3\subset \mathbb{R}^4$. This example plays an important role in twisted holography \cite{costelloGaiotto2018twisted}.

In \Secref{subsec:complex_structures_on_principal_bundles} we consider integrability of the aforementioned almost complex structure and give a novel proof of its equivalence with the flatness of the associated $(1,0)$-form as stated without explicit proof in \cite{Herfray:2016std}. This provides a more streamlined and less technical proof of the equivalent result proven in \cite{zentner2013integrable}. Lastly, we also show that the resulting complex manifold is automatically a Calabi--Yau manifold with a holomorphic volume form and a K\"ahler metric. For example, $SL(2,\mathbb{C})$ becomes a Calabi--Yau manifold known as the deformed conifold \cite{Candelas:1989js,costelloGaiotto2018twisted}.

\subsection{Basics of \texorpdfstring{$SU(2)$}{SU(2)} principal bundles}\label{subsec:principal_bundles}

In \Secref{sec:KS_review}, we presented the theory of almost complex structures and their deformations on general even-dimensional differentiable manifolds. We will now specialize to six-dimensional manifolds, which are $SU(2)$ principal (fiber) bundles over a three-dimensional base, and study their complex structures. To this end, let us consider a six-dimensional differentiable manifold $P_6$ which we assume to be the total space of an $SU(2)$ principal bundle with base $B_3$ and a horizontal projector $\pi\colon P_6\to B_3$. In addition, $u\in SU(2)$ acts on $P_6$ via the right-action diffeomorphism $\mathcal{R}_u\colon P_6\rightarrow P_6$ which leaves the projector $\pi\circ \mathcal{R}_u = \pi$ invariant. The pre-image of the projection $\pi^{-1}(p)$ at a point $p\in B_3$ is called the fiber of the bundle and in the present case is isomorphic to $SU(2) = S^3$. Therefore the total space $P_6$ factorizes locally to a direct product $B_3\times S^3$, but not necessarily globally. However, since $SU(2)$ is compact and simply connected, it turns out that all $SU(2)$ principal bundles over a three-dimensional base are trivial \cite{Dijkgraaf:1989pz}, so that $P_6 = B_3\times S^3$ also globally.

The base space of a principal bundle can be identified with a homogeneous space since the horizontal projection $\pi$ defines an equivalence relation on $P_6$ under which two points are in the same equivalence class if they are projected to the same point in $B_3$. In fact, these equivalence classes correspond to $SU(2)$ orbits induced by the right action diffeomorphism, i.e.~the fibers. Since the total space is a product space, we can label each fiber uniquely by a point on the base. Thus, the quotient of the $SU(2)$ action on the total space is the base space, $B_3 = P_6/SU(2)$. See \cite{Nakahara:2003nw} for a text book introduction to principal bundles.

\paragraph{Vertical vector fields.} On any principal bundle, the projection can be used to define the vertical tangent space $V _p\subset T_pP$ and tangent bundle $V$ which consists of vector fields $v^a$ whose push-forward under $\pi$ to the base vanishes,
\begin{equation}
    \pi_{*}(v) = 0\,,\quad v\in V\,.\label{eq:vertical_tangent_space_def}
\end{equation}
Because the fibers over a point are isomorphic to $SU(2)$, the vertical tangent space is isomorphic to the Lie algebra $\mathfrak{su}(2)$. In fact there exists a natural map from a Lie algebra element $a$ to a vertical vector field $a^\#$, called the \textit{fundamental vector field} of $a$. At a point $p\in P_6$, the tangent vector $a^\#_p\in V_p$ of this vector field lies in the direction of the flow defined by the diffeomorphism $\mathcal{R}_{e^{a t }}(p)$, with $t\in \mathbb{R}$, and acts on a function $S\colon P_6\to \mathbb{R}$ as
\begin{equation}
    a^\#_p (S(p)) \equiv \frac{\d}{\d t} S(\mathcal{R}_{e^{a t}}(p))\bigg\vert_{t=0}\,.\label{eq:fundamental_vector_field_def}
\end{equation}
Since the projection is invariant under right action, one can verify $\pi_{*}(a^\#)=0$ for any fundamental vector field. Moreover, under right-action $(\mathcal{R}_u)_*a^\# = (u^{-1}au)^\#$ \cite{Nakahara:2003nw} so that for any $a,b\in\mathfrak{su}(2)$ the fundamental vector fields satisfy the identity
\begin{equation}
    [a^\#,b^\#]= [a,b]^\#\,, \label{eq:Lie_alg_iso}
\end{equation}
where the bracket on the left hand side denotes the Lie bracket on the tangent space of $P_6$ and on the right hand side the one on the Lie algebra, thus making the map a Lie algebra isomorphism.

\paragraph{Horizontality.} On any principal bundle $P_6$ it is possible to define horizontal 1-forms whose contractions with all vertical vector fields vanish. Intuitively, these are the 1-forms with form legs in the direction of the base $B_3$. Consider a basis $\mathcal{E}^i$ of three such horizontal 1-forms and define the horizontal $\mathfrak{su}(2)$-valued 1-form
\begin{equation}
    \mathcal{E}_a = \mathcal{E}_a^i\,\tau_i\,,
\end{equation}
where $\tau_i$ are the generators \eqref{eq:su2_lie_algebra} of $\mathfrak{su}(2)$. For any fundamental vector field $b^\#$, it satisfies
\begin{equation}
    (b^\#)^a\,\mathcal{E}_a  = 0\,.\label{eq:upliftframe_fundamental_vfield}
\end{equation}
We further assume that our basis $\mathcal{E}^i$ is such that $\mathcal{E}$ is equivariant under the right-action of $SU(2)$ on $P_6$, in other words, $\mathcal{E}$ satisfies
\begin{equation}
    \widetilde{\mathcal{E}}_a(x)  = u^{-1}\,\mathcal{E}_a(x)\,u\,.
    \label{eq:E_adjoint_transformation}
\end{equation}
Such a horizontal $\mathfrak{su}(2)$-valued 1-form is called a 1-form of adjoint-type and its existence is an additional structure on $P_6$.

In contrast to horizontal 1-forms, there does not a priori exist a notion of horizontality for vector fields on general principal bundles. Thus we assume that $P_6$ is equipped with a connection which is an assignment of a horizontal tangent space $H_p$ complement to $V_p$ such that $T_pP = V_p \oplus H_p$ and $(\mathcal{R}_u)_* H_p = H_{\mathcal{R}_u(p)} $. Elements of the horizontal tangent bundle are then called horizontal vector fields. As a result, there exists a connection 1-form
\begin{equation}
    \mathcal{W}_a = \mathcal{W}^i_a\,\tau_i\,,
\end{equation}
which is an $\mathfrak{su}(2)$-valued 1-form that has the following two properties: it annihilates all horizontal vectors $h\in H$ as $\mathcal{W}_a\,h^a = 0$ and its contraction with any fundamental vector field $b^\#$ is given by

\begin{equation}
    (b^\#)^a\,\mathcal{W}_a  = b\,.\label{eq:connection_fundamental_vfield}
\end{equation}
From these two properties it follows that $\mathcal{W}$ transforms equivariantly under the right-action
\begin{equation}
    \widetilde{\mathcal{W}}_a(x)  = u^{-1}\,\mathcal{W}_a(x)\,u\,.
    \label{eq:connection_adjoint_transformation}
\end{equation}
Together, the 1-forms $\mathcal{E}^i$ and $\mathcal{W}^i$ form a basis for the cotangent bundle on $P_6$. Let $\mathcal{E}_i$ and $\mathcal{W}_i$ denote the vector fields that form the dual basis for the vector bundle, in other words, they satisfy
\begin{equation}
    \mathcal{E}_a^i\,\mathcal{E}^{a}_j = \delta^{i}_j\,,\quad \mathcal{W}_a^i\,\mathcal{W}^{a}_j = \delta^{i}_j\,,\quad \mathcal{W}^{a}_i\,\mathcal{E}_a^j =\mathcal{W}_{a}^i\,\mathcal{E}^a_j = 0\,.
    \label{eq:orthonormality_E_W}
\end{equation}
It follows that $\mathcal{E}_i$ are horizontal and $\mathcal{W}_i$ are vertical vector fields.

\paragraph{$SL(2,\mathbb{C})$ as a principal bundle.} As a benchmark example we discuss the group manifold of $SL(2,\mathbb{C})$ which can be written as an $SU(2)$ principal bundle over $\mathbb{H}_3\subset \mathbb{R}^4$ \cite{jensen2016surfaces}. To make this explicit, we consider the group manifold $P_6=SL(2,\mathbb{C})$, which can be defined via the embedding into $\mathbb{C}^4$ as
\begin{equation}
    SL(2,\mathbb{C})=\left\{\mathcal{M}=\begin{pmatrix}
        v_1 & v_2\\
        v_3 & v_4
    \end{pmatrix}\bigg\vert\,  v_1v_4-v_2v_3=1\,,\,(v_1,v_2,v_3,v_4)\in\mathbb{C}^4\right\}\,.
    \label{eq:deformed_conifold}
\end{equation}
The hyperbolic space $\mathbb{H}_3$ is the subset of $\mathbb{R}^4$ defined by
\begin{equation}
    \mathbb{H}_3 = \{(x_1,x_2,x_3,x_4)\in\mathbb{R}^4 \,\vert\, x_1^2+x_2^2+x_3^2-x_4^2=-1 \}\,.
    \label{eq:hyperbolic_plane}
\end{equation}
When $\mathbb{R}^4$ is endowed with the flat Euclidean metric, the induced metric on $\mathbb{H}_3$ has constant negative curvature equal to $-1$, corresponding to AdS radius $\ell=1$ and justifying the nomenclature of $\mathbb{H}_3$. A space with general AdS radius may be obtained by rescaling the coordinates with a positive $\ell >0$. However, in what follows, we do not assume the presence of any metric and simply understand $\mathbb{H}_3$ as a subset of $\mathbb{R}^4$.

Note that $\mathbb{H}_3$ is isomorphic to the space of Hermitian matrices $H_+(2,\mathbb{C}) $ with unit determinant defined by
\begin{equation}
    H_+(2,\mathbb{C}) = \{\mathcal{H}\in GL(2,\mathbb{C})\,\vert\,\mathcal{H}^\dagger = \mathcal{H}\,,\, \det{\mathcal{H}} = 1\}\,.
\end{equation}
Parametrizing Hermitian matrices as
\begin{equation}
    \mathcal{H} = \begin{pmatrix}
        x_1 & x_2+ix_3\\
        x_2-ix_3 & -x_1+\frac{x_4^2}{x_1}
    \end{pmatrix}\,,\quad (x_1,x_2,x_3,x_4) \in \mathbb{R}^4\,,
\end{equation}
the isomorphism with \eqref{eq:hyperbolic_plane} follows from
\begin{equation}
    \det\,{\mathcal{H}} = -x_1^2-x_2^2-x_3^2+x_4^2 = 1\,.
\end{equation} 
We define the horizontal projection of the principal bundle, i.e.~a map from $SL(2,\mathbb{C})$ to $\mathbb{H}_3$ via
\begin{equation}
    \pi\colon P_6\to H_+(2,\mathbb{C}) \cong \mathbb{H}_3\,,\quad \pi(\mathcal{M})\equiv \mathcal{M}\mathcal{M}^\dagger\,.
\end{equation}
It is straightforward to show that $\mathcal{M}\mathcal{M}^\dagger$ is Hermitian, has positive determinant and, that any Hermitian matrix with positive determinant can be represented in this way by $\mathcal{M}\in SL(2,\mathbb{C})$. In constructing the horizontal projection, we have established the first ingredient for the fiber bundle structure.

The second ingredient is the right-action of $SU(2)$ which is obtained by identifying $ u \in SU(2)$ as an element of $SL(2,\mathbb{C})$ that satisfies $u^\dagger u = u u^\dagger = 1$. The right-action of $SU(2)$ is then simply defined as
\begin{equation}
    \widetilde{\mathcal{M}} \equiv \mathcal{M}\,u
    \label{eq:right_action_on_SL2C}
\end{equation}
and we see it respects the projection $\pi(\widetilde{\mathcal{M}}) = \mathcal{M}\,uu^\dagger\, \mathcal{M}^\dagger =  \pi(\mathcal{M})$ as required. The fiber, i.e.~the orbit of the right-action, is thus isomorphic to $SU(2) = S^3$. In addition, the base is
\begin{equation}
    H_+(2,\mathbb{C})=SL(2,\mathbb{C})/SU(2)\cong\mathbb{H}_3\,.
\end{equation}
Thus we have seen that $P_6=SL(2,\mathbb{C})$ can be regarded as a six-dimensional principal bundle with $SU(2)\cong S^3$ fiber and base $\mathbb{H}_3$. In fact this is true globally as mentioned above, so the total manifold is the product $SL(2,\mathbb{C})=\mathbb{H}_3\times S^3$. This is clear from the polar decomposition $\mathcal{M} = \Delta^{1\slash 2} \,u$, with $\Delta = \mathcal{M}\mathcal{M}^\dagger \in H_+(2,\mathbb{C})$ and $u\in SU(2)$, which provides the mapping $\mathcal{M}\mapsto (\Delta,u)\in \mathbb{H}_3\times S^3$.

Lastly, by the embedding \eqref{eq:deformed_conifold} into $\mathbb{C}^4$, $SL(2,\mathbb{C})$ inherits a natural complex structure $J_{\text{ind}}$ obtained by pulling back the form components of the standard complex structure on $\mathbb{C}^4$ and restricting their action on functions on $SL(2,\mathbb{C})$ (see \Appref{app:induced_complex_structure}). The complex coordinates corresponding to $J_{\text{ind}}$ can be chosen to be the three coordinates $(w_1,w_2,w_3)=(v_1,v_2,v_3)$ of $\mathbb{C}^4$ (or their holomorphic reparametrizations) since the embedding equation $v_1v_4-v_2v_3 = 1$ is holomorphic. The fact that these are compatible with $J_{\text{ind}}$ is shown in detail in \Appref{app:induced_complex_structure}.

\subsection{\texorpdfstring{$SU(2)$}{SU(2)} principal bundles as complex manifolds}\label{subsec:complex_structures_on_principal_bundles}

Let $P_6$ be an $SU(2)$ principal bundle whose properties we reviewed in detail in the previous section. There we introduced all the ingredients necessary to construct an almost complex structure on $P_6$: the $SU(2)$-equivariant horizontal 1-form $\mathcal{E} = \mathcal{E}^i\tau_i$, the connection $\mathcal{W} = \mathcal{W}^i\tau_i$ 1-form and their associated dual horizontal $\mathcal{E}_i$ and vertical $\mathcal{W}_i$ vector fields respectively. The almost complex structure is then defined by \cite{Herfray:2016std}
\begin{equation}
    J^b_a  =\mathcal{W}^i_a\, \mathcal{E}^{b}_i-\mathcal{E}^i_a\,\mathcal{W}^{b}_i\,, 
\label{eq:complex_structure_bundle}
\end{equation}
which due to the relations \eqref{eq:orthonormality_E_W} indeed satisfies $J^2 = -1 $.

The almost complex structure defined in \eqref{eq:complex_structure_bundle} has two important properties. First, it maps vertical vector fields to horizontal ones and vice versa, which follows from that fact that $J(\mathcal{W}_i) = \mathcal{E}_i$ and $J(\mathcal{E}_i) = -\mathcal{W}_i$ and since $\mathcal{W}_i$ and $\mathcal{E}_i$ span the vertical and horizontal tangent bundles respectively, as long as they are linearly independent. Thus, there is no linear subspace of the vertical tangent space that is left invariant by the complex structure. Using the definition of vertical vector fields \eqref{eq:vertical_tangent_space_def}, for any $v\in V$ we write this property as
\begin{equation}
    \pi_{*}J (v) \neq 0\,.\label{eq:no_vertical_inv_subspace}
\end{equation}
The same property is shared by other almost complex structures on fiber bundles that have appeared in the literature \cite{Dombrowski1962,aguilar1996isotropic,zentner2013integrable}. In \Appref{app:no_invariant_subspace} and below in \Secref{subsec:reduction} we prove that the condition \eqref{eq:no_vertical_inv_subspace} is equivalent to a certain $3\times 3$ block of the components of $J$ being invertible.\footnote{To our knowledge, the equivalence between \eqref{eq:no_vertical_inv_subspace} and invertibility of certain components of $J$ has not been previously noticed in the literature.}

Second, due to $SU(2)$-equivariance \eqref{eq:E_adjoint_transformation} and \eqref{eq:connection_adjoint_transformation} of the 1-forms $\mathcal{E}^i$ and $\mathcal{W}^i$, the almost complex structure \eqref{eq:complex_structure_bundle} is $SU(2)$-invariant, in other words, under the right-action diffeomorphism $\mathcal{R}_u$ of $SU(2)$ on $P_6$ it transforms as
\begin{equation}
    \widetilde{J}^a_b(x) = J^a_b(x)\,.
    \label{eq:su2_invariance}
\end{equation}
where the left-hand side is given by \eqref{eq:J_tilde} with $\mathcal{D} = \mathcal{R}_u$. This follows from the fact that equivariance translates to
\begin{equation}
    \widetilde{\mathcal{E}}^i(x) = \Lambda^{i}_{\;\;j}\,\mathcal{E}^j(x)\,,\quad \widetilde{\mathcal{E}}_i(x) = (\Lambda^{-1})^{j}_{\;\;i}\,\mathcal{E}_j(x)
\end{equation}
where $\Lambda \in SO(3)$ is such that $u^{-1} \tau_i u = \Lambda^{j}_{\;\;i}\,\tau_j$. The connection $\mathcal{W}^i$ and its dual vector field $\mathcal{W}_i$ transform similarly. If $\mathcal{E}$ is not $SU(2)$-equivariant \eqref{eq:E_adjoint_transformation} then $J$ still defines an almost complex structure, but it is no longer $SU(2)$-invariant.

Now, we can define the complex valued 1-forms \cite{Herfray:2016std}
\begin{equation}
    \mathcal{A}^i = \mathcal{W}^i+i\,\mathcal{E}^i\,,\quad \overbar{\mathcal{A}}^i = \mathcal{W}^i-i\,\mathcal{E}^i\,,
    \label{eq:A_forms_W_E}
\end{equation}
and the dual vector fields
\begin{equation}
    \mathcal{V}_i = \frac{1}{2}\,(\mathcal{W}_i-i\,\mathcal{E}_i)\,,\quad \overbar{\mathcal{V}}_i =\frac{1}{2}\,(\mathcal{W}_i+i\,\mathcal{E}_i)\,.
    \label{eq:dual_vector_fields_W_E}
\end{equation}
They allow to express the complex structure \eqref{eq:complex_structure_bundle} in the form \eqref{eq:complex_structure_A_V}. It follows that  $\mathcal{A}^i$ and $\overbar{\mathcal{A}}^i$ are a set of $(1,0)$- and $(0,1)$-forms with respect to the almost complex structure \eqref{eq:complex_structure_bundle} respectively. Similarly, $\mathcal{V}_i$ and $\overbar{\mathcal{V}}_i$ are $(1,0)$- and $(0,1)$-vector fields respectively. 

From the $(1,0)$- and $(0,1)$-forms of the above almost complex structure, we can define the $\mathfrak{sl}(2,\mathbb{C})$-valued 1-forms
\begin{equation}
    \mathcal{A} = \mathcal{A}^i\,\tau_i\,,\quad \overbar{\mathcal{A}} = \overbar{\mathcal{A}}^i\,\tau_i\,.
\end{equation}
Furthermore, because of \eqref{eq:E_adjoint_transformation} and \eqref{eq:connection_adjoint_transformation}, $\mathcal{A}$ transforms equivariantly under right action,
\begin{equation}
    \widetilde{\mathcal{A}}(x) = u^{-1}\,\mathcal{A}(x)\,u\,.\label{eq:equivariance_of_A}
\end{equation}

\paragraph{Flatness and integrability.} The 1-form $\mathcal{A}$ on $P_6$ is defined to be flat when its curvature 2-form $F_\mathcal{A}$ vanishes
\begin{equation}\label{eq:flatness_connection}
    F_\mathcal{A} \equiv \d\mathcal{A}+\mathcal{A}\wedge \mathcal{A}=0\,,
\end{equation}
where the exterior derivative and the wedge product are six-dimensional. For $\mathcal{E}$ and $\mathcal{W}$, this is equivalent to the equations
\begin{equation}
    \d \mathcal{W} + \mathcal{W}\wedge \mathcal{W} - \mathcal{E}\wedge \mathcal{E} = 0\,,\quad \d \mathcal{E} + \mathcal{W}\wedge \mathcal{E} + \mathcal{E}\wedge \mathcal{W} = 0\,.
    \label{eq:flatness_connection_2}
\end{equation}
In \cite{Herfray:2016std}, it is claimed without explicit proof that \eqref{eq:flatness_connection} is equivalent to the almost complex structure \eqref{eq:complex_structure_bundle} being integrable. An explicit proof can be found in \cite{zentner2013integrable} where the equivalence of \eqref{eq:flatness_connection_2} and integrability is shown. Here we will present an alternative and less technical proof of the equivalence working directly at the level of the $(1,0)$-form $\mathcal{A}$. Certain details of the proof are relegated to \Appref{app:flatness_integrability}.

To start, we rewrite flatness \eqref{eq:flatness_connection} in an alternative form. Notice that since the vector fields \eqref{eq:dual_vector_fields_W_E} span the tangent bundle of $P_6$, flatness \eqref{eq:flatness_connection} is equivalent to the $\mathfrak{sl}(2,\mathbb{C})$-valued equations
\begin{equation}
    F_\mathcal{A}(\mathcal{V}_i,\mathcal{V}_j) = F_\mathcal{A}(\overbar{\mathcal{V}}_i,\overbar{\mathcal{V}}_j)  = F_\mathcal{A}(\mathcal{V}_i,\overbar{\mathcal{V}}_j) = 0\,.
\end{equation}
For two arbitrary vector fields $X,Y$ on $P_6$ we have the general formula \cite{Nakahara:2003nw}
\begin{equation}
    F_\mathcal{A}(X,Y) = X(\mathcal{A}(Y))-Y(\mathcal{A}(X))-\mathcal{A}([X,Y])+[\mathcal{A}(X),\mathcal{A}(Y)]\,.
\end{equation}
Using $\mathcal{A}(\overbar{\mathcal{V}}_i) = 0$, we obtain
\begin{align}
   F_\mathcal{A}(\overbar{\mathcal{V}}_i,\overbar{\mathcal{V}}_j) = -\mathcal{A}([\overbar{\mathcal{V}}_i,\overbar{\mathcal{V}}_j])\,,\quad F_\mathcal{A}(\mathcal{V}_i,\overbar{\mathcal{V}}_j) = -\mathcal{A}([\mathcal{V}_i,\overbar{\mathcal{V}}_j])\,.\label{eq:flatness_aa_ha_comp}
\end{align}
Using also $\mathcal{A}(\mathcal{V}_i) = \tau_i$, which follows from the duality relations \eqref{eq:orthonormality} of the components, and the Lie algebra \eqref{eq:su2_lie_algebra}, we obtain similarly
\begin{equation}
    F_\mathcal{A}(\mathcal{V}_i,\mathcal{V}_j) = -\mathcal{A}([\mathcal{V}_i,\mathcal{V}_j]-\varepsilon_{ij}^{\;\;\;k}\,\mathcal{V}_k)\,.\label{eq:flatness_hh_comp}
\end{equation}
Equations \eqref{eq:flatness_aa_ha_comp} and \eqref{eq:flatness_hh_comp} imply that
\begin{equation}
    \Pi([\mathcal{V}_i,\mathcal{V}_j])=\varepsilon  _{ij}^{\;\;\;k}\,\mathcal{V}_k \,,\quad \Pi([\overbar{\mathcal{V}}_i,\overbar{\mathcal{V}}_j]) =  \Pi([\mathcal{V}_i,\overbar{\mathcal{V}}_j]) = 0\,,
    \label{eq:flatness_V_midstep}
\end{equation}
where the projector is defined in \eqref{eq:projectors}. Taking the complex conjugate of the last equation gives $ \overbar{\Pi}([\mathcal{V}_i,\overbar{\mathcal{V}}_j]) = 0$ so that $[\mathcal{V}_i,\overbar{\mathcal{V}}_j] = 0$. The complex conjugate of the second equation on the other hand implies $\overbar{\Pi}([\mathcal{V}_i,\mathcal{V}_j]) = 0$ so that the $\Pi$-projection in the first equation can be removed. Therefore flatness \eqref{eq:flatness_connection} of $\mathcal{A}$ is equivalent to
\begin{equation}
    [\mathcal{V}_i,\mathcal{V}_j] = \varepsilon_{ij}^{\;\;\;k}\,\mathcal{V}_k\,,\quad [\mathcal{V}_i,\overbar{\mathcal{V}}_j] = 0\,.
    \label{eq:flatness_V}
\end{equation}
Due to the absence of an invariant vertical subspace \eqref{eq:no_vertical_inv_subspace}, we can write \eqref{eq:dual_vector_fields_W_E} (see \Appref{app:flatness_integrability} for the derivation)
\begin{equation}
    \mathcal{V}_i = \Pi( \tau_i^{\#})\,,\label{eq:V_fundamental_vector_field}
\end{equation}
where $\tau_i$ is an $\mathfrak{su}(2) $ generator. Substituting to \eqref{eq:flatness_V}, we find flatness is equivalent to
\begin{equation}
    [\Pi(\tau_i^{\#}),\Pi(\tau_j^{\#})] - \Pi([\tau_i^{\#},\tau_j^{\#}]) = 0\,,\quad [\Pi(\tau_i^{\#}),\overbar{\Pi}(\tau_j^{\#})] = 0\,,
    \label{eq:flatness_alt}
\end{equation}
where we have used that fundamental vector fields give a Lie algebra isomorphism \eqref{eq:Lie_alg_iso}. We prove in \Appref{app:flatness_integrability} that for an $SU(2)$-invariant almost complex structure satisfying \eqref{eq:su2_invariance}, we have
\begin{equation}
    [\Pi(\tau_i^{\#}),\Pi(\tau_j^{\#})] - \Pi([\tau_i^{\#},\tau_j^{\#}]) = \frac{1}{4}\,N(\tau_i^{\#},\tau_j^{\#})\,,\quad [\Pi(\tau_i^{\#}),\overbar{\Pi}(\tau_j^{\#})] = -\frac{1}{4}\,N(\tau_i^{\#},\tau_j^{\#})\,,\label{eq:brackets_fundamental_vector_fields}
\end{equation}
where $N(X,Y)$ is the Nijenhuis tensor \eqref{eq:vanishing_Nijenhuis} contracted with two vector fields $X,Y$. Thus flatness \eqref{eq:flatness_alt} is equivalent to
\begin{equation}
    N(\tau_i^{\#},\tau_j^{\#}) = 0\,.
    \label{eq:flatness_integrability}
\end{equation}
Since by assumption there is no invariant vertical subspace \eqref{eq:no_vertical_inv_subspace}, the vertical tangent space is spanned by $\tau_i^\#$ and the horizontal subspace by $J(\tau_i^{\#})$. Combining with the fact that the Nijenhuis tensor satisfies $N(J\tau_i^{\#},\tau_j^{\#}) = -JN(\tau_i^{\#},\tau_j^{\#})$, \eqref{eq:flatness_integrability} is equivalent to the vanishing of all components of the Nijenhuis tensor $N(X,Y) = 0$ for all vector fields $X,Y$. Hence flatness \eqref{eq:flatness_connection} of the $(1,0)$-form $\mathcal{A}$ is equivalent to the integrability of the almost complex structure \eqref{eq:complex_structure_bundle}.

Note that the absence of an invariant vertical subspace \eqref{eq:no_vertical_inv_subspace} and the $SU(2)$-invariance \eqref{eq:su2_invariance} of the almost complex structure are fundamental for the proof. Without them, flatness only implies integrability, but not the other way around. This is because integrability concerns the closure of the Lie bracket of the vector fields $\mathcal{V}_i$ requiring that $[\mathcal{V}_i,\mathcal{V}_j] = Q_{ij}^{\;\;\;k}\,\mathcal{V}_k$ with some set of structure functions $Q_{ij}^{\;\;\;k} = Q_{ij}^{\;\;\;k}(x)$. Flatness \eqref{eq:flatness_V} clearly implies that the Lie bracket closes, but it also requires that the structure functions to coincide with $\mathfrak{su}(2)$ structure constants $Q_{ij}^{\;\;\;k}(x) = \varepsilon_{ij}^{\;\;\;k}$. The latter is only true if the almost complex structure has the aforementioned properties.

\paragraph{Complex coordinates.} When the $SU(2)$-equivariant 1-form $\mathcal{A} = \mathcal{W}+i\mathcal{E} $ is flat, the $SU(2)$-invariant almost complex structure $J$ given by \eqref{eq:complex_structure_bundle} is integrable. Due to $SU(2)$-invariance, the right-action diffeomorphism $\mathcal{R}_u$ takes a simple form in the complex coordinates $(w^i,\overbar{w}^i)$ of $J$. Under $\mathcal{R}_u$, the complex coordinates transform as scalars as in \eqref{eq:complex_coordinates_are_scalars}. However, since $J$ is invariant under these diffeomorphisms, the transformed coordinates must be holomorphically related to the previous ones in order to solve their defining equation \eqref{eq:definition_holomorphic_coordinates}. In other words, the action of $\mathcal{R}_u$ from the right becomes an action of a holomorphic-antiholomorphic diffeomorphism $(f^i_u,\overbar{f}^i_u)$ from the left:
\begin{equation}
    w^i(\mathcal{R}_u(x)) = f_u^i(w)\,,\quad \overbar{w}^i(\mathcal{R}_u(x)) = \overbar{f}_u^i(\overbar{w})\,.
\end{equation}
This implies that when the right-action $\mathcal{R}_u$ is written in complex coordinates it takes the simple form
\begin{equation}
    R_u^i(w,\overbar{w}) = f_u^i(w)\,,\quad \overbar{R}_u^i(w,\overbar{w}) = \overbar{f}_u^i(\overbar{w})\,.
    \label{eq:right_action_complex_coordinates}
\end{equation}
Thus the invariance of $J$ under right-action is due to the fact that $\mathcal{R}_u$ is a holomorphic-antiholomorphic diffeomorphism.

The complex coordinates $(w^i,\overbar{w}^i)$ of $J$ also become complex coordinates on $SL(2,\mathbb{C}) $ in the following manner. First, note that the flatness equation \eqref{eq:flatness_connection} for $\mathcal{A}$ has local solutions of the form
\begin{equation}
    \mathcal{A} = \mathcal{M}^{-1}\d \mathcal{M} \,,\quad \overbar{\mathcal{A}} = \overbar{\mathcal{M}}^{-1}\d\overbar{\mathcal{M}}\,,
    \label{eq:solutions_flat_sl2c_connection}
\end{equation}
where $\mathcal{M}(x),\overbar{\mathcal{M}}(x) \in SL(2,\mathbb{C})$ are related in such a way to ensure that $\overbar{\mathcal{A}}^i = (\mathcal{A}^i)^* $. We see that a flat $\mathcal{A}$ on $P_6$ defines a natural map $\mathcal{M}\colon P_6\rightarrow SL(2,\mathbb{C})$ from the principal bundle to $SL(2,\mathbb{C})$. The $SU(2)$-equivariance of $\mathcal{A}$ implies that under right-action we have
\begin{equation}
    \widetilde{\mathcal{M}}(x) = \mathcal{M}(\mathcal{R}_u(x)) = \mathcal{M}(x)\,u
    \label{eq:curly_M_right_action}
\end{equation}
so that the right-action diffeomorphism on $P_6$ maps to the right-action of $SU(2)$ on $SL(2,\mathbb{C})$ defined in \eqref{eq:right_action_on_SL2C}.

As discussed above in \Secref{subsec:complex_structures_intro}, the 1-forms $\mathcal{A}^i$ are $(1,0)$-forms respect to the complex structure $J$ which by \eqref{eq:solutions_flat_sl2c_connection} amounts to
\begin{equation}
    \overbar{\Pi}^b_a\, \partial_b\mathcal{M} = \frac{1}{2}\,(\partial_a +i\,J^b_a\,\partial_b)\,\mathcal{M} = 0\,.
    \label{eq:M_is_holomorphic}
\end{equation}
Thus, by \eqref{eq:definition_holomorphic_coordinates}, $\mathcal{M}= \mathcal{M}(w)$ is a holomorphic function so that the complex coordinates of $J$ on $P_6$ define complex coordinates on $SL(2,\mathbb{C})$, in other words, $J $ defines a complex structure on $SL(2,\mathbb{C})$. These coordinates on $P_6$ are related to the ones on $SL(2,\mathbb{C})$ defined in the previous section and \Appref{app:induced_complex_structure} by a holomorphic map. Thus the complex structure here coincides $J = J_{\text{ind}}$ with the structure inherited from the complex structure of $\mathbb{C}^4$ via the embedding $SL(2,\mathbb{C})\subset \mathbb{C}^4$. 

Now since $\mathcal{M}$ is a holomorphic with respect to $J$, we obtain
\begin{equation}
    \mathcal{A}^i  =(\mathcal{M}^{-1}\, \partial_j \mathcal{M})^i\, \d w^j\equiv \Sigma^i_j\, \d w^j\,,\label{eq:holomorphic_sigma}
\end{equation}
where $\Sigma^i_j =\Sigma^i_j(w) $ are holomorphic functions. Similarly, $\overbar{\mathcal{A}}^i = \overbar{\Sigma}^i_j\,\d \overbar{w}^j$ with $\overbar{\Sigma}^i_j =\overbar{\Sigma}^i_j(\overbar{w}) $ are antiholomorphic. It follows that under right-action \eqref{eq:right_action_complex_coordinates}, we have the transformation
\begin{equation}
    \widetilde{\Sigma}^i_j(w) = \partial_j f^k_u(w)\,\Sigma^i_k(f_u(w)) = \Lambda^i_{\;\;k}\,\Sigma^k_j(w)\,.
    \label{eq:Sigma_right_action}
\end{equation}
where $\Lambda \in SO(3)$ such that $u^{-1}\tau_i u = \Lambda^j_{\;\;i}\,\tau_j$ and we used the transformation law \eqref{eq:Sigma_transformation_C}. A similar equation holds for $\overbar{\Sigma}^i_j$.

\paragraph{Calabi--Yau structure.} Given an $\mathfrak{sl}(2,\mathbb{C})$-valued 1-form $\mathcal{A}$ satisfying \eqref{eq:flatness_connection}, the complex manifold $(P_6,J)$ automatically admits a holomorphic volume form and a K\"ahler metric making it a Calabi--Yau manifold. This shows that Kodaira--Spencer gravity can be defined on any six-dimensional $SU(2)$ principal bundle equipped with the above complex structure (see the discussion in \Secref{subsec:deformations_review}). We are not aware of a proof of this statement in the literature so let us show it here explicitly.

The holomorphic volume form is explicitly given by \cite{Herfray:2016std}
\begin{equation}
      \Omega =-4 \tr{(\mathcal{A}\wedge\mathcal{A}\wedge\mathcal{A})}=\varepsilon_{ijk}\,\mathcal{A}^i\wedge\mathcal{A}^j\wedge\mathcal{A}^k\,,
\end{equation}
where we used $\tr{(\tau_i\tau_j\tau_k)}= -\frac{1}{4}\,\varepsilon_{ijk}$. This is indeed a $(3,0)$-form, as $\mathcal{A}^i$ are $(1,0)$-forms. To show that $\Omega$ is non-degenerate, we write it in the complex coordinates $w^i$ associated with $J$ using \eqref{eq:connection_sigma_def},
\begin{equation}
    \Omega = \varepsilon_{ijk}\, \Sigma^i_l\, \Sigma^j_m \Sigma^k_n \,\d w^l\wedge\d w^m \wedge \d w^n = \frac{1}{3!}\det{\Sigma} \,\d w^1\wedge\d w^2\wedge\d w^3\,.
\end{equation}
Since $\Sigma^i_j$ form a $GL(3,\mathbb{C})$ element, we find that $\det{\Sigma}\neq 0$. In addition, since $\mathcal{A}$ is flat, $\Sigma^i_j$ and $\det{\Sigma}$ are holomorphic functions as discussed below \eqref{eq:holomorphic_sigma}. Thus $\Omega$ defines a non-degenerate $(3,0)$-form with holomorphic components making it a holomorphic volume form.  Note further, that $\Omega$ is invariant under the right action diffeomorphism, which follows from \eqref{eq:equivariance_of_A}, or equivalently, from \eqref{eq:Sigma_right_action} together with $\det{\Lambda} = 1$ for $\Lambda \in SO(3)$. Furthermore, we have $|\det{\Sigma}\,| >0$, such that $\Omega\wedge\overbar{\Omega}$ is a real positive volume form.

Now, we turn towards the discussion of the K\"ahler metric. To this end, we define the real function
\begin{equation}
    \kappa(x)=\tr{(\mathcal{M}\mathcal{M}^\dagger)}\,,
\end{equation}
where $\mathcal{M}$ is defined by \eqref{eq:solutions_flat_sl2c_connection} and its existence follows from the flatness of $\mathcal{A}$ assuming $\mathcal{A}$ has no holonomies. Note that $\kappa$ is $SU(2)$-invariant, since $\mathcal{M}$ transforms as \eqref{eq:curly_M_right_action} under right-action. For any scalar field $K(\kappa(x))$, we obtain a Hermitian metric via
\begin{equation}
    h= h_{ij}\,\d w ^i\otimes \d \overbar{w}^j + h_{ji} \,\d \overbar{w}^i\otimes \d w^j\,,
\end{equation}
where the components are defined as
\begin{equation}
    h_{ij} \equiv \partial_i \overbar{\partial}_j K(\kappa) = K''(\kappa)\tr{(\partial_i \mathcal{M} \,\mathcal{M}^\dagger)} \tr{(\mathcal{M}\,\overbar{\partial}_j\mathcal{M}^\dagger)}+K'(\kappa) \tr{(\partial_i \mathcal{M}\,\overbar{\partial}_j \mathcal{M}^\dagger )}\,.\label{eq:Kahler_metric}
\end{equation}
Here, we have used the fact that $\mathcal{M}$ is holomorphic and $\mathcal{M}^\dagger$ is antiholomorphic. The Hermitian metric $h$ constructed this way is positive definite only for a special class of functions called plurisubharmonic functions. Any function $K(\kappa)$ which satisfies this constraint is called K\"ahler potential. Moreover, the corresponding $(1,1)$-form\footnote{In particular, $K(\kappa)= \kappa$ is a valid choice of K\"ahler potential and yields $\omega_{ij}=\tr{(\partial_{i} \mathcal{M}\,\overbar{\partial}_{j} \mathcal{M}^\dagger)}$. The resulting K\"ahler form differs from the one given in \cite{costelloGaiotto2018twisted} defined as $\omega_{ab}=-\frac{1}{2}\delta_{ij}\,\mathcal{A}^i_{[a}(\mathcal{A}^\dagger)^j_{b]}$ which is not closed as can be seen from the fact that $\mathcal{A}$ is flat instead of closed.}
\begin{equation}
    \omega = i\,h_{ij}\,\d w^i\wedge\d\overbar{w} ^j\,
\end{equation}
is closed. Furthermore, it is possible to find a K\"ahler potential $K(\kappa)$ such that the associated Hermitian metric \eqref{eq:Kahler_metric} is Ricci flat, \cite{Candelas:1989js}
\begin{equation}
    K'(\kappa)=\left(\frac{3}{8}\right)^{\frac{1}{3}}\frac{(\sinh{(2\arcosh{\frac{\kappa}{2}})}-2\arcosh{\frac{\kappa}{2}})^{\frac{1}{3}}}{\sinh{\tau}}\,.
    \label{eq:Ricci_flat_K}
\end{equation}
However, in the case of $SL(2,\mathbb{C})$ the metric obtained for this choice of $K$ does not pull-back\footnote{The pull-back onto the base must be done along a section which will be defined below in \Secref{sec:uplift}.\label{footnote:pullback}} to a negatively curved AdS metric on the base $B_3$ of the bundle, but instead to a flat metric. In fact, this metric is the metric of a deformed conifold, with the cone direction given by $\kappa = \tr{(\mathcal{M}\mathcal{M}^\dagger)}$. When the AdS radius is small, $\ell\to0$, the metric reduces to that of a cone over $S^3\times S^2$ and for constant slices of the cone coordinate, equals the Sasaki--Einstein metric $T^{1,1}$ \cite{Klebanov:2000hb, Candelas:1989js, Ohta:1999we}.

For completeness, we now discuss another natural Hermitian metric that can be defined on an $SU(2)$ principal bundle, but which does not play a role for our construction in this paper and for Kodaira--Spencer gravity. The second metric $\hat{h}$ is defined by $\hat{h}_{ab} = \delta_{ij}\,\mathcal{A}^i_a\,\overbar{\mathcal{A}}^j_b$ \cite{Herfray:2016std}. The corresponding $(1,1)$-form $\hat{\omega}$ has the components $\hat{\omega}_{ab} = \frac{i}{2}\,\delta_{ij} \mathcal{A}^i_{[a} \overbar{\mathcal{A}}^j_{b]} = \delta_{ij}\,\mathcal{W}^i_{[a}\mathcal{E}^i_{b]}$, but it is not closed since $\mathcal{A}$ is flat so that $\hat{\omega}$ is not a K\"ahler form. Therefore, the associated Hermitian metric is not K\"ahler, a fact pointed out in \cite{Herfray:2016std}. The correct K\"ahler metric on the principal bundle is instead \eqref{eq:Kahler_metric}. On fundamental vector fields $a^\#$ and $b^\#$ the metric $\hat{h}$ coincides with the Killing form $-2\tr{(ab)}$ on the $\mathfrak{su}(2)$ Lie algebra. In fact, the metric $\hat{h}$ is a continuation of the Killing form by acting on horizontal vector fields with the complex structure. The pullback\footnote{See footnote \ref{footnote:pullback}.} of this metric onto the base gives the metric \eqref{eq:metric_vielbein}.

\section{Uplift of classical Einstein gravity to Kodaira--Spencer gravity}\label{sec:uplift}

In the previous section, we introduced the construction \cite{zentner2013integrable,Herfray:2016std} of an integrable almost complex structure on a six-dimensional $SU(2)$ principal bundle $B_3\times S^3$ from a flat $\mathfrak{sl}(2,\mathbb{C})$-valued 1-form. In this section, we present our main result and apply the construction to relate the fundamental field of three-dimensional Einstein gravity in the first-order formulation, an $\mathfrak{sl}(2,\mathbb{C})$-valued three-dimensional gauge field, to the field of six-dimensional Kodaira--Spencer gravity, a deformation of a background complex structure. This is an off-shell mapping of the degrees of freedom of the two theories which on-shell reduces to a relation between classical solutions: a pair of flat three-dimensional gauge fields on the base map to an integrable complex structure deformation on the bundle.

We begin in \Secref{subsec:uplift} by outlining the general (and standard) procedure which uplifts $SU(2)$ gauge fields on the base to $SU(2)$-equivariant 1-forms on the principal bundle \cite{Nakahara:2003nw}. This uplift has been explicitly outlined in \cite{Herfray:2016std} in the present context to relate three-dimensional Einstein gravity to six-dimensional Hitchin gravity. Here we instead use the same uplift to relate Einstein gravity to Kodaira--Spencer gravity which requires performing the uplift twice. This entails first uplifting a reference solution of Einstein gravity to define a background integrable almost complex structure on $B_3 \times S^3$. The uplift of a second (off-shell) gauge field defines a deformation of the background complex structure.

In \Secref{subsec:reduction} we determine when Kodaira--Spencer gravity on $B_3\times S^3$ can be consistently dimensionally reduced to three-dimensional Einstein gravity on the base $B_3$. We derive necessary conditions for a complex structure deformation to reduce to a unique gauge field on the base which include $SU(2)$-invariance of the deformation under the right-action of $SU(2)$ and the absence of an invariant vertical subspace. Together with the uplift these results provide an equivalence between the two theories on the classical level.

\subsection{Dimensional uplift}\label{subsec:uplift}

\begin{figure}
    \centering
    \def\svgwidth{0.4\columnwidth}
    %% Creator: Inkscape 1.1.2 (b8e25be833, 2022-02-05), www.inkscape.org
%% Accompanies image file 'Principal_bundle.pdf' (pdf, eps, ps)
\begingroup%
  \makeatletter%
  \providecommand\color[2][]{%
    \errmessage{(Inkscape) Color is used for the text in Inkscape, but the package 'color.sty' is not loaded}%
    \renewcommand\color[2][]{}%
  }%
  \providecommand\transparent[1]{%
    \errmessage{(Inkscape) Transparency is used (non-zero) for the text in Inkscape, but the package 'transparent.sty' is not loaded}%
    \renewcommand\transparent[1]{}%
  }%
  \providecommand\rotatebox[2]{#2}%
  \newcommand*\fsize{\dimexpr\f@size pt\relax}%
  \newcommand*\lineheight[1]{\fontsize{\fsize}{#1\fsize}\selectfont}%
  \ifx\svgwidth\undefined%
    \setlength{\unitlength}{522.46339813bp}%
    \ifx\svgscale\undefined%
      \relax%
    \else%
      \setlength{\unitlength}{\unitlength * \real{\svgscale}}%
    \fi%
  \else%
    \setlength{\unitlength}{\svgwidth}%
  \fi%
  \global\let\svgwidth\undefined%
  \global\let\svgscale\undefined%
  \makeatother%
  \begin{picture}(1,0.58834851)%
    \lineheight{1}%
    \setlength\tabcolsep{0pt}%
    \put(0.65688878,0.18608487){\makebox(0,0)[lt]{\lineheight{1.25}\smash{\begin{tabular}[t]{l}$B_3$\end{tabular}}}}%
    \put(0,0){\includegraphics[width=\unitlength,page=1]{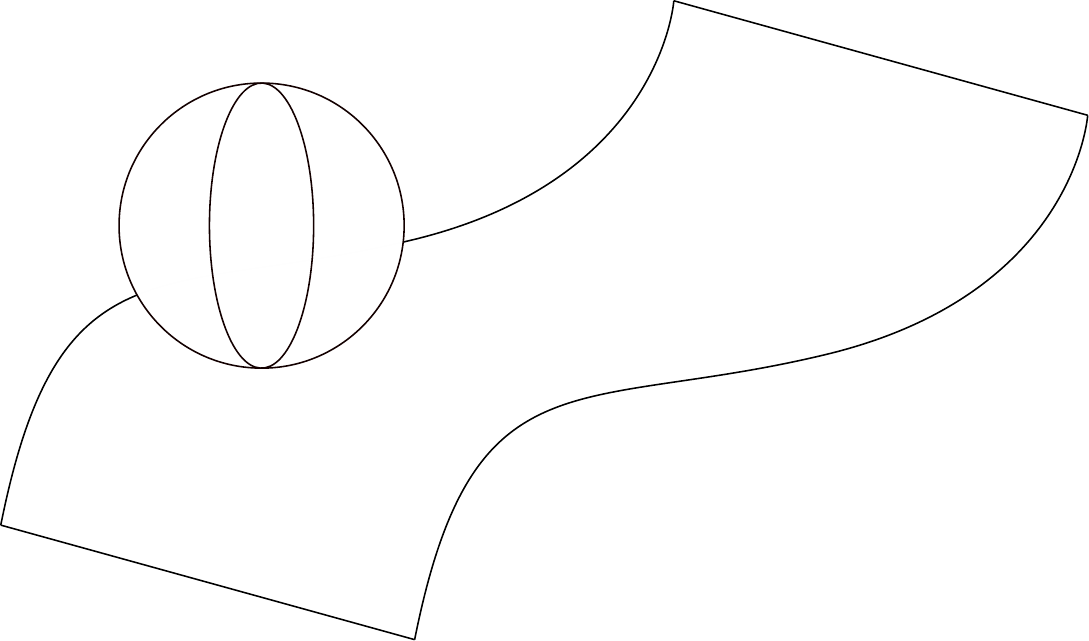}}%
    \put(0.19948891,0.53269736){\makebox(0,0)[lt]{\lineheight{1.25}\smash{\begin{tabular}[t]{l}$S^3$\end{tabular}}}}%
    \put(0.15684982,0.14821109){\makebox(0,0)[lt]{\lineheight{1.25}\smash{\begin{tabular}[t]{l}$y$\end{tabular}}}}%
    \put(0,0){\includegraphics[width=\unitlength,page=2]{Principal_bundle.pdf}}%
    \put(0.29519835,0.35325197){\makebox(0,0)[lt]{\lineheight{1.25}\smash{\begin{tabular}[t]{l}$\theta$\end{tabular}}}}%
  \end{picture}%
\endgroup%

    \caption{Principal bundle geometry $P_6$ used in the uplift of geometric data from $B_3$. $y$ denotes a set of coordinates of $B_3$ and $\theta$ a set of coordinates of the compact $S^3$ directions.}
    \label{fig:principal_bundle}
\end{figure}

Let us consider a three-dimensional manifold $B_3$ parametrized by coordinates $y^\mu$. To perform the uplift to a six-dimensional $SU(2)$ principal bundle $P_6$, we need to introduce a section $s = s(y)\in SU(2)$ which associates to each point on the base an $SU(2)$ element.\footnote{Here, our nomenclature differs from the standard literature: we call $s\colon B_3\rightarrow SU(2)$ a section while a section is normally defined as a map to $P_6$.} In addition, we introduce a parametrization $g = g(\theta)$ of $SU(2)$ where $\theta^i$ are coordinates on an $S^3$. Letting $x^a$ be six-dimensional coordinates covering $P_6$, together $y = y(x)$ and $\theta = \theta(x)$ define a (global) trivialization of $P_6$ under which it appears as a direct product $P_6 = B_3\times S^3$. This setup is illustrated in \Figref{fig:principal_bundle}.

\paragraph{Uplift of three-dimensional fields.} The base $B_3$ is equipped with a dreibein and a spin connection which can be encoded in the $SU(2)$ gauge field $W$ and $\mathfrak{su}(2)$-valued 1-form $E$ via \eqref{eq:omega_w_defs} and \eqref{eq:W_E_defs}. We can now uplift $W$ to a connection $\mathcal{W}$ and the dreibein $E$ to a horizontal 1-form $\mathcal{E}$ on $P_6$ via
\begin{equation}
	\mathcal{W} \equiv G^{-1}W G + G^{-1}\d G\,,\quad \mathcal{E} \equiv \frac{1}{\ell}\,G^{-1}EG\,,
    \label{eq:uplift_definition}
\end{equation}
where $\ell$ will be related to the cosmological constant of Einstein gravity, $\d$ denotes the six-dimensional exterior derivative in $x$ and
\begin{equation}
    G(y,\theta) \equiv s(y)\,g(\theta)\in SU(2)\,.\label{eq:uplift_u_def}
\end{equation}
By definition, the 1-forms $\d y^\mu$ are horizontal (see also \Appref{app:no_invariant_subspace}) so that $\mathcal{E}$ is also horizontal annihilating all vertical vector fields. In addition, $\mathcal{W}$ as a connection defines the horizontal vector bundle consisting of vector fields on $P_6$ that are annihilated by $\mathcal{W}$ \cite{Nakahara:2003nw}.

Under six-dimensional diffeomorphisms acting on $x = (y,\theta)$, the connection $\mathcal{W}$ defined by the uplift \eqref{eq:uplift_definition} transforms as a 1-form since $s(y)$ and $g(\theta)$ transform as scalars. In the coordinates (trivialization) chosen above, the right-action of $SU(2)$ on $P_6$ is a diffeomorphism acting only on the fiber directions $\mathcal{R}_u(y,\theta) = (y,R_u^\theta(\theta))$. Under the right-action, $g(\theta)$ transforms as
\begin{equation}
    \widetilde{g}(\theta)  \equiv g(R_u^\theta(\theta)) = g(\theta)\,u\,,
\end{equation}
where $u \in SU(2)$ is a coordinate independent constant. It follows that $\widetilde{G}(y,\theta) = G(y,\theta)\,u $, and since $\widetilde{s}(y) = s(y)$, \eqref{eq:uplift_definition} implies that
\begin{equation}
    \widetilde{\mathcal{W}}_a(x) = u^{-1}\,\mathcal{W}_a(x)\,u\,,\quad \widetilde{\mathcal{E}}_a(x) = u^{-1}\,\mathcal{E}_a(x)\,u\,.
    \label{eq:SU2_W_E}
\end{equation}
Hence $\mathcal{W}$ is $SU(2)$-equivariant under the right-action of $SU(2)$, as required for a connection, and so is $\mathcal{E}$. Thus, with $\mathcal{E}$, the uplift produces automatically a horizontal 1-form which is $SU(2)$-equivariant.

In addition, there are six-dimensional $SU(2)$-equivariant diffeomorphisms $\Phi_U\colon P_6\rightarrow P_6 $ under which $\widetilde{G}(y,\theta) = G(\Phi_U(y,\theta)) \equiv U(y)\,G(y,\theta) $ where $U=U(y)\in SU(2)$ depends only on the base coordinates. Under such diffeomorphisms, the uplifted connection transforms as
\begin{equation}
    \widetilde{\mathcal{W}} = \widetilde{G}^{-1}W\,\widetilde{G} + \widetilde{G}^{-1}\d \widetilde{G} = G^{-1}\,\widehat{W}\,G + G^{-1}\d G\,,
\end{equation}
which is the uplift of the gauge transformed field $\widehat{W}$ \eqref{eq:W_hat_E_hat} under $U(y)$. Hence gauge transformations on the base (local $SO(3)$ rotations) correspond to such six-dimensional diffeomorphisms $\Phi_U$ on the principal bundle.

Let us then consider the $\mathfrak{sl}(2,\mathbb{C})$-valued gauge field $A = W + \frac{i}{\ell}\,E$ which is the fundamental field \eqref{eq:A_Abar_definitions} of three-dimensional Einstein gravity in the first-order formulation. From \eqref{eq:uplift_definition} we obtain that its uplift is given by
\begin{equation}
    \mathcal{A} = G^{-1}AG + G^{-1}\d G = \mathcal{W} + i\mathcal{E}\,,
    \label{eq:curly_A_uplift}
\end{equation}
which is $SU(2)$-equivariant because of the equivariance \eqref{eq:SU2_W_E} of $\mathcal{W}$ and $\mathcal{E}$. Now, as reviewed in \Secref{subsec:3D_gravity_review}, solutions of three-dimensional Einstein gravity with a negative cosmological constant in its first-order formulation are equivalent to the gauge field $A = W + \frac{i}{\ell}\,E$ being flat
\begin{equation}
    dA + A\wedge A  = 0\,.
    \label{eq:A_flatness_again}
\end{equation}
Using \eqref{eq:curly_A_uplift}, we see that this equation is equivalent to the six-dimensional connection being flat
\begin{equation}
    d\mathcal{A} + \mathcal{A}\wedge \mathcal{A} = G^{-1}\,(dA + A\wedge A)\,G = 0\,.\label{eq:flatness_conjugation_relation}
\end{equation}
Since a flat \eqref{eq:A_flatness_again} gauge field can be written locally as $A = M^{-1}\,\d M$ for $M(y)\in SL(2,\mathbb{C})$, it follows from \eqref{eq:curly_A_uplift} that $\mathcal{A} = \mathcal{M}^{-1}\,\d \mathcal{M}$ where $\mathcal{M}(x)\in SL(2,\mathbb{C})$ is given by
\begin{equation}
    \mathcal{M}(y,\theta) = M(y)\, G(y,\theta)\,.\label{eq:M_uplift}
\end{equation}
Thus, it follows by the proof of \Secref{subsec:complex_structures_on_principal_bundles} and \cite{zentner2013integrable,Herfray:2016std} that a solution of Einstein gravity in its first-order formulation on $B_3$ defines an integrable almost complex structure $J$ \eqref{eq:complex_structure_bundle} on the bundle $P_6$. Because $\mathcal{A}$ is $SU(2)$-equivariant, $J$ is also $SU(2)$-invariant \eqref{eq:su2_invariance}. In addition, the uplifted almost complex structure of the form \eqref{eq:complex_structure_bundle} by construction has no vertical invariant subspace \eqref{eq:no_vertical_inv_subspace}. The explicit expression for the uplifted $J$ in terms of the components of the dreibein $E^i$ and of $W^i$ is given below in \Secref{subsec:reduction}.

\paragraph{Uplifted complex structure deformation.} With the above ingredients, we can relate gauge fields of three-dimensional Einstein gravity to complex structure deformations of six-dimensional Kodaira--Spencer gravity. First, let $E $ and $W$ define a reference solution of Einstein's equations \eqref{eq:Palatini_EOM_forms} corresponding to a flat \eqref{eq:A_flatness_again} gauge field $A = W + \frac{i}{\ell}\,E$ on the base $B_3$. Then $E,W$ uplift to a horizontal 1-form and a connection $\mathcal{E},\mathcal{W}$ respectively and $A$ uplifts to a flat \eqref{eq:flatness_conjugation_relation} $\mathfrak{sl}(2,\mathbb{C})$-valued 1-form $\mathcal{A} = \mathcal{W} + i\mathcal{E}$. As a result $\mathcal{E},\mathcal{W}$ define a background complex structure $J$ on $P_6$ via \eqref{eq:complex_structure_bundle} which is integrable by the flatness of $\mathcal{A}$. Thus $P_6$ is equipped with complex coordinates
\begin{equation}
    (w^i,\overbar{w}^i) = (w^i(y,\theta),\overbar{w}^i(y,\theta))
    \label{eq:background_complex_coordinates}
\end{equation}
obtained as solutions of the equations \eqref{eq:definition_holomorphic_coordinates}.

Consider then a different set $E',W'$ on the three-dimensional base, which uplift to $\mathcal{E}',\mathcal{W}'$ on $P_6$. For now, we will consider $E'$ and $W'$ to be off-shell, i.e.~to not satisfy Einstein's equation. From this data, we now define a pair of non-flat gauge fields $A'=W'+\frac{i}{\ell}\,E'$ and $\overbar{A}'=W'-\frac{i}{\ell}\,E'$ which uplift to non-flat 1-forms $\mathcal{A}' = \mathcal{A}'^i\tau_i$ and $\overbar{\mathcal{A}}' = \overbar{\mathcal{A}}'^i\tau_i $ that we decompose in complex coordinates \eqref{eq:background_complex_coordinates} as in \eqref{eq:Aprime_definition},
\begin{equation}
    \mathcal{A}'^i = \Sigma'^i_j\,(\d w^j + \alpha_{k}^{\;\,\, j}\d\overbar{w}^{k})\,,\quad \overbar{\mathcal{A}}'^i = \overbar{\Sigma}'^i_j\,(\d \overbar{w}^j + \overbar{\alpha}_{k}^{\;\,\, j}\d w^{k})\,.
\end{equation}
Here $\alpha$ is the off-shell fundamental field of Kodaira--Spencer gravity, a deformation of the background complex structure $J$, and $\overbar{\alpha}$ is its complex conjugate. Explicitly, from the uplift \eqref{eq:curly_A_uplift}, the relation between the Einstein gravity gauge field $A$ and the KS field $\alpha$ becomes
\begin{equation}
    \Sigma'^i_j(w,\overbar{w})\,(\d w^j + \alpha_{k}^{\;\,\, j}(w,\overbar{w})\,\d\overbar{w}^{k}) = G^{-1}(y,\theta)\,A'_\mu(y)\, G(y,\theta) \,\d y^\mu + G^{-1}(y,\theta)\,\d G(y,\theta)\,,
    \label{eq:explicit_uplift}
\end{equation}
where the complex coordinates are given by \eqref{eq:background_complex_coordinates}.

The uplifted forms $\mathcal{E}'$ and $\mathcal{W}'$ define an almost complex structure $J'$ via the construction \eqref{eq:complex_structure_bundle} which can be understood as a deformation of the background structure $J$ parametrized by $\alpha$ and $\Sigma'$.\footnote{Since $J'$ is not necessarily integrable, it depends on both $\alpha$ and $\Sigma'$ when written in complex coordinates of $J$. When $J'$ is integrable, the $\Sigma'$ dependence cancels and it is given by \eqref{eq:Jprime_explicitly}.} By construction, $J'$ has no vertical invariant subspace \eqref{eq:no_vertical_inv_subspace} and is $SU(2)$-invariant \eqref{eq:su2_invariance}. These properties descend to the deformation $\alpha$ as follows.

First, the $SU(2)$-invariance of $J'$ and $J$ descends to $SU(2)$-invariance of the deformation $\alpha$ arising from the uplift \eqref{eq:explicit_uplift}. First, we use that due to $SU(2)$-invariance of $J$, the right-action $\mathcal{R}_u$ with $u \in SU(2)$ in complex coordinates \eqref{eq:background_complex_coordinates} is a holomorphic-antiholomorphic diffeomorphism \eqref{eq:right_action_complex_coordinates}. Then the invariance of $J'$ implies that\footnote{Invariance also implies $\widetilde{\Sigma}'^i_j(w,\overbar{w})  = \Lambda^i_{\;\;k}\,\Sigma'^k_j(w,\overbar{w})$.}
\begin{equation}
    \widetilde{\alpha}_{i}^{\;\,\, j}(w,\overbar{w})  = \alpha_{i}^{\;\,\, j}(w,\overbar{w})\,,
    \label{eq:alpha_SU2_invariance}
\end{equation}
where the left-hand side is given by \eqref{eq:alpha_transformation_C} with $f^i(w) = f_u^i(w)$ and $\overbar{f}^i(\overbar{w}) = \overbar{f}_u^i(\overbar{w})$. Thus the uplifted complex structure deformation is also $SU(2)$-invariant.

Since $J'$ satisfies \eqref{eq:no_vertical_inv_subspace}, an integrable $\alpha$ satisfies separately for $i=1,2,3$ (see \Appref{app:no_invariant_subspace} for a derivation)
\begin{equation}
    2\,\text{Re}\,(\Sigma_l^i\,\overbar{\sigma}^k_j\,\alpha^{\;\;\,l}_k) \neq -\delta_j^i\,.\label{eq:alpha_no_inv_subpace}
\end{equation}
Therefore the uplifted KS fields are not arbitrary, but a special subset of the moduli space satisfying the constraints \eqref{eq:alpha_SU2_invariance} and \eqref{eq:alpha_no_inv_subpace}. The numbers of degrees of freedom in the two theories match as discussed in the next section.

\paragraph{Uplifted solutions.} When $E',W'$ solve the three-dimensional Einstein's equations, the gauge field $A'$ and its uplift $\mathcal{A}'$ are flat. Thus the deformed almost complex structure $J'$ is integrable and takes the simple form \eqref{eq:Jprime_explicitly}. In \Secref{subsec:deformations_review}, we have shown that integrability of $J'$ is equivalent to the Kodaira--Spencer equation
\begin{equation}
    \overbar{\partial}_{[j}\, \alpha_{k]}^{\;\;\,i}-\alpha_{[j}^{\;\;\,l}\partial_{\vert l \vert}\,\alpha_{k]}^{\;\;\,i} = 0\,.
\end{equation}
Thus, a solution of three-dimensional Einstein's equations $A' = W'+\frac{i}{\ell}\,E'$ uplifts to an $SU(2)$-invariant solution of the Kodaira--Spencer equation on the six-dimensional principal bundle $P_6$ via \eqref{eq:explicit_uplift}. The solution is both $SU(2)$-invariant and satisfies the condition \eqref{eq:alpha_no_inv_subpace}.

We see that the uplifted KS field depends on a reference solution $A$ of Einstein gravity that specifies a background complex structure $J$. This corresponds to the dependence of KS gravity on a chosen background complex structure. In other words, KS gravity is only defined on a complex manifold, or more precisely, on a Calabi--Yau manifold. This is unlike three-dimensional gravity which is defined on any differentiable manifold without additional structure.

\subsection{Dimensional reduction}\label{subsec:reduction}

Thus, the three-dimensional gauge field $A'$ determines a unique $SU(2)$-invariant complex structure deformation $\alpha$ in six-dimensions, such that the deformed complex structure $J'$ satisfies \eqref{eq:no_vertical_inv_subspace}, via the uplift procedure outlined in the previous section. The direction can also be reversed: an $SU(2)$-invariant $\alpha$ on an $SU(2)$ principal bundle, together with the condition \eqref{eq:no_vertical_inv_subspace} for $J'$, dimensionally reduces to a unique $A'$ in three-dimensions. This can be seen by a direct counting of the number of degrees of freedom as follows.

A general almost complex structure in six dimensions (a real $6\times 6$ matrix that squares to minus the identity) is parametrized by $18$ real functions of six variables (see \Appref{app:J_DOFs} for a proof). Then, $SU(2)$-invariance \eqref{eq:su2_invariance} imposes a constraint on these functions which is easiest to understand by expanding $J$ in a basis of six 1-forms $P^K = (\d y^\mu,\Theta^i)$, where $\Theta = \d G G^{-1} = \Theta^i\,\tau_i$ are the three right-invariant 1-forms of $SU(2)$, and six dual vector fields $P_K= (\partial_{y^\mu}, \Theta_i)$ defined by $P_K(P^L) = \delta^L_K$, as
\begin{equation}
    J = \mathcal{J}^K_L(y,\theta)\, P_K\otimes P^L\,.
    \label{eq:J_right_invariant_basis}
\end{equation}
The 1-forms $\Theta^i$ are invariant under the right-action diffeomorphism $\mathcal{R}_u$, and since $\d y^\mu$ are also invariant, both $P^K$ and $P_K$ are right-invariant. Thus right-invariance of $J$ reduces to a constraint on the components $\mathcal{J}^K_L(y,R_u^\theta(\theta)) = \mathcal{J}^K_L(y,\theta)$ in this basis. Since $\mathcal{R}_u$ acts on $\theta$ transitively, it follows that the components are independent of the $S^3$ coordinates,
\begin{equation}
    \mathcal{J}^K_L(y,\theta) = \mathcal{J}^K_L(y)\,.
\end{equation}
In other words, an $SU(2)$-invariant $J$ is parametrized by $18$ real functions of three variables (the base coordinates $y^\mu$). This is exactly the same number of functions as in an $\mathfrak{sl}(2,\mathbb{C})$-valued gauge field $A$ on the base.

The counting also applies to the second gauge field $A'$ and to the deformed almost complex structure $J'$: the $SU(2)$-invariant deformation $\alpha$ satisfying \eqref{eq:alpha_SU2_invariance} consists of $3\times 3 = 9$ complex functions of $y$ matching with the $18$ real functions of $A'$. However, notice that general non-invariant deformations $\alpha$ on $P_6$ contain more degrees of freedom than just a three-dimensional gauge field. These are the Kaluza--Klein (KK) modes of $\alpha$ which are obtained by expanding it in spherical harmonics of $S^3$. The $SU(2)$-invariant zero mode reduces to $A$ on $B_3$, but higher KK modes produce additional massive fields which are not described by pure three-dimensional Einstein gravity. Decomposition of $\alpha$ into higher KK modes are considered in \cite{Zeng:2023qqp}, but without showing the connection to three-dimensional gravity.

As was pointed out in \cite{Herfray:2016std,zentner2013integrable}, $SU(2)$ invariance of an almost complex structure $J$ alone does not guarantee a consistent dimensional reduction to a gauge field $A$ in three dimensions. The reason for this is that $J$ may have a vertical invariant subspace in which case it does not define a connection that is needed for the definition of horizontal vector fields $\mathcal{E}^i$.\footnote{On $SU(2)$ principal bundles, the notion of vertical vector fields is always defined even without a connection. If $J$ has an invariant vertical subspace, there exist vertical vectors that get mapped to vertical vectors under the action of $J$. When this happens, $J$ alone is not sufficient to define a set of horizontal vector fields $\mathcal{E}^i$ that span the horizontal tangent space and are linearly independent from the $\mathcal{W}^i$.} In fact, the absence of the vertical invariant subspace \eqref{eq:no_vertical_inv_subspace} is equivalent to the $\Theta^i\otimes \partial_{y^\mu}$ components $\mathcal{J}_i^\mu$ of $J$ being invertible
\begin{equation}
    \det{\mathcal{J}_i^\mu} \neq 0\,.
    \label{eq:invertibility_of_J}
\end{equation}
We can understand where this condition comes from the three-dimensional perspective by considering the uplift. The almost complex structure \eqref{eq:complex_structure_bundle} obtained using the uplift \eqref{eq:uplift_definition} can be written in the right-invariant basis \eqref{eq:J_right_invariant_basis} as (see \Appref{app:no_invariant_subspace} for a more detailed derivation)
\begin{equation}
\begin{split}
    J &= - W^i_\mu\,E^\nu_i\, \d y^\mu\otimes \partial_{y^\nu}- E_i^\mu\, \Theta^i\otimes \partial_{y^\mu}\\
    &\;\;\;\;+(E^i_\mu+W^i_\nu\,E^\nu_j\, W^j_\mu)\,\d y^\mu\otimes  \Theta_i+\, E^\mu_i\,W^j_\mu \,\Theta^i \otimes \Theta_j\,,\label{eq:dim_reduction_explicit}
\end{split}
\end{equation}
where $E_i^\mu$ is the inverse dreibein satisfying \eqref{eq:inverse_vielbein}. Thus we see that $\mathcal{J}_i^\mu = -E_i^\mu$ and the condition \eqref{eq:invertibility_of_J} follows from the invertibility $\det{E^i_\mu} \neq 0$ of the dreibein.

\section{Example:~uplift of the Ba\~nados geometry} \label{sec:banados_uplift}

In this section, we give an explicit demonstration of the results of \Secref{sec:uplift} by uplifting the Ba\~nados solution \eqref{eq:banados_sol_gauge_field} of three-dimensional Einstein gravity with a negative cosmological constant to a six-dimensional solution of Kodaira--Spencer gravity. The Ba\~nados solution is the holographic dual of an excited state of the dual two-dimensional CFT in which the stress tensor expectation value is given by a Schwarzian derivative. The uplift thus provides the Fefferman--Graham expansion for the complex structure deformation and allows to identify where the stress tensor expectation value is sitting.

As reviewed in \Secref{sec:solutions_in_FG_coordinates}, the gauge field on the three-dimensional base can be expressed as $A = M^{-1}\d M$ in terms of a $SL(2,\mathbb{C})$ valued function $M(r,\hat{y})$ as in \eqref{eq:3d_flatness_solution}. Here, $r$ is the radial coordinate, $\hat{y}$ are the boundary coordinates.
Moreover, $M$ is given in terms of a $SL(2,\mathbb{C})$ valued function $h(\hat{y})$ of the boundary coordinates as in \eqref{eq:M_base}. The Ba\~nados solution \cite{Banados:2002ey} amounts to the choices \eqref{eq:banados_sol1}, \eqref{eq:banados_sol2} and is explicitly given by
\begin{align}
    M(r,\hat{y}) &= e^{f(z)\, J_-}e^{-\log{f'(z)} \,J_3}e^{-\frac{1}{2} (\log{f'(z)})'\, J_+}\,e^{r J_3}\,,\nonumber\\
    \overbar{M}(r,\hat{y}) &= e^{-\overbar{f}(\overbar{z})\, J_+}e^{\log{\overbar{f}'(\overbar{z})}\,J_3}e^{\frac{1}{2}(\log{\overbar{f}'(\overbar{z})})'\,J_-}\,e^{-r J_3}\,,\label{eq:M_base_def}
\end{align}
where $f(z)$ and $\overbar{f}(\overbar{z})$ are two functions of the boundary complex coordinates $z = z(\hat{y}) $ and $\overbar{z} = \overbar{z}(\hat{y})$ respectively. The corresponding gauge field obtained from \eqref{eq:3d_flatness_solution} is given in \eqref{eq:banados_sol_gauge_field}.

Next, in order to define the uplifted gauge fields $\mathcal{A}$ from \eqref{eq:curly_A_uplift}, we need to specify a section $s(y)$ along which to uplift to define $G(y,\theta)$ as in \eqref{eq:uplift_u_def}. It is advantageous to choose the section $s(y) = \1 \in SU(2)$, since in this case $G(y,\theta) = g(\theta) $ is only a function of the angular coordinates $\theta_i$. As angular coordinates, we chose the parametrization of the $SU(2)$ element in terms of the Euler angles,
\begin{equation}
    g(\theta)=e^{\theta_1 \tau_3}e^{\theta_2\tau_1}e^{\theta_3\tau_3}=\begin{pmatrix}
        e^{-i\frac{\theta_1+\theta_3}{2}}\cos\frac{\theta_2}{2} & - i e^{-i\frac{\theta_1-\theta_3}{2}}\sin\frac{\theta_2}{2}\\
        -i e^{i \frac{\theta_1-\theta_3}{2}}\sin\frac{\theta_2}{2} & e^{i\frac{\theta_1+\theta_3}{2}}\cos\frac{\theta_2}{2}
    \end{pmatrix}\in SU(2)\,.\label{eq:euler_angle_def}
\end{equation}
We will now construct a background complex structure $J$, which we choose to be the uplift of the gauge field $A$ where we set the functions $f(z)=z$ and $\overbar{f}(\overbar{z})=\overbar{z}$. As discussed in \Secref{sec:solutions_in_FG_coordinates}, this choice corresponds to the Poincar\'e AdS solution. We construct the uplifted gauge field $\mathcal{A}$ by \eqref{eq:curly_A_uplift} where $\mathcal{M}$ is given by \eqref{eq:M_uplift}. As discussed below \eqref{eq:M_is_holomorphic}, the components of $\mathcal{M} = \mathcal{M}(w) $ are holomorphic functions of the complex coordinates $w^i $ of $J$. We choose the parametrization
\begin{equation}
    \mathcal{M}=\begin{pmatrix}
        w_1 & w_2\\
        w_1 w_3 & \frac{1}{w_1}+w_2w_3
    \end{pmatrix}\,,
\end{equation}
which ensures that $\det{\mathcal{M}}=1$ and we denote $w^i \equiv w_i$ from now on to simplify the notation below (similarly $\overbar{w}^i \equiv \overbar{w}_i$). By using \eqref{eq:M_uplift}, as well as \eqref{eq:M_base_def} and \eqref{eq:euler_angle_def}, we obtain that relation between complex coordinates $w_i$ and coordinates $(r,z,\overbar{z}, \theta_i)$ is given by
\begin{equation}
\begin{split}
    w_1 &= e^{\frac{r}{2}}e^{-i\frac{\theta_1+\theta_3}{2}} \cos\frac{\theta_2}{2}\,,\\
    w_2 &= -i e^{\frac{r}{2}}e^{-i\frac{\theta_1-\theta_3}{2}} \sin\frac{\theta_2}{2}\,,\\
    w_3 &= z- i e^{-r}e^{i\theta_1} \tan\frac{\theta_2}{2}\,. \label{eq:undeformed_coords}
\end{split}
\end{equation}
Defining $\overbar{w}_i$ as the conjugates of $w_i$, we obtain the relations
\begin{equation}
    z = w_3+\frac{1}{||w||^2} \frac{\overbar{w}_2}{w_1}\,,\quad e^{r\slash2}=\sqrt{|w_1|^2+|w_2|^2}\equiv ||w||\,.\label{eq:real_coords_complex}
\end{equation}
Similar relations can also be defined for $\theta_i$. We note that these relations between holomorphic and real coordinates are the same as the ones derived in \cite{costelloGaiotto2018twisted}, up to a holomorphic-antiholomorphic diffeomorphism which maps $w_3$ to $w_3\slash w_1$ and $\overbar{w}_3$ to $\overbar{w}_3\slash \overbar{w}_1$. This shows that the complex structure we obtain from our construction is the same as the one obtained from calculating the backreaction of B-branes in the topological B-model on $\mathbb{C}^3$ \cite{costelloGaiotto2018twisted}.

Now, we repeat the uplift procedure for a gauge field $A'$ where $f(z)$ and $\overbar{f}(\overbar{z})$ are arbitrary functions again. We obtain a family of uplifted gauge fields $\mathcal{A}'$ parametrized by $f$ and $\overbar{f}$. Let $\mathcal{M}'$ be associated to $\mathcal{A}'$ as in \eqref{eq:curly_A_uplift}. We choose the same parametrization
\begin{equation}
    \mathcal{M}' = \begin{pmatrix}
        \hat{w}_1 & \hat{w}_2\\
        \hat{w}_1 \hat{w}_3 & \frac{1}{\hat{w}_1}+\hat{w}_2 \hat{w}_3
    \end{pmatrix}\,
\end{equation}
to obtain a family of complex coordinates $\hat{w}_i$ distinct from the $w_i$ defined in \eqref{eq:undeformed_coords}. By explicit calculation, we find
\begin{equation}
\begin{gathered}
    \hat{w}_1 = \frac{w_1}{f'(z)^{1\slash 2}}+ \frac{\overbar{w}_2}{||w||^2}\frac{f''(z)}{2f'(z)^{3\slash 2}}\,,\qquad \hat{w}_2 = \frac{w_2}{f'(z)^{1\slash 2}} - \frac{\overbar{w}_1}{||w||^2} \frac{f''(z)}{2f'(z)^{3\slash 2}}\,,\\
    \hat{w}_3 = f(z)-
    \frac{1}{||w||^2}\frac{\overbar{w}_2}{w_1}\frac{f'(z)}{1 +\frac{1}{||w||^2} \frac{\overbar{w}_2}{w_1} \frac{f''(z)}{2f'(z)}}\,.\label{eq:deformed_coordinates}
\end{gathered}
\end{equation}
Here, we understand $z$ as a function of $w_i$ and $\overbar{w}_i$ as in \eqref{eq:real_coords_complex}. The $\hat{w}_i$ define a family of integrable complex structures by \eqref{eq:J_simple_form}. Namely, \eqref{eq:deformed_coordinates} defines a diffeomorphism, whose associated complex structure deformation is given by \eqref{eq:alpha_diffeo}. We obtain
\begin{align}
        \alpha_i^{\;\,\, j}(w,\overbar{w})\,\d \overbar{w}^i\otimes \partial_j &= \frac{T(z)}{||w||^{6}}\,\overbar{w}_1\overbar{w}_2\,\biggl(\d \overbar{w}_2 \otimes \partial_1+ \d\overbar{w}_1  \otimes \partial_2-\frac{\overbar{w}_2}{\overbar{w}_1}\,\d \overbar{w}_1  \otimes\partial_1-\frac{\overbar{w}_1}{\overbar{w}_2}\, \d \overbar{w}_2  \otimes\partial_2\biggr)\nonumber\\
        &+\frac{T(z)}{||w||^{8}}\frac{\overbar{w}_2^2}{w_1^2}\left(\overbar{w}_1\, \d \overbar{w}_2\otimes \partial_3-\overbar{w}_2 \,\d \overbar{w}_1\otimes\partial_3\right)\,.\label{eq:alpha_explicit}
\end{align}
Solutions of this form also satisfy the cohomology condition \eqref{eq:cohomology_condition} as can be checked by explicit computation.

Moreover, we check that \eqref{eq:alpha_explicit} is right-invariant under the action of $SU(2)$ as in \eqref{eq:alpha_SU2_invariance}. Let $u=e^{u_1 \tau_3}e^{u_2\tau_1}e^{u_3\tau_3}\in SU(2)$, we define the right action on $\mathcal{M}$ as in \eqref{eq:right_action_on_SL2C}. Using the parametrization
\begin{equation}
    \widetilde{\mathcal{M}}=\begin{pmatrix}
        \widetilde{w}_1 & \widetilde{w}_2\\
        \widetilde{w}_1 \widetilde{w}_3 & \frac{1}{\widetilde{w}_1}+\widetilde{w}_2 \widetilde{w}_3
    \end{pmatrix}\,,
\end{equation}
defines a holomorphic-antiholomorphic diffeomorphism \eqref{eq:right_action_complex_coordinates} given by
\begin{equation}
\begin{split}
    \widetilde{w}_1 = f^1_u(w) &= e^{-\frac{i}{2}(u_1+u_3)}\biggl(w_1 \cos{\frac{u_2}{2}}-i e^{i u_1} w_2 \sin\frac{u_2}{2}\biggr)\,,\\
    \widetilde{w}_2 = f^2_u(w) &= e^{-\frac{i}{2}(u_1-u_3)}\biggl(e^{i u_1} w_2 \cos\frac{u_2}{2}- i w_1 \sin\frac{u_2}{2}\biggr)\,,\\
    \widetilde{w}_3 = f^3_u(w) &= \frac{w_1w_3 \cos\frac{u_2}{2}-i e^{i u_1}(\frac{1}{w_1}+ w_2 w_3) \sin\frac{u_2}{2}}{w_1 \cos\frac{u_2}{2}-i e^{i u_1} w_2 \sin\frac{u_2}{2}}\,,
\end{split}
\label{eq:right_action_banados}
\end{equation}
parametrized by the constants $u_i$. Under this holomorphic-antiholomorphic diffeomorphism $\alpha$ transforms as in \eqref{eq:alpha_transformation_C}. By explicit calculation of the Jacobians of the map \eqref{eq:right_action_banados} we find that $\alpha$ is $SU(2)$-invariant \eqref{eq:alpha_SU2_invariance}.

Furthermore, we see by explicit calculation that the complex structure deformation \eqref{eq:alpha_explicit} satisfies \eqref{eq:alpha_no_inv_subpace} for any $T(z)$, such that the deformed complex structure does not have a vertical invariant subspace. To perform this calculation, we first calculate $\Sigma^i_j$ by using \eqref{eq:holomorphic_sigma} to obtain
\begin{equation}
    \Sigma^i_j = \begin{pmatrix}
        \frac{iw_2}{w_1^2} & \frac{i}{w_1} & i\,(w_1^2-w_2^2)\\
        -\frac{w_2}{w_1^2} & -\frac{1}{w_1} & w_1^2 + w_2^2\\
        \frac{2i}{w_1} & 0 & -2iw_1w_2
    \end{pmatrix}\,.
\end{equation}
Contracting with \eqref{eq:alpha_explicit}, we find that \eqref{eq:alpha_no_inv_subpace} is satisfied. The complex structure deformation \eqref{eq:alpha_explicit} thus provides the explicit realization of the map from solutions of three-dimensional Einstein's equation with negative cosmological constant to $SU(2)$ invariant solutions of six-dimensional KS equation. We note that from \eqref{eq:undeformed_coords}, it is straightforward to obtain an expansion of the uplifted complex structure deformation \eqref{eq:alpha_explicit} as an expansion in powers of $e^r$, similar to the Fefferman--Graham expansion of the metric \eqref{eq:Banados_metric}.

\section{Discussion and outlook}\label{sec:outlook}

We proved a new relationship between 3D Einstein gravity in its Chern--Simons formulation on a base manifold $B_3$ and the $SU(2)$-invariant sector of six-dimensional Kodaira--Spencer gravity on the principal bundle $B_3\times S^3$. Moreover, we derived an off-shell relation between classical fields of the two theories by an uplift procedure and showed that the equations of motion are equivalent for negative cosmological constant in Euclidean metric signature. We conclude this paper by discussing some open questions and possible future directions of our work.

\paragraph{Lorentzian signature and de Sitter.}

Our construction is performed in Euclidean signature and for a negative cosmological constant $\lambda \equiv -1\slash \ell^2 < 0$. In this case, 3D Einstein gravity in the first-order formulation has a local $SO(3)$ rotation symmetry that leaves the flat Euclidean metric invariant. In the CS formulation, this is realized as $SU(2)$ gauge symmetry via the double cover. There is a subtle interplay between the sign of the cosmological constant $\lambda$, the signature of the metric, the gauge group needed to reproduce Einstein's equations and whether the $\mathfrak{su}(2)$ components $A^i,\overbar{A}^i$ of the three-dimensional gauge fields are real or complex.

For $\lambda>0$ in Euclidean signature, the same symmetry remains, but the rewriting of the Einstein action in terms of CS actions requires the introduction of two copies of $SU(2)$: there are two gauge fields $A = A^i\,\tau_i$, $\overbar{A} = \overbar{A}^i\,\overbar{\tau}_i$ where $[\tau_i,\overbar{\tau}_j] = 0$ are two commuting generators of $\mathfrak{su}(2)$ and it is the diagonal $SU(2)\subset  SU(2)\times SU(2)\cong SO(4)$ that implements the local rotation group. A similar story applies to the $\lambda < 0$ case, however in Lorentzian signature. In this case, the local symmetry of the flat Minkowski metric is $SO(1,2)$, whose double cover is $SL(2,\mathbb{R})$. The local symmetry is realized as $SL(2,\mathbb{R})$ gauge symmetry for two independent gauge fields $A = A^i\,\hat{\tau}_i$ and $\overbar{A} = \overbar{A}^i\,\hat{\overbar{\tau}}_i$ where $[\hat{\tau}_i,\hat{\overbar{\tau}}_j] = 0$ are two commuting generators of $\mathfrak{sl}(2,\mathbb{R})$. Local Lorentz transformations are implemented as the diagonal $SL(2,\mathbb{R})\subset SL(2,\mathbb{R})\times SL(2,\mathbb{R}) \cong SO(2,2)$. In both of these cases, the gauge fields are given in terms of the spin connection and the dreibein as $A^i = W^i + \sqrt{\vert \lambda\vert}\,E^i $, $\overbar{A}^i = W^i - \sqrt{\vert \lambda\vert}\,E^i$ which are real valued.\footnote{This means that for Euclidean $\lambda<0$, the gauge fields are $\mathfrak{su}(2)$-valued, while for Lorentzian $\lambda< 0$, they are $\mathfrak{sl}(2,\mathbb{R})$-valued.} Einstein's equations are equivalent to their flatness.

Because $A^i,\overbar{A}^i$ are real valued instead of complex for Euclidean $\lambda>0$ and Lorentzian $\lambda < 0$, they cannot be uplifted to a six-dimensional almost complex structure.\footnote{However, the uplift of $W^k+i E^k$ still results in an almost complex structure whose integrability is not connected to Einstein's equation on the base.} However, in this case the uplift procedure leads to another type of structure, called almost paracomplex structure \cite{Herfray:2016std}, i.e.~a real vector valued one-form $I^a_b$ that squares to the identity, $I^a_c I^c_b=\delta_b^a$. There also exists a notion of integrability of almost paracomplex structures similar to the vanishing of the Nijenhuis tensor. It will be interesting to investigate the relation between this integrability condition and Einstein's equations for positive cosmological constant.

However, it turns out that our construction also applies to $\lambda >0$ in Lorentzian signature. In this case, Einstein's equations are reproduced by simply replacing the $\mathfrak{su}(2)$ generators $\tau_i$ in \Secref{subsec:3D_gravity_review} with $\mathfrak{sl}(2,\mathbb{R})$ generators $\hat{\tau}_i$. The components of the gauge fields are given by the same expressions $A^i = W^i + \frac{i}{\ell}\,E^i$ and $\overbar{A}^i = W^i - \frac{i}{\ell}\,E^i$, however, $\ell$ is now identified with the cosmological constant as $\lambda = 1\slash \ell^2>0 $. Since the complexification of $\mathfrak{sl}(2,\mathbb{R})$ is also $\mathfrak{sl}(2,\mathbb{C})$, the gauge fields $A = A^i\,\hat{\tau}_i$, $\overbar{A} = A^i\,\hat{\tau}_i$ are $\mathfrak{sl}(2,\mathbb{C})$-valued and can be uplifted to a six-dimensional almost complex structure as in \Secref{subsec:uplift}.\footnote{The fact that Lorentzian 3D Einstein gravity with $\lambda>0$ is described by $SL(2,\mathbb{C})$ CS theory is already pointed out in \cite{Witten:1989ip}.} It follows that integrability of the almost complex structure is equivalent to Lorentzian Einstein's equations with a positive cosmological constant. The six-dimensional manifold is again $SL(2,\mathbb{C})$. Compared to Euclidean $\lambda<0$, the fibers are $SL(2,\mathbb{R})$ instead of $SU(2)$. This potentially provides an embedding of Lorentzian de Sitter gravity into topological string theory, and an approach complementary to \cite{Collier:2025lux} to the quantization of de Sitter gravity.

\paragraph{Extension to the quantum level.} Our analysis is purely classical and the next step is to extend it to the quantum level. This involves relating the well-established path integral of Chern--Simons theory to the known loop expansion of Kodaira--Spencer gravity. This raises some interesting questions.

First, KS gravity is not a background independent theory, but depends on a background complex structure parametrized as a deformation $(t,\overbar{t})$ of some reference structure $J_0$ \cite{Witten:1993ed}. At the quantum level, KS gravity suffers from a holomorphic anomaly where its path integral depends not only on the holomorphic deformation $t$, but also on $\overbar{t}$ \cite{Bershadsky:1993ta,Bershadsky:1993cx}. The anomaly arises from the use of a Kähler metric (that knows about $\overbar{t}$) to perform gauge fixing in perturbation theory. The natural question is whether the holomorphic anomaly of KS gravity on the principal bundle is realized by CS theory on the base. 

The second question concerns the configuration spaces over which the path integration should be performed. To translate 3D Einstein gravity to the CS formulation requires the dreibein to be invertible, $\det{E} \neq 0$, which is equivalent to the metric being non-degenerate. However, after relaxing the torsion-free constraint and considering the connection to be an independent field, the path integral of CS theory has to be performed also over gauge fields with non-invertible dreibeins \cite{Witten:1988hc,Witten:1989ip}. Even though dreibeins with $\det{E} = 0$ form a set of measure zero, the path integrals of Einstein gravity and CS theory differ \cite{Witten:1988hc}. A direct quantization in the metric formulation has been proposed in \cite{Collier:2023fwi}.

The deformed almost complex structures $J'$ on $B_3\times S^3$ we construct in this paper have no vertical invariant subspace $\pi_*J'(v) \neq 0$. We prove that this corresponds to a certain block of components of $J'$ being invertible, and in fact, this block coincides exactly with the dreibein $E'$ upon dimensional reduction. Thus only invertible dreibeins $\det{E'} \neq 0$ arise from the almost complex structures we consider. KS gravity also contains deformed structures $J'$ satisfying $\pi_*J'(v) = 0$ for some vertical vectors $v$ which potentially must be summed over in the path integral. However, they have no direct dimensional reduction to CS theory. It is interesting to note that the dimensional reduction fails exactly when the three-dimensional metric becomes degenerate. The implications of the relation between the KS gravity and CS theory path integrals is left for future work.

The first step in answering quantum questions is to uplift the classical action of 3D Einstein gravity to the action of KS gravity. Given that we show the equations of motion of the two theories are related, we expect such an uplift to exist. We leave a detailed study of this for future work.

\paragraph{Embedding in twisted holography.} We uplift a reference solution $A$ of 3D Einstein gravity with a negative cosmological constant on $B_3$ to a Calabi--Yau structure $(J,\Omega,h)$ on $B_3\times S^3$, and demonstrate that the uplift of a second off-shell gauge field $A'$ defines Kodaira--Spencer theory on this background Calabi--Yau three-fold. When $B_3 = \mathbb{H}_3\subset \mathbb{R}^4$ is the hyperbolic space, we obtain $ \mathbb{H}_3\times S^3\cong SL(2,\mathbb{C})$ which, as we show, coincides with the deformed conifold \cite{Candelas:1989js} for the unique Ricci flat Kähler metric $h$.

Kodaira--Spencer theory on the deformed conifold was conjectured to be the holographic dual gravitational theory of a two-dimensional chiral Euclidean conformal field theory living on the asymptotic boundary of $\mathbb{H}_3\subset SL(2,\mathbb{C})$ \cite{costelloGaiotto2018twisted}. This duality is known as twisted holography and it is conjectured to describe a topological subsector of the holographic duality between $\mathcal{N}=4$ supersymmetric Yang--Mills (SYM) theory and type IIB superstring theory on AdS$_5\times S^5$ \cite{Maldacena:1997,Witten:1998qj,Gubser:1998bc}. In fact, the chiral CFT in question is a gauged $U(N)$ beta-gamma system $\mathfrak{A}_N$ which is the holomorphic twist of $\mathcal{N}=4$ SYM \cite{Beem:2013sza}, while Kodaira--Spencer theory is conjectured to be the twist of type IIB supergravity on the gravity side \cite{costello2016twisted}. The duality can also be derived by a brane argument \cite{costello2016twisted} similar to the original derivation of the AdS/CFT correspondence \cite{Maldacena:1997}. The starting point is to consider the topological B-model on $\mathbb{C}^3\cong \mathbb{R}^6$ and a stack of $N$ B-branes extending along a complex line $\mathbb{C}$ \cite{costelloGaiotto2018twisted}. The worldvolume theory of the branes coincides with $\mathfrak{A}_N$ \cite{Witten:1992fb,costelloGaiotto2018twisted} while the target space gravity theory of the B-model is Kodaira--Spencer theory \cite{Bershadsky:1993ta,Bershadsky:1993cx}. In the large-$N$ limit, the backreaction of the branes deforms the canonical complex structure of $\mathbb{C}^3\subset \mathbb{C}^4$ to the deformed conifold solution $SL(2,\mathbb{C})\subset \mathbb{C}^4$ of Kodaira--Spencer theory \cite{costelloGaiotto2018twisted}. As we show in \Secref{sec:banados_uplift}, this same solution arises as the uplift of the Poincar\'e AdS solution on $\mathbb{H}_3$.

While twisted holography captures a topological subsector of the familiar AdS$_5$/CFT$_4$ duality, it also stands on its own right as an AdS$_3$/CFT$_2$ correspondence. In the usual way, there is a matching of symmetries: the isometries of $\mathbb{H}_3\subset SL(2,\mathbb{C})$ capture the global conformal group of the dual beta-gamma system while the $SU(2)$ isometry of the $S^3\subset SL(2,\mathbb{C})$ captures the $R$-symmetry group. The curvature radius $\ell = N$ of $\mathbb{H}_3$ determines the central charge $-3\,(N^2-1)$ of $\mathfrak{A}_N$ which is negative. The main difference to AdS$_5$/CFT$_4$ duality is that there is no coupling constant in $\mathfrak{A}_N$ that controls stringy effects in the bulk. In fact, $1\slash N$-corrections of the CFT are conjectured to be captured equivalently by quantum corrections to Kodaira--Spencer theory and by the string worldsheet genus expansion \cite{costelloGaiotto2018twisted}.

Since the beta-gamma system contains a Virasoro stress tensor \cite{Beem:2013sza} indicates that 3D Einstein gravity is contained in Kodaira--Spencer theory. Our uplift shows exactly how 3D Einstein gravity is embedded in an $SU(2)$-invariant subsector of twisted holography. The uplift of the Ba\~nados solutions presented in \Secref{sec:banados_uplift} provides the first example of a Fefferman--Graham type expansion in this context. It identifies the location of the stress tensor expectation value in the subleading terms of the complex structure deformation. The derivation of the full generating functional of stress tensor correlation functions \cite{Polyakov:1987zb} as an on-shell action of Kodaira--Spencer theory is left for future work.

The $SU(2)$-invariant complex structure deformations considered in this work are zero-modes of general non-invariant deformations in an expansion in spherical harmonics of the $S^3$. Upon dimensional reduction, higher harmonics give raise to higher Kaluza--Klein modes in three dimensions which are additional fields on top of 3D Einstein gravity with increasing spin. Combined with the mathematical technology of \cite{Zeng:2023qqp}, our approach may be used to determine the equation of motion of these modes. This will shed light on the field content needed to UV complete AdS$_3$/CFT$_2$ duality in the topological string. Since the beta-gamma system is well-defined for any finite value of $N$, this approach is of interest for answering questions about non-perturbative effects in three-dimensional quantum gravity such as the need for sum over topologies and ensemble averaging.

\paragraph{Inclusion of holonomies and black holes.}
In the analysis presented here, we assume that the gauge fields $A$ and $\mathcal{A}$ have trivial holonomies. In general, solutions of the Chern-Simons equation of motion of the form \eqref{eq:3d_flatness_solution} are only solutions in an open neighborhood, whereas global solutions have to include holonomies \cite{Banados:2002ey, Banados:1998gg}. Some solutions of this more general form describe black hole solutions of Einstein's equations \cite{Banados:1992wn,Banados:1998gg}. Using the uplift procedure, it is interesting to study KS gravity on backgrounds corresponding to three-dimensional black hole solutions. For this analysis, a rigorous treatment of the uplift of holonomies is needed. It is then natural to ask about the nature of the theory dual to KS gravity on these backgrounds or whether such a theory even exists. Besides adding to the twisted holographic dictionary, this construction might enable a counting of black hole microstates in terms of the beta-gamma system.

One possible tool to diagnose black holes in KS gravity from the boundary perspective is to compute the torus partition function of the beta-gamma system as the Schur limit of the superconformal index of $\mathcal{N} = 4$ SYM theory \cite{Beem:2013sza}. The superconformal index was shown to encompass a wide range of large-$N$ behaviors depending on the scaling of the global $U(1)$ charge \cite{Gaiotto:2021xce}. In particular, if the charge scales quadratically in $N$, the superconformal index decomposes into an expansion in $e^{-N^2}$, capturing the entropy of non-perturbative degrees of freedom constituting a black hole in the bulk \cite{Gaiotto:2021xce}. It will be interesting to understand whether this holds for the beta-gamma system.

\paragraph{Ryu--Takayanagi formula in KS gravity.}

The theory dual to KS gravity is the chiral half of a non-unitary CFT, namely, a gauged beta-gamma system. This theory has a known negative central charge. A central question we would like to answer is whether the Maldacena--Lewkowycz--Dong type argument \cite{Lewkowycz:2013nqa,Dong:2016fnf} is applicable to KS gravity, i.e.~if it yields an expression similar to an RT formula \cite{Ryu:2006bv,Ryu:2006ef} which may then be checked against the entanglement entropy in the beta-gamma 2D CFT.  As we know how to calculate holographic entanglement in 3D Chern--Simons gravity from the expectation value of Wilson loops, we will start by adapting the methods of \cite{Ammon:2013hba} to uplift the semiclassical approach for a single interval to the KS setup. This should yield an independent derivation of the prefactor of the KS action.

\acknowledgments{We would like to thank Kasia Budzik, Nana Cabo Bizet, Kevin Costello, Giuseppe Di Giulio, Lorenz Eberhardt, Davide Gaiotto, Elliott Gesteau, Victor Godet, Thomas Kögel, Adrián López-Raven, Dominik Neuenfeld, Boris Pioline, Ingmar Saberi and Bo Sundborg for useful discussions. We acknowledge financial support  by the Deutsche Forschungsgemeinschaft (DFG, German Research Foundation) through the German-Israeli Project Cooperation (DIP) grant ‘Holography and the Swampland’, as well as under Germany’s Excellence Strategy through the W\"{u}rzburg-Dresden Cluster of Excellence on Complexity and Topology in Quantum Matter - ct.qmat (EXC 2147, project-id 390858490).}

\begin{appendix}

\section{Integrability conditions for almost complex structures}\label{app:integrability_conditions}

In this appendix we show that a necessary condition for the existence of complex coordinates is the vanishing of the Nijenhuis tensor. In addition, we prove that the vanishing of the Nijenhuis tensor is equivalent to the Kodaira--Spencer equation.

\paragraph{Alternative expression for the Nijenhuis tensor.} Let us consider the contraction of the Nijenhuis tensor with two vector fields
\begin{equation}
    N^c(X,Y) \equiv N_{ab}^c\,X^aY^b\,,
\end{equation}
where the right hand side is defined as
\begin{equation}
    N_{ab}^c = 2\,\bigl(J^c_d\,\partial_{[a}J_{b]}^d-J_{[a}^d\,\partial_{\vert d\vert} J_{b]}^c\bigr)\,.
    \label{eq:Nijenhuis_app}
\end{equation}
We will now prove that
\begin{equation}
	N(X,Y) = [X,Y]+J([X,J(Y)])+J([J(X),Y])-[J(X),J(Y)]\,,
    \label{eq:Nijenhuis_more_standard}
\end{equation}
where $J^a(X) = J^a_b\,X^b$ and $[X,Y]^a = X^b\,\partial_b Y^a - Y^b\,\partial_b X^a$ is the Lie bracket of vector fields. To this end, we simply substitute the definitions to obtain
\begin{align}
    N^c(X,Y) &=X^a\,\partial_a Y^c - Y^a\,\partial_a X^c+J^c_a\,(X^b\,\partial_b (J^a_d\,Y^d) - J^b_d\, Y^d\,\partial_b X^a)\nonumber\\
    &+J^c_a\,(J^b_d\,X^d\,\partial_b Y^a - Y^b\,\partial_b (J^a_d\,X^d))-(J^a_b\,X^b\,\partial_a (J^c_d\,Y^d)- J^a_b\,Y^b\,\partial_a (J^c_d\,X^d))\,.
    \label{eq:Nijenhuis_more_standard_midstep}
\end{align}
Here the second term is
\begin{equation}
    J^c_a\,(X^b\,\partial_b (J^a_d\,Y^d) - J^b_d\, Y^d\,\partial_b X^a) = -X^b\,\partial_b Y^c + X^b\,Y^d\,J^c_a\,\partial_b J^a_d- J^c_a\,J^b_d\, Y^d\,\partial_b X^a\,,
\end{equation}
the third term is
\begin{equation}
    J^c_a\,(J^b_d\,X^d\,\partial_b Y^a - Y^b\,\partial_b (J^a_d\,X^d)) = J^c_a\,J^b_d\,X^d\,\partial_b Y^a +Y^b\,\partial_b X^c  - X^d\,Y^b\,J^c_a\,\partial_b J^a_d\,,
\end{equation}
and the fourth term is
\begin{align}
    &-(J^a_b\,X^b\,\partial_a (J^c_d\,Y^d)- J^a_b\,Y^b\,\partial_a (J^c_d\,X^d))\nonumber\\
    &= -J^a_b\,J^c_d\,X^b\,\partial_a Y^d -X^b\,Y^d\,J^a_b\,\partial_a J^c_d +J^a_b\,J^c_d\,Y^b\,\partial_a X^d+X^d\,Y^b\,J^a_b\,\partial_a J^c_d\,.
\end{align}
We see that all terms involving derivatives of $X$ and $Y$ cancel in \eqref{eq:Nijenhuis_more_standard_midstep} and we are left with
\begin{align}
    N^c(X,Y) &= X^b\,Y^d\,J^c_a\,\partial_b J^a_d- X^d\,Y^b\,J^c_a\,\partial_b J^a_d-X^b\,Y^d\,J^a_b\,\partial_a J^c_d+X^d\,Y^b\,J^a_b\,\partial_a J^c_d\\
    &= (J^c_a\,\partial_b J^a_d-J^c_a\,\partial_d J^a_b-J^a_b\,\partial_a J^c_d+J^a_d\,\partial_a J^c_b)\,X^b\,Y^d\,.
\end{align}

\paragraph{Vanishing of the Nijenhuis tensor.} From \eqref{eq:definition_holomorphic_coordinates} it follows that
\begin{equation}
    \overbar{\pounds}_{[a}\overbar{\pounds}_{b]}\,w^i = \bigl(i\, \partial_{[a} J_{b]}^{c} - J_{[a}^{d}\,\partial_{\vert d\vert} J_{b]}^{c}\big)\, \partial_c w^i\,,\quad \pounds_{[a}\pounds_{b]}\,\overbar{w}^i = -\bigl(i\, \partial_{[a} J_{b]}^{c} + J_{[a}^{d}\,\partial_{\vert d\vert} J_{b]}^{c}\big)\, \partial_c \overbar{w}^i\,.
    \label{eq:integrability_1}
\end{equation}
Notice that in terms of the projectors \eqref{eq:projectors}, equation \eqref{eq:definition_holomorphic_coordinates} takes the form \eqref{eq:complex_coordinates_Pi}. This implies that $\partial_a w^i$ and $\partial_a \overbar{w}^i $ are $(1,0)$- and $(0,1)$-forms respectively, and assuming they form a complete basis, they project the right-hand sides of \eqref{eq:integrability_1} to the respective tangent spaces. Thus the integrability conditions $ \overbar{\pounds}_{[a}\overbar{\pounds}_{b]}\,w^i = \pounds_{[a}\pounds_{b]}\,\overbar{w}^i = 0$ are equivalent to
\begin{equation}
    (\delta^c_d-i\,J^c_d)\,\bigl(i\, \partial_{[a} J_{b]}^{d} - J_{[a}^{e}\,\partial_{\vert e\vert} J_{b]}^{d}\big) = 0= (\delta^c_d+i\,J^c_d)\,\bigl(i\, \partial_{[a} J_{b]}^{d} + J_{[a}^{e}\,\partial_{\vert e\vert}  J_{b]}^{c}\big)\,,
    \label{eq:holomorphic_int}
\end{equation}
where we have used that the projectors are given by \eqref{eq:projectors}.

First, we expand
\begin{align}
    &(\delta^c_d-i\,J^c_d)\,\bigl(i\, \partial_{[a} J_{b]}^{d} - J_{[a}^{e}\,\partial_{\vert e\vert}J_{b]}^{d}\big)\label{eq:firstline_exp}\\
    &=\bigl(- J_{a}^{e}\,\partial_{e}J_{b}^{c} + J_{b}^{e}\,\partial_{e}J_{a}^{c}+J^c_d\,\partial_{a} J_{b}^{d} -J^c_d\, \partial_{b} J_{a}^{d}\bigr)+i\,\bigl(\partial_{a} J_{b}^{c} - \partial_{b} J_{a}^{c}+J^c_d\,J_{a}^{e}\,\partial_{e}J_{b}^{d} -J^c_d\,J_{b}^{e}\,\partial_{e}J_{a}^{d}\bigr)\,.\nonumber
\end{align}
In terms of the Nijenhuis tensor \eqref{eq:Nijenhuis_app} it follows by using $J^2 = -1$ that the second line of \eqref{eq:firstline_exp} can be written as
\begin{equation}
    (\delta^c_d-i\,J^c_d)\,\bigl(i\, \partial_{[a} J_{b]}^{d} - J_{[a}^{e}\,\partial_{\vert e\vert}J_{b]}^{d}\big) = \frac{1}{2}(\delta^c_d-i\,J^c_d)\,N_{ab}^d \,,
\end{equation}
and similarly, we obtain
\begin{equation}
    (\delta^c_d+i\,J^c_d)\,\bigl(i\, \partial_{[a} J_{b]}^{d} + J_{[a}^{e}\,\partial_{\vert e\vert}J_{b]}^{d}\big) = \frac{1}{2}(\delta^c_d+i\,J^c_d)\,N_{ab}^d\,.
\end{equation}
Therefore the equations \eqref{eq:holomorphic_int} are equivalent to
\begin{equation}
    \Pi^c_d\,N_{ab}^d = \overbar{\Pi}^c_d\,N_{ab}^d = 0\,,
\end{equation}
which imply $N_{ab}^c = 0$.

\paragraph{Integrability as the closure of Lie brackets.} Using the expression \eqref{eq:Nijenhuis_more_standard} for the Nijenhuis tensor, we consider
\begin{equation}
	N(\overbar{\mathcal{V}}_{i},\overbar{\mathcal{V}}_{j}) = [\overbar{\mathcal{V}}_{i},\overbar{\mathcal{V}}_{j}]+J([\overbar{\mathcal{V}}_{i},J(\overbar{\mathcal{V}}_{j})])+J([J(\overbar{\mathcal{V}}_{i}),\overbar{\mathcal{V}}_{j}])-[J(\overbar{\mathcal{V}}_{i}),J(\overbar{\mathcal{V}}_{j})]\,,
\end{equation}
where the vector field $\mathcal{\overbar{V}}_i$ is dual to $\overbar{\mathcal{A}}_i$ and defined in \eqref{eq:orthonormality}. Using
\begin{equation}
	J(\mathcal{V}_{i}) = i\,\mathcal{V}_i\,,\quad J(\overbar{\mathcal{V}}_{i}) = -i\,\overbar{\mathcal{V}}_{i}\,,
\end{equation}
which follow from the definition of $J$ \eqref{eq:complex_structure_A_V} and the orthonormality relations \eqref{eq:orthonormality}, we obtain
\begin{equation}
	N(\overbar{\mathcal{V}}_{i},\overbar{\mathcal{V}}_{j}) = [\overbar{\mathcal{V}}_{i},\overbar{\mathcal{V}}_{j}]-iJ([\overbar{\mathcal{V}}_{i},\overbar{\mathcal{V}}_{j}])-iJ([\overbar{\mathcal{V}}_{i},\overbar{\mathcal{V}}_{j}])+[\overbar{\mathcal{V}}_{i},\overbar{\mathcal{V}}_{j}]= 4\,\Pi([\overbar{\mathcal{V}}_{j},\overbar{\mathcal{V}}_{k}])\,,
\end{equation}
where the projector $\Pi $ is defined in \eqref{eq:projectors}. Similarly, we find $N(\mathcal{V}_{i},\mathcal{V}_{j}) = 4\,\overbar{\Pi}([\mathcal{V}_i,\mathcal{V}_j])$. Since $\Pi + \overbar{\Pi} = 1$, the vanishing of the Nijenhuis tensor is equivalent to
\begin{equation}
    \Pi([\overbar{\mathcal{V}}_{i},\overbar{\mathcal{V}}_{j}]) = \overbar{\Pi}([\mathcal{V}_i,\mathcal{V}_j]) = 0\,.
    \label{eq:closure_of_brackets}
\end{equation}Thus we have proven that the Nijenhuis tensor vanishes $N = 0$ if and only if the the algebra of $(1,0)$ vector fields closes under the Lie bracket, or in other words, if and only if the Lie bracket of any two $(1,0)$ vector fields is a $(1,0)$ vector field $[\mathcal{V}_i,\mathcal{V}_j] \in T^{(1,0)}P$. This gives an alternative definition of integrability of $J$.

Given a basis of dual one-forms \eqref{eq:10_forms}, the equations \eqref{eq:closure_of_brackets} are equivalent to
\begin{equation}
    \mathcal{A}^k([\overbar{\mathcal{V}}_{i},\overbar{\mathcal{V}}_{j}]) = \overbar{\mathcal{A}}^k([\mathcal{V}_i,\mathcal{V}_j]) = 0\,,
    \label{eq:integrability_V_commutator_2}
\end{equation}
for all $k = 1,2,3$. This follows from the fact that $\mathcal{A}^i(\Pi(X)) = \mathcal{A}^i(X) = \Pi(\mathcal{A}^i(X)) $ and $\overbar{\mathcal{A}}^i(\overbar{\Pi}(X)) = \overbar{\mathcal{A}}^i(X) = \overbar{\Pi}(\overbar{\mathcal{A}}^i(X)) $ where $X$ is an arbitrary vector field.

\paragraph{The Kodaira--Spencer equation.} We will now apply this to consider the integrability of the almost complex structure $J'$ defined in \eqref{eq:Jprime_explicitly}. By using the expressions \eqref{eq:dualvectors_primed} for the dual vector fields, we obtain
\begin{align}
    [\overbar{\mathcal{V}}_{i}',\overbar{\mathcal{V}}_{j}'] &= \overbar{\sigma}^k_i\overbar{\sigma}^l_j\,[(\overbar{\partial}_{k}-\alpha_{k}^{\;\;\, n}\,\partial_n),(\overbar{\partial}_{l}-\alpha_{l}^{\;\;\, m}\,\partial_m)]\nonumber\\
    &+(\overbar{\mathcal{V}}_{i}'\overbar{\sigma}^l_j)\,(\overbar{\partial}_{l}-\alpha_{l}^{\;\;\, m}\,\partial_m)-(\overbar{\mathcal{V}}_{j}'\overbar{\sigma}^k_i)\,(\overbar{\partial}_{k}-\alpha_{k}^{\;\;\, n}\,\partial_n)\,,\label{eq:Vbarprime_commutator}
\end{align}
where the remaining commutator is explicitly
\begin{equation}
    [(\overbar{\partial}_{k}-\alpha_{k}^{\;\;\, n}\,\partial_n),(\overbar{\partial}_{l}-\alpha_{l}^{\;\;\, m}\,\partial_m)] = -2\,(\overbar{\partial}_{[k}\alpha_{l]}^{\;\;\, n} - \alpha_{[k}^{\;\;\, m}\,\partial_{\vert m\vert}\alpha_{l]}^{\;\;\, n})\,\partial_n\,.
\end{equation}
Thus \eqref{eq:Vbarprime_commutator} can be written as
\begin{equation}
    [\overbar{\mathcal{V}}_{i}',\overbar{\mathcal{V}}_{j}'] = -2\,\overbar{\sigma}^k_i\overbar{\sigma}^l_j\,(\overbar{\partial}_{[k}\alpha_{l]}^{\;\;\, n} - \alpha_{[k}^{\;\;\, m}\,\partial_{\vert m\vert}\alpha_{l]}^{\;\;\, n})\,\partial_n+(\overbar{\mathcal{V}}_{i}'\overbar{\sigma}^k_j)\,\overbar{\Sigma}_k^n\,\overbar{\mathcal{V}}'_n-(\overbar{\mathcal{V}}_{j}'\overbar{\sigma}^k_i)\,\overbar{\Sigma}_k^n\,\overbar{\mathcal{V}}_n'\,,
\end{equation}
where $\overbar{\Sigma}$ is the inverse of $\overbar{\sigma}$ and we used again \eqref{eq:dualvectors_primed}. Here we write
\begin{equation}
    \partial_n = (K^{-1})_n^j (\Sigma^k_j \mathcal{V}'_k+\overbar{\alpha}_j^{\;\;\, p}\overbar{\Sigma}_p^k \overbar{\mathcal{V}}'_k)\,,
\end{equation}
where we have defined $K^i_j = \delta^i_j -\overbar{\alpha}_j^{\;\;\, p}\alpha_p^{\;\;\, i}$. Because of \eqref{eq:determinant_condition} the inverse of $K$ always exists.
We obtain
\begin{equation}
\begin{split}
    [\overbar{\mathcal{V}}_{i}',\overbar{\mathcal{V}}_{j}'] = &-2\,\overbar{\sigma}^k_i\overbar{\sigma}^l_j\,(K^{-1})_n^p
\Sigma^q_p\,(\overbar{\partial}_{[k}\alpha_{l]}^{\;\;\, n} - \alpha_{[k}^{\;\;\, m}\,\partial_{\vert m\vert}\alpha_{l]}^{\;\;\, n})\,\mathcal{V}_{q}'\\
&-2\,\overbar{\sigma}^k_i\overbar{\sigma}^l_j\,(K^{-1})_n^r
\overbar{\alpha}_r^{\;\;\, p}\overbar{\Sigma}_p^q\,(\overbar{\partial}_{[k}\alpha_{l]}^{\;\;\, n} - \alpha_{[k}^{\;\;\, m}\,\partial_{\vert m\vert}\alpha_{l]}^{\;\;\, n})\, \overbar{\mathcal{V}}'_q+2\,(\overbar{\Sigma}_l^n\,\overbar{\mathcal{V}}_{[i}'\overbar{\sigma}^l_{j]}))\,\overbar{\mathcal{V}}_n'\,.
\end{split}
\end{equation}
Since $\mathcal{A}'(\overbar{\mathcal{V}}_{i}') = 0$ and $\mathcal{A}'^i(\mathcal{V}'_q) = \delta_q^i$ due to \eqref{eq:Aprime_definition}, we obtain
\begin{equation}
    \mathcal{A}'^q([\overbar{\mathcal{V}}_{i}',\overbar{\mathcal{V}}_{j}']) = -2\,\overbar{\sigma}^k_i\overbar{\sigma}^l_j\,(K^{-1})_n^p
\Sigma^q_p\,(\overbar{\partial}_{[k}\alpha_{l]}^{\;\;\, n} - \alpha_{[k}^{\;\;\, m}\,\partial_{\vert m\vert}\alpha_{l]}^{\;\;\, n})\,.
\end{equation}
Similarly, we obtain
\begin{equation}
    \overbar{\mathcal{A}}'^q([\mathcal{V}_{i}',\mathcal{V}_{j}']) = -2\,\sigma^k_i\sigma^l_j\,(\overbar{K}^{-1})_n^p\overbar{\Sigma}_p^q\,(\partial_{[k}\overbar{\alpha}_{l]}^{\;\;\, n} - \overbar{\alpha}_{[k}^{\;\;\, m}\,\overbar{\partial}_{\vert m\vert}\overbar{\alpha}_{l]}^{\;\;\, n})\,.
\end{equation}
Here, we have defined $\overbar{K}_j^i = \delta_j^i -\alpha_j^{\;\;\, p}\overbar{\alpha}_p^{\;\;\, i}$.
Since $\sigma$, $\overbar{\sigma}$ are invertible, the integrability \eqref{eq:integrability_V_commutator_2} of $J'$ is equivalent to the Kodaira--Spencer equations \eqref{eq:KS_equation_definition}.

\section{Diffeomorphism covariance of Kodaira--Spencer equation}\label{app:diffeos}

In this appendix, we derive the transformation behavior of the complex structure deformation under diffeomorphisms and show that the Kodaira--Spencer equation transforms covariantly \cite{Bandelloni:1998vp}. 

Let us consider a diffeomorphism $\mathcal{D}^a = \mathcal{D}^a(x)$ which in complex coordinates of $J$ has components $\mathcal{D}=(D^i(w,\overbar{w}),\overbar{D}^i(w,\overbar{w}))$. Under this diffeomorphism, the components of the connection $\mathcal{A}'^i$ as defined in \eqref{eq:Aprime_definition} transform as 1-forms,
\begin{equation}
\begin{split}
    \widetilde{\mathcal{A}}'^i (x) &\equiv ({\Sigma}^i_j\circ \mathcal{D})\,(\d \widetilde{w}^j + ({\alpha}_{k}^{\;\,\, j}\circ \mathcal{D})\,\d\widetilde{\overbar{w}}^k) \\
    &= ({\Sigma}^i_j\circ\mathcal{D}) \,\left[\left(\frac{\partial D^j}{\partial w^m}+({\alpha}_{k}^{\;\,\, j}\circ\mathcal{D})\,\frac{\partial \overbar{D}^k}{\partial w^m}\right) \d w^m+\left(\frac{\partial D^j}{\partial \overbar{w}^m}+({\alpha}_{k}^{\;\,\, j}\circ\mathcal{D})\,\frac{\partial \overbar{D}^k}{\partial \overbar{w}^m}\right) \d \overbar{w}^m\right] \,,\label{eq:transformation_connection}
\end{split}
\end{equation}
where $\widetilde{w}^i$ and $\widetilde{\overbar{w}}^i$ are defined in \eqref{eq:complex_coordinates_are_scalars}. Now, we define,
\begin{equation}
	N_{j}^{i} \equiv \partial_{j}D^{i}+(\alpha_{k}^{\;\;\,i}\circ \mathcal{D})\,\partial_{j}\overbar{D}^{k}\,,\quad \overbar{N}^{i}_{j} \equiv \overbar{\partial}_{j}D^{i}+(\alpha_{k}^{\;\;\,i}\circ \mathcal{D})\,\overbar{\partial}_{j}\overbar{D}^{k}\,.\label{eq:N_def}
\end{equation}
Using these definitions, we rewrite \eqref{eq:transformation_connection} as
\begin{equation}
\begin{split}
    \widetilde{\mathcal{A}}'^i (x) &= ({\Sigma}^i_j\circ \mathcal{D})\,(N^j_k\, \d w^k+ \overbar{N}^j_k\,\d \overbar{w}^k)\\
    &=({\Sigma}^i_j\circ \mathcal{D})\, N^j_k \,(\d w^k+ (N^{-1})^k_n\, \overbar{N}^n_m\,\d \overbar{w}^m)\,.
\end{split}
\end{equation}
Here, we have assumed that the inverse of $N$ exists. Thus, $\widetilde{\mathcal{A}}'^i (x)$ has the form \eqref{eq:Aprime_definition} and we read off the transformation of $\Sigma^i_j$ and $\alpha_{i}^{\;\,\, j}$,
\begin{equation}
    \widetilde{\alpha}_{i}^{\;\,\, j} = (N^{-1})^j_k\, \overbar{N}^k_i\,,\quad \widetilde{\Sigma}^i_j = ({\Sigma}^i_k\circ \mathcal{D})\, N^k_j\,.
    \label{eq:alpha_trans_app}
\end{equation}
We note that for a holomorphic-antiholomorphic diffeomorphisms $\mathcal{C}_J$ as defined in \eqref{eq:hol_antihol_def}, the definitions of $N^i_j$ and $\overbar{N}^i_j$ simplify,
\begin{equation}
    N^i_j = \partial_j f^i\,,\quad
    \overbar{N}^i_j  = (\alpha_{k}^{\;\,\, i}\circ \mathcal{C}_J)\,\overbar{\partial}_j\overbar{f}^k\,,
\end{equation}
which proves \eqref{eq:alpha_transformation_C} and \eqref{eq:Sigma_transformation_C}. Analogous statements hold for $\overbar{\Sigma}^i_j$ and $\overbar{\alpha}_{i}^{\;\,\, j}$.

Next, we investigate the transformation of the Kodaira-Spencer equation \eqref{eq:KS_equation_definition}. The transformed holomorphic Kodaira--Spencer equation is given by
\begin{equation}
    \widetilde{E}^i_{jk}= \overbar{\partial}_{[j}\, \widetilde{\alpha}_{k]}^{\;\;\,i}-\widetilde{\alpha}_{[j}^{\;\;\,l}\,\partial_{\vert l \vert}\,\widetilde{\alpha}_{k]}^{\;\;\,i} = (\overbar{\partial}_{[j}-\widetilde{\alpha}_{[j}^{\;\;\,l}\,\partial_{\vert l \vert})\,\widetilde{\alpha}_{k]}^{\;\;\,i}\,.
\end{equation}
Using the transformation of the complex structure deformation \eqref{eq:alpha_trans_app}, we find
\begin{equation}
\begin{split}
    \widetilde{E}^i_{jk} &= (\overbar{\partial}_{[j}-(N^{-1})^l_n\, \overbar{N}^n_{[j}\,\partial_{\vert l \vert})\,(\overbar{N}^m_{k]}\,(N^{-1})^i_m )\\
    &= (N^{-1})^i_m(\overbar{\partial}_{[j}-(N^{-1})^l_n\, \overbar{N}^n_{[j}\,\partial_{\vert l \vert})\,\overbar{N}^m_{k]} +\overbar{N}^m_{[k}\,(\overbar{\partial}_{j]}-\overbar{N}^n_{j]}\,(N^{-1})^l_n\, \partial_{\vert l \vert})\,(N^{-1})^i_m \\
    &= (N^{-1})^i_m(\overbar{\partial}_{[j}-(N^{-1})^l_n\, \overbar{N}^n_{[j}\,\partial_{\vert l \vert})\,\overbar{N}^m_{k]} - (N^{-1})^i_p(N^{-1})^q_m \overbar{N}^m_{[k} (\overbar{\partial}_{j]}-\overbar{N}^n_{j]}(N^{-1})^l_n  \partial_{\vert l \vert})N^p_q \,,
\end{split}
\end{equation}
where in the last step, we have used the inverse derivative rule. After using the definitions \eqref{eq:N_def} and some laborious manipulations, we obtain
\begin{equation}
    \widetilde{E}^i_{jk} = (N^{-1})_m^i\,\bigl((\overbar{\partial}_{[j}-\widetilde{\alpha}_{[j}^{\;\;\,p}\,\partial_{\vert p \vert})\,\overbar{D}^n\bigr)\,\bigl( (\overbar{\partial}_{k]}-\widetilde{\alpha}_{k]}^{\;\;\,q}\partial_{\vert q \vert})\,\overbar{D}^l\bigr)\, (E_{nl}^m\, \circ \mathcal{D})\,.
\end{equation}
Thus, the Kodaira--Spencer equation is diffeomorphism invariant in the sense that the deformation $\widetilde{\alpha}_{i}^{\;\;\,j}$ is integrable if $\alpha_{i}^{\;\;\,j}$ is. See also \cite{Bandelloni:1998vp}.

\section{Equivalence of flatness and integrability of the almost complex structure}\label{app:flatness_integrability}

In this appendix, we provide details on the proof of equivalence between flatness of $\mathcal{A}$ and integrability of the $SU(2)$ invariant almost complex structure. In particular we give proofs for \eqref{eq:V_fundamental_vector_field} and \eqref{eq:brackets_fundamental_vector_fields} in the main text.

\paragraph{Derivation of \eqref{eq:V_fundamental_vector_field}.} We decompose $\mathcal{V}_i$ into vertical $v_i\in V$ and horizontal $h_i\in H$ vector fields as
\begin{equation}
    \mathcal{V}_i = v_i + h_i\,.
\end{equation}
The vertical tangent space is spanned by fundamental vector fields. In addition, since by assumption the almost complex structure $J$ has no invariant vertical subspace, the horizontal tangent bundle is spanned by the action of $J$ on the vertical tangent bundle. Thus there exist Lie algebra elements $a_i,b_i\in \mathfrak{su}(2)$ such that
\begin{equation}
    \mathcal{V}_i = a_i^\#+ J(b_i^\#)\,,
\end{equation}
which in components is explicitly
\begin{equation}
    \mathcal{V}_i^a = (a_i^\#)^a+ J_b^a\, (b_i^\#)^a\,.
\end{equation}
The vector fields $\mathcal{V}_i$ are defined by \eqref{eq:orthonormality} and they are $(1,0)$-vector fields that satisfy $J(\mathcal{V}_i) = i\,\mathcal{V}_i$. Thus it follows that
\begin{equation}
    \overbar{\Pi}_b^a\, \mathcal{V}_i^b = 0\,,
\end{equation}
where the projector is defined in \eqref{eq:projectors}. It follows that $b_i ^\# =- i a_i^\# $. Thus we obtain
\begin{equation}
    \mathcal{V}_i^a = (\delta_b^a-i J_b^a)(a_i^\#)^b = 2\,\Pi_b^a\, (a_i^\#)^b\,.
\end{equation}
Next, we have
\begin{equation}
    \mathcal{A}(\mathcal{V}_i ) = \mathcal{A}^j_a\, \mathcal{V}^a_i\, \tau_j = \tau_i\,.
    \label{eq:A_V_action}
\end{equation}
Using $\mathcal{A} = \mathcal{W} + i\mathcal{E} $ \eqref{eq:A_forms_W_E} together with the identities \eqref{eq:upliftframe_fundamental_vfield} and \eqref{eq:connection_fundamental_vfield}, equation \eqref{eq:A_V_action} implies $a_i = \frac{1}{2}\,\tau_i$ thus proving \eqref{eq:V_fundamental_vector_field}.

\paragraph{Proof of \eqref{eq:brackets_fundamental_vector_fields}.} We start by rewriting
\begin{equation}
   4\,[\Pi (\tau_i^{\#}),\Pi (\tau_j^{\#})]  = [\tau_i^{\#},\tau_j^{\#}]-i\,[\tau_i^{\#},J(\tau_j^{\#})]-i\,[J(\tau_i^{\#}),\tau_j^{\#}]-[J(\tau_i^{\#}),J(\tau_j^{\#})]\,,
   \label{eq:Pi_Pi_commutator}
\end{equation}
where we have used the definition of the projectors \eqref{eq:projectors}.
By the formula \eqref{eq:Nijenhuis_more_standard} for the Nijenhuis tensor, we obtain
\begin{equation}
    [J(\tau_i^{\#}),J(\tau_j^{\#})] = -N(\tau_i^{\#},\tau_j^{\#})+[\tau_i^{\#},\tau_j^{\#}]+J([\tau_i^{\#},J(\tau_j^{\#})])+J([J(\tau_i^{\#}),\tau_j^{\#}])\,.
    \label{eq:J_J_commutator_app}
\end{equation}
Substituting to \eqref{eq:Pi_Pi_commutator} gives
\begin{equation}
    4\,[\Pi (\tau_i^{\#}),\Pi (\tau_j^{\#})]  = N(\tau_i^{\#},\tau_j^{\#})-2i\,\Pi([\tau_i^{\#},J(\tau_j^{\#})]+[J(\tau_i^{\#}),\tau_j^{\#}])\,.\label{eq:Lie_bracket_proj_hol_hol}
\end{equation}
Similarly, we obtain from \eqref{eq:projectors} that
\begin{equation}
    4\,[\Pi (\tau_i^{\#}),\overbar{\Pi} (\tau_j^{\#})]  = [\tau_i^{\#},\tau_j^{\#}]+i\,[\tau_i^{\#},J(\tau_j^{\#})]-i\,[J(\tau_i^{\#}),\tau_j^{\#}]+[J(\tau_i^{\#}),J(\tau_j^{\#})]\,.
\end{equation}
Substituting \eqref{eq:J_J_commutator_app} gives
\begin{equation}
    4\,[\Pi (\tau_i^{\#}),\overbar{\Pi} (\tau_j^{\#})]  =-N(\tau_i^{\#},\tau_j^{\#})+2\,[\tau_i^{\#},\tau_j^{\#}]-2i\,\overbar{\Pi}([J(\tau_i^{\#}),\tau_j^{\#}])+2i\,\Pi([\tau_i^{\#},J(\tau_j^{\#})])\,.\label{eq:Lie_bracket_proj_hol_antihol}
\end{equation}
Now, we consider a curve $g(t)=e^{t\tau_i}\in SU(2)$ that generates the fundamental vector field $\tau_i^\#$ to write the Lie bracket
\begin{equation}
    [\tau_i^{\#},J(\tau_j^{\#})] = \mathcal{L}_{\tau_i^{\#}}J(\tau_j^{\#}) = \lim_{t\rightarrow 0}\frac{(\mathcal{R}_{g(t)})_{*} J (\tau_j^\#) - J(\tau_j^\#)}{t}\,,
    \label{eq:Lie_bracket_fund}
\end{equation}
where $\mathcal{L}$ denotes the Lie derivative. When the Lie bracket is evaluated at a point $p$, the vector field $J (\tau_j^\#)$ in the first term in the numerator is understood to be evaluated at $\mathcal{R}_{g(-t)}(p)$. This ensures that $(\mathcal{R}_{g(t)})_{*} J (\tau_j^\#)$ is an element of the tangent space at $p$ and the subtraction is well defined. Under the right-action diffeomorphism induced by $g(-t) = g(t)^{-1}$, $J$ transforms as
\begin{equation}
    \widetilde{J} = (\mathcal{R}_{g(t)})_* (\mathcal{R}_{g(-t)})^* J\,.
\end{equation}
Here, if we evaluate the left-side of the equation at a point $p$, we have to evaluate $J$ on the right-hand side at $\mathcal{R}_{g(-t)}(p)$. Thus, it follows that the first term in the numerator of \eqref{eq:Lie_bracket_fund} is given by
\begin{equation}
    (\mathcal{R}_{g(t)})_{*} J (\tau_j^\#) = \widetilde{J}((\mathcal{R}_{g(t)})_{*}\tau_j^\#)\,.
\end{equation}
Thus, if $J$ is $SU(2)$-invariant and \eqref{eq:su2_invariance} holds, we find
\begin{equation}
    [\tau_i^{\#},J(\tau_j^{\#})] = J\left(\lim_{t\rightarrow 0} \frac{(\mathcal{R}_{g(t)})_{*} \tau_j^\# - \tau_j^\#}{t}\right) = J \left(\mathcal{L}_{\tau_i^{\#}}\tau_j^{\#} \right)= J ([ \tau_i^{\#},\tau_j^{\#}])\,.\label{eq:Lie_bracket_complex_structure}
\end{equation}
Substituting \eqref{eq:Lie_bracket_complex_structure} into \eqref{eq:Lie_bracket_proj_hol_hol} and using the definition of the projectors \eqref{eq:projectors} as well as the property \eqref{eq:almost_complex_structure_def}, we obtain
\begin{equation}
    4\,[\Pi (\tau_i^{\#}),\Pi (\tau_j^{\#})] - 4\,\Pi([\tau_i^{\#},\tau_j^{\#}])  = N(\tau_i^{\#},\tau_j^{\#})\,,
\end{equation}
and substituting to \eqref{eq:Lie_bracket_proj_hol_antihol}, we obtain
\begin{equation}
    4\,[\Pi (\tau_i^{\#}),\overbar{\Pi} (\tau_j^{\#})]  =-N(\tau_i^{\#},\tau_j^{\#})\,,
\end{equation}
thus proving \eqref{eq:brackets_fundamental_vector_fields}.

\section{Induced complex structure on \texorpdfstring{$SL(2,\mathbb{C})$}{SL(2,C)}}\label{app:induced_complex_structure}

In this appendix, we give a more detailed derivation of the complex structure on $P_6= SL(2,\mathbb{C})$ induced by the embedding in $\mathbb{C}^4$ as in \eqref{eq:deformed_conifold}. Let $(v^{\hat{i}},\overbar{v}^{\hat{i}})$, with $\hat{i} = 1,2,3,4$, be complex coordinates on $\mathbb{C}^4$ where each $v^{\hat{i}}$ parametrizes a factor of $\mathbb{C}$ in the standard way $\mathbb{C}\cong \mathbb{R}^2$. The corresponding complex structure $J\colon T\mathbb{C}^4\rightarrow T\mathbb{C}^4$ on $\mathbb{C}^4$ takes the form
\begin{equation}
    J= i\, \d v^{\hat{i}}\otimes \hat{\partial}_{\hat{i}} - i\,\d \overbar{v}^{\hat{i}}\otimes\hat{\overbar{\partial}}_{\hat{i}}\,.
\end{equation}
Here, we denote $\hat{\partial}_{\hat{i}} = \frac{\partial}{\partial v^{\hat{i}}}$ and $\hat{\overbar{\partial}}_{\hat{i}} = \frac{\partial}{\partial \overbar{v}^{\hat{i}}}$. The hypersurface $P_6\subset \mathbb{C}^4$ is defined by the two equations
\begin{equation}
    f(w) \equiv v^1 v^4-v^2 v^3 -1 = 0\,,\quad \overbar{f}(\hat{\overbar{v}}) \equiv \overbar{v}^1 \overbar{v}^4-\overbar{v}^2 \overbar{v}^3 -1 = 0\,.
\end{equation}
The kernel of $\d f$ defines a notion of vectors tangent to the embedding hypersurface. Explicitly calculating $\d f$, we find three linearly independent vectors spanning its kernel,
\begin{align}
    b_1 &= \hat{\partial}_1 - \frac{v_4}{v_1}\,\hat{\partial}_4\,,\\
    b_2 &= \hat{\partial}_2 +\frac{v_3}{v_1}\,\hat{\partial}_4\,,\\
    b_3 &= \hat{\partial}_3 + \frac{v_2}{v_1}\,\hat{\partial}_4\,.
\end{align}
Similarly, the kernel of $\d \overbar{f}$ is spanned by the complex conjugates $\overbar{b}_i = (b_i)^*$ and together $(b_i,\overbar{b}_i)$ span the linear subspace of $T_p\mathbb{C}^4$ tangent to $P_6$.

Now, we introduce a parametrization $w^i$ of $P_6$, with $i = 1,2,3$, and write the embedding $E\colon P_6\rightarrow \mathbb{C}^4$ as $v^{\hat{i}} = E^{\hat{i}}(w)$ and $\overbar{v}^{\hat{i}} = \overbar{E}^{\hat{i}}(\overbar{w})$. By definition, the embedding satisfies $f(E(w)) = 0$, $\overbar{f}(\overbar{E}(\overbar{w})) = 0$ and it is explicitly
\begin{equation}
    v^{1,2,3} = E^{1,2,3}(w) = w^{1,2,3} \,,\quad v^4 = E^4(w)= \frac{1+w^2 w^3}{w^1}\,.
    \label{eq:embedding_E_app}
\end{equation}
Pulling back the form index of $J$ onto $P_6$ using this embedding gives by an explicit calculation
\begin{equation}
    E^*J = i\, \d w^i\otimes b_i-i\, \d \overbar{w}^i\otimes \overbar{b}_i\,.
\end{equation}
This is a map $E^*J\colon TP_6\rightarrow T\mathbb{C}^4$ and does not yet define a complex structure on $P_6$ whose image is $T P_6$. In other words, for $X\in TP_6$, $(E^*J)(X)$ is a vector field on $\mathbb{C}^4$ which is defined completely by its action on functions in $\mathbb{C}^4$. Notice that the action of $(E^*J)(X)$ on functions restricted to $P_6$ defines a vector field on $P_6$. Thus we can define an induced complex structure $J_{\text{ind}}\colon TP_6\rightarrow TP_6$ from $E^*J$ via restriction as
\begin{equation}
    J_{\text{ind}}(X)(S(w)) \equiv (E^*J)(X)(\hat{S}(v))\vert_{v = E(w)}
\end{equation}
where $S(w)$ is a function on $P_6$ obtained as the restriction of a function $\hat{S}(v)$ on $\mathbb{C}^4$ via $S(w) \equiv \hat{S}(E(w))$. It follows that the complex coordinates associated to $J_{\text{ind}}$ on $P_6$ coincide with $w^i$. To see this, we use the fact that
\begin{equation}
   \partial_i S(w) = \partial_i \hat{S}(E(w)) = \partial_iE^{\hat{i}}(w)\,\hat{\partial}_{\hat{i}} \hat{S}(v) \vert_{v = E(w)}\,.
\end{equation}
By using the embedding \eqref{eq:embedding_E_app}, we obtain
\begin{equation}
    \partial_i S(w) = b_i(\hat{S}(v))\vert_{v = E(w)}
    \label{eq:partialtilde_i}
\end{equation}
so that we can write the induced complex structure as
\begin{equation}
    J_{\mathrm{ind}} = i\, \d w^i\otimes  \partial_i - i\, \d \overbar{w} ^i\otimes \overbar{\partial}_i\,.
\end{equation}
To summarize, there is an induced complex structure on $SL(2,\mathbb{C})$ inherited by the embedding into $\mathbb{C}^4$ under which $SL(2,\mathbb{C})$ inherits three of the four holomorphic coordinates of $\mathbb{C}^4$.

\section{Degrees of freedom of an almost complex structure}\label{app:J_DOFs}

In this appendix, we present a counting of the independent degrees of freedom of an almost complex structure. To do so, it is advantageous to reframe \eqref{eq:almost_complex_structure_def} in terms of matrices.

An almost complex structure on a $2n$-dimensional manifold
acts on the tangent space as a vector valued 1-form with
\begin{equation}
    J^2=-\mathrm{id}\,.\label{eq:comples_structure_squares_m_id}
\end{equation}
Here $\mathrm{id}$ is the identity endomorphism. We represent the almost complex structure as a matrix. The minimal polynomial of $J$ is
\begin{equation}
    \chi^2+1=(\chi+i)(\chi-i)=0\,.
\end{equation}
Since the minimal polynomial has no duplicate roots, $J$ is complex diagonalizable with eigenvalues $\pm i$. For $k=1,\ldots,n$ we denote by $v_k$ an eigenvector of $J$ with eigenvalue $+i$. But since $J$ is real, it follows that
\begin{equation}
    J v_k = i v_k\quad \Rightarrow\quad J\overbar{v}_k=-i\overbar{v}_k\,,\label{eq:J_conjugate_eigenvalues}
\end{equation}
i.e.~the eigenvectors occur in conjugate pairs and there exists a real basis $(x_i, y_i)$ with
\begin{equation}
    v_i = x_i +i y_i\,,\qquad \overbar{v}_i = x_i - i y_i\,.
\end{equation}
By comparing coefficients in \eqref{eq:J_conjugate_eigenvalues}, we find that
\begin{equation}
    J x_i = -y_i\,, \qquad J y_i = x_i\,.
\end{equation}
Thus, we construct the real matrix
\begin{equation}
    P = (x_1,\dots,x_n, y_1,\dots,y_{n})\in GL(2n,\mathbb{R})\,,
\end{equation}
such that
\begin{equation}
    J = P J_0 P^{-1}\,,
    \label{eq:complex_structure_sl6r_parametrization}
\end{equation}
where $J_0$ is a reference complex structure defined as
\begin{equation}
    J_0 = \begin{pmatrix}
        0 & \1_n \\
        -\1_n & 0
    \end{pmatrix}.
\end{equation}
Here, $\1_n \equiv \text{diag}\,(1,\ldots,1)$ is the $n$-dimensional unit matrix.

Since $P$ is a real matrix, we find that any $J$ satisfying \eqref{eq:comples_structure_squares_m_id} is real conjugate to $J_0$. It follows that any two $J$ and $J'$ are real conjugate so that all possible almost complex structures lie in a $GL(2n,\mathbb{R})$ orbit of $J_0$. To find the dimensionality of this orbit, we have to find the stabilizer, i.e.~the subset $C\subset GL(2n,\mathbb{R}) $ defined by $S\in GL(2n,\mathbb{R})$ satisfying
\begin{equation}
    S J_0 S^{-1} = J_0 \quad \Leftrightarrow\quad  [S, J_0]=0\,.
\end{equation}
By explicit computation, we find that $S$ has to be of the block diagonal form
\begin{equation}
    S = \begin{pmatrix}
        A & -B \\
        B & A
    \end{pmatrix}\,, \qquad A,B\in \mathbb{R}^{n\times n}\,,\label{eq:centralizer_form}
\end{equation}
such that the inverse $S^{-1}$ exists. Note that $S$ is conjugate to the complex matrix
\begin{equation}
    K = \begin{pmatrix}
        A+i B & 0\\
        0 & A-i B
    \end{pmatrix}\,, \quad A\pm i B \in \mathbb{C}^{n\times n}\,,
\end{equation}
such that
\begin{equation}
    \det S = \det K = \det(A+iB) \det(A-i B)\,.
\end{equation}
Thus $S$ is invertible iff the complex matrices $A+i B$ and $A-i B$ are invertible. The map
\begin{equation}
    C\to GL(n,\mathbb{C}),\qquad S= \begin{pmatrix}
        A & -B \\
        B & A
    \end{pmatrix}\mapsto A+i B\,.
\end{equation}
defines a group isomorphism between the stabilizer $C$ and the group $GL(n,\mathbb{C})$, since the map preserves group multiplication. We check this by calculating
\begin{equation}
    S \cdot S' = \begin{pmatrix}
        A A' - B B'& -(A B'+B A')\\
        AB'+BA' & A A'- B B'
    \end{pmatrix}\,
\end{equation}
and
\begin{equation}
    (A+i B )(A'+i B') = A A'- B B'+i\, (A B'+B A')\,.
\end{equation}
The fact that the stabilizer is isomorphic to $GL(n,\mathbb{C})$ is not surprising, as we know that complex structures are equivalent if they are related by a holomorphic-antiholomorphic diffeomorphism as defined in \eqref{eq:hol_antihol_def}, which preserve the span of the $+i$ eigenvectors and act locally as $GL(n ,\mathbb{C})$. 

The action of the stabilizer defines an equivalence relation on $GL(2n,\mathbb{R})$,
\begin{equation}
    P\sim P' \quad\Leftrightarrow\quad \exists S \in C, P'=S P\,,
\end{equation}
since the left action by elements of the centralizer leaves the complex structure \eqref{eq:complex_structure_sl6r_parametrization} invariant. Thus, the number of independent degrees of freedom is
\begin{equation}
    \mathrm{dim}\,GL(2n, \mathbb{R})-\mathrm{dim}\,GL(n, \mathbb{C}) = (2n)^2-2n^2 = 2n^2 \,.
\end{equation}
Thus, a general almost complex structure on a $2n$-dimensional manifold is parametrized by $2n^2$ independent functions of all coordinates. For a six-dimensional manifold with $n=3$, this makes $18$ independent functions.

\section{Absence of an invariant vertical subspace}\label{app:no_invariant_subspace}

In this appendix, we give a more detailed discussion of \eqref{eq:no_vertical_inv_subspace}, i.e.~the absence of a vertical invariant subspace of the almost complex structure.

\paragraph{Derivation of \eqref{eq:dim_reduction_explicit}.} We now derive the explicit expression \eqref{eq:dim_reduction_explicit} of the uplifted almost complex structure in the right-invariant basis. Note that the right- and left-invariant Maurer--Cartan forms are conjugate,
\begin{equation}
    G^{-1}\d G = G^{-1}\d G\, G^{-1} G = G^{-1}\,\Theta\, G\,,
\end{equation}
where we used the definition $\Theta = \d G G^{-1}$ as in the main text. Using this fact, we rewrite the definitions \eqref{eq:uplift_definition} of the uplifted $\mathcal{W}$,
\begin{equation}
    \mathcal{W}= G^{-1} WG+G^{-1}\,\Theta\, G = G^{-1}(W+\Theta)\, G\,.
\end{equation}
Expanding $W=W^i\tau_i$ and $\Theta= \Theta^i\tau_i$ and using \eqref{eq:adjoint_action}, we obtain
\begin{equation}
    \mathcal{W}^i = \Lambda^i_{\;\;j}\, (W^j+\Theta^j)= \Lambda^i_{\;\;j}\, (W^j_\mu\,\d y^\mu+\Theta^j)\,.
    \label{eq:Wi_Theta}
\end{equation}
Similarly, from \eqref{eq:uplift_definition}, we find
\begin{equation}
    \mathcal{E}^i = \Lambda^i_{\;\; j}\, E^j= \Lambda^i_{\;\; j}\, E^j_\mu\, \d y^\mu\,.
    \label{eq:Ei_Theta}
\end{equation}
In both of the above equations, we have suppressed the dependence of $\Lambda^i_{\;\;j} = \Lambda^i_{\;\;j}(x)$ on the coordinates $x^\mu$. Note that both $E^i$ and $W^i$ have only form legs in the $\d y^\alpha$ directions, since they live on the base. Thus, we have expressed $\mathcal{W}^i$ and $\mathcal{E}^i$ in the basis $P^K = (\d y^\mu,\Theta^i)$. As described in the main text, we define a dual basis $P_K =(\partial_{y^\mu},\Theta_i)$. The vector fields $\mathcal{W}_i$ in this basis take the form
\begin{equation}
    \mathcal{W}_i =(\Lambda^{-1})^j_{\;\;i}\, \Theta_j\,,
    \label{eq:Wi_inverse_Theta}
\end{equation}
and similarly $\mathcal{E}_i$ are
\begin{equation}
    \mathcal{E}_i = (\Lambda^{-1})^j_{\;\;i}E_j^\mu\, (\partial_{y^\mu}- W_\mu^k\, \Theta_k)\,.\label{eq:vielbein_right_invariant_basis}
\end{equation}
Note that $E_j^\mu$ is defined by \eqref{eq:inverse_vielbein}. We construct the almost complex structure from \eqref{eq:complex_structure_bundle} and obtain
\begin{equation}
\begin{split}
    J &= - W^i_\mu\,E^\nu_i\, \d y^\mu\otimes \partial_{y^\nu}- E_i^\mu\, \Theta^i\otimes \partial_{y^\mu}\\
    &\;\;\;\;+(E^i_\mu+W^i_\nu\,E^\nu_j\, W^j_\mu)\,\d y^\mu\otimes  \Theta_i+\, E^\mu_i\,W^j_\mu \,\Theta^i \otimes \Theta_j\,.\label{eq:app_complex_structure_right_inv_basis}
\end{split}
\end{equation}

\paragraph{Consistency of dimensional reduction.} The property \eqref{eq:invertibility_of_J} is vital for the consistency of the dimensional reduction, since it ensures that in the right-invariant basis the $\Theta^i\otimes \partial_{y^\mu}$ components form an invertible matrix. These are exactly the almost complex structures that arise from an uplift via \eqref{eq:app_complex_structure_right_inv_basis} where it is assumed that the dreibein is invertible. We now prove that \eqref{eq:invertibility_of_J} follows from \eqref{eq:no_vertical_inv_subspace}.

We denote the components in the right invariant basis by
\begin{equation}
    J = \mathcal{J}_\mu^\nu\, \d y^\mu\otimes \partial_{y^\nu}+\mathcal{J}_i^\mu\, \Theta^i\otimes \partial_{y^\mu}+\mathcal{J}^i_\mu\,\d y^\mu\otimes  \Theta_i+\mathcal{J}_i^j \,\Theta^i \otimes \Theta_j\,.
\end{equation}
To investigate \eqref{eq:no_vertical_inv_subspace}, we first calculate $J(\tau_i^\#)$. Using the definition \eqref{eq:fundamental_vector_field_def} of fundamental vector fields and the fact that the basis $P^K$ is right invariant, we calculate
\begin{equation}
    \d y^\mu(\tau_i^\#) = \tau_i^\#(y^\mu(p))= \frac{d}{dt} y^\mu(\mathcal{R}_{e^{t\tau_i}}(p))\bigg\vert_{t=0}= \frac{d}{dt} y^\mu(p)\bigg\vert_{t=0}=0\,,
\end{equation}
where we used that $y^\mu(\mathcal{R}_{u}(p)) = y^\mu(p)$. Furthermore, using the transformation $\widetilde{G}(p) = G(\mathcal{R}_u(p))= G(p)\,u $ of $G$ under right-action, we find
\begin{equation}
    \Theta(\tau_i^\#) =  \frac{\d}{\d t} G(\mathcal{R}_{e^{t\tau_i}}(p))\, G^{-1}(p)\bigg\vert_{t=0}= G (p)\,\frac{\d}{\d t} e^{t\tau_i} \bigg\vert_{t=0}G^{-1} (p)= G\, \tau_i\,G^{-1} = (\Lambda^{-1})^j_{\;\;i}\tau_j\,.
\end{equation}
Therefore, we have $\Theta^j(\tau_i^\#) = (\Lambda^{-1})^j_{\;\;i}$. It follows that in the right invariant basis, the fundamental vector fields take the form
\begin{equation}
    \tau_i^\# = (\Lambda^{-1})^j_{\;\;i}\,\Theta_j\,.
\end{equation}
Thus, we obtain
\begin{equation}
    J(\tau_i^\#) = (\Lambda^{-1})^j_{\;\;i} J(\Theta_j) = (\Lambda^{-1})^j_{\;\;i}\,\mathcal{J}_j^\mu\, \partial_{y^\mu}+(\Lambda^{-1})^j_{\;\;i}\,\mathcal{J}_j^k\, \Theta_k\,.
\end{equation}
We notice that 
\begin{equation}
    \pi_* \Theta_i = \Lambda^j_{\;\; i}\, \pi_* \tau_j^\#=0\,,\quad \pi_*\partial_{y^\mu} = \partial_{y^\mu}\,,\label{eq:pullback_right_invariant_base}
\end{equation}
so that we obtain
\begin{equation}
    \pi_* J(\tau_i^\#) =(\Lambda^{-1})^j_{\;\;i}\mathcal{J}_j^\mu\, \partial_{y^\mu}\,.
\end{equation}
Since we can expand any vertical vector $a^\#$ as $a^\#=a^i\tau_i^\#$, the condition \eqref{eq:no_vertical_inv_subspace} requires that $(\Lambda^{-1})^j_{\;\;i}\mathcal{J}_j^\mu\,a^i\,\partial_{y^\mu}\neq 0$ for arbitrary $a^i$, i.e.~$(\Lambda^{-1})^j_{\;\;i}\mathcal{J}_j^\mu$ has a trivial kernel. Thus the product is invertible. Since $(\Lambda^{-1})^j_{\;\;i}$ is invertible by construction, this implies that $\mathcal{J}_i^\mu$ is invertible i.e.~there exists $\mathcal{J}^i_\mu$ such that $\mathcal{J}^i_\mu\, \mathcal{J}^\mu_j = \delta_j^i$. Thus, we have proven \eqref{eq:invertibility_of_J}.

\paragraph{Derivation of \eqref{eq:alpha_no_inv_subpace}.}
Now, we turn to discuss the implications of \eqref{eq:no_vertical_inv_subspace} on a complex structure deformation. We denote by $J'$ of the form \eqref{eq:Jprime_explicitly} an integrable deformation of the complex structure $J$, which we assume to satisfy \eqref{eq:no_vertical_inv_subspace}. We will now derive the condition for $\alpha$ under which $J'$ also satisfies \eqref{eq:no_vertical_inv_subspace}. For any vertical vector $v\in V$,
\begin{equation}
    0 \neq \pi_* J'(v) = \pi_* J(v)+\pi_* \alpha(v)\,,
\end{equation}
where we use the shorthand
\begin{equation}
    \alpha \equiv 2i\,(\alpha_{j}^{\;\;\,i}\,d\overbar{w}^{j}\otimes \partial_i-\overbar{\alpha}_{j}^{\;\;\,i}\,dw^{j}\otimes \overbar{\partial}_i)
    \label{eq:alpha_shorthand}
\end{equation}
in the holomorphic coordinates of the background complex structure $J$. The above equation is linear, so we can instead write
\begin{equation}
    \pi_* \alpha(\tau_i^\#) \neq - \pi_* J (\tau_i^\#) = - \pi_* \mathcal{E}_i\,.
\end{equation}
Here, we used $\mathcal{W}_i = \tau_i^\#$ and $J(\mathcal{W}_i)= \mathcal{E}_i$. Using \eqref{eq:vielbein_right_invariant_basis} and \eqref{eq:pullback_right_invariant_base}, we calculate
\begin{equation}
    \pi_* \mathcal{E}_i = (\Lambda^{-1})_i^{\;\; j}E_j^\mu\, \partial_{y^\mu}\,.
\end{equation}
Expanding $\alpha$ as
\begin{equation}
    \alpha = (\alpha_1)_j^i\, \mathcal{W}^j\otimes\mathcal{E}_i +(\alpha_2)_j^i\, \mathcal{E}^j\otimes\mathcal{W}_i +(\alpha_3)_j^i\,  \mathcal{E}^j\otimes\mathcal{E}_i+(\alpha_4)_j^i \,\mathcal{W}^j\otimes\mathcal{W}_i
    \label{eq:alpha_W_E}
\end{equation}
and using \eqref{eq:Wi_Theta} - \eqref{eq:vielbein_right_invariant_basis} together with \eqref{eq:pullback_right_invariant_base}, we find 
\begin{equation}
    \pi_* \alpha(\tau_i^\#) \neq  - \pi_* \mathcal{E}_i\,\quad\Leftrightarrow\quad (\alpha_1)^i_j \neq-\delta_j^i\,.\label{eq:condition_alpha}
\end{equation}
In order to translate this inequality into the holomorphic coordinates of the background complex structure, we invert \eqref{eq:connection_sigma_def} and use \eqref{eq:dual_vector_fields_W_E} to obtain
\begin{equation}
    \partial_i = \frac{1}{2}\,\Sigma_i^j\,(\mathcal{W}_j-i\mathcal{E}_j)\, ,\quad \overbar{\partial}_i = \frac{1}{2}\, \overbar{\Sigma}_i^j\, (\mathcal{W}_j+i\mathcal{E}_j)\,.
\end{equation}
Similarly, using \eqref{eq:A_forms_W_E}, we find
\begin{equation}
    \d w^i = \sigma^i_j\, (\mathcal{W}^j+i \mathcal{E}^j)\,,\quad \d \overbar{w}^i = \overbar{\sigma}^i_j\, (\mathcal{W}^j-i \mathcal{E}^j)\,.
\end{equation}
Substituting to \eqref{eq:alpha_shorthand} and comparing with \eqref{eq:alpha_W_E}, we obtain
\begin{equation}
    (\alpha_1)_j^i = \Sigma_l^i\,\overbar{\sigma}^k_j\,\alpha^{\;\;\,l}_k\,  +\overbar{\Sigma}_l^i\,\sigma^k_j\,\overbar{\alpha}^{\;\;\,l}_k = 2\,\text{Re}\,(\Sigma_l^i\,\overbar{\sigma}^k_j\,\alpha^{\;\;\,l}_k)\,. 
\end{equation}
Thus, using \eqref{eq:condition_alpha} we arrive at the condition
\begin{equation}
    2\,\text{Re}\,(\Sigma_l^i\,\overbar{\sigma}^k_j\,\alpha^{\;\;\,l}_k) \neq -\delta_j^i\,.
\end{equation}
If $\alpha$ satisfies this condition and is integrable, the deformed complex structure $J'$ satisfies \eqref{eq:no_vertical_inv_subspace}.

\end{appendix}

\addcontentsline{toc}{section}{References}
\bibliography{bib.bib}
\bibliographystyle{JHEP}

\end{document}